\documentclass[preprintnumbers,amsmath,amssymb,nofootinbib]{revtex4}
\usepackage{graphicx,color}

\begin{document}

\begin{center}
{\Large {\bf  Comprehensive Calculations on the OZI-forbidden
Nonleptonic Decays of Orthoquarkonia $J/\psi(\Upsilon)\to \pi\pi,
\rho\pi$}}
\end{center}

\vspace*{0.3cm}
\begin{center}
{Tong Li$^1$, Shu-Min Zhao$^{1,2}$ and Xue-Qian
Li$^1$}\\
\vspace*{0.3cm}
$^1${\it Department of Physics, Nankai University, Tianjin 300071, China}\\

$^2${\it Department of Physics and Technology, Hebei University,
Baoding 071002, China}
\end{center}

\vspace*{0.5cm}

\begin{center}
\begin{minipage}{12cm}

\noindent Abstract:\\
In this work, we calculate the decay rates of the OZI-forbidden
processes $J/\psi(\Upsilon)\to \pi\pi, \rho\pi$ at the order of
the leading-twist distribution amplitude. The process of
$J/\psi(\Upsilon)\to \pi^+ \pi^-$ violates isospin conservation
and the amplitude is explicitly proportional to the isospin
violation factor $m_u-m_d$, our numerical results on their decay
rates are consistent with the data. The process
$J/\psi(\Upsilon)\to \rho\pi$ violates the hadronic helicity
conservation and should be suppressed, as indicated in literature,
its decay rate can only be proportional to $m_q^2$ at the order of
leading twist. Our theoretical evaluation confirms this statement
that the theoretical evaluation on $\Gamma(J/\psi(\Upsilon)\to
\rho\pi)$ is almost one order smaller than the data unless the
model parameters take certain extreme values. It may imply that
the sizable branching ratio of $J/\psi(\Upsilon)\to \rho\pi$
should be explained by either higher twist contributions or other
mechanisms.
\end{minipage}

\end{center}
\newpage

\section{Introduction}
It is generally believed that the narrowness of the ground states
of heavy quarkonia $J/\psi$ and $\Upsilon$ is due to the so called
OZI suppression\cite{OZI}. This OZI rule demands that if there are
no quark lines connecting the initial and final hadron states, the
processes are suppressed. At beginning, it seemed to be a
phenomenological principle, however, further studies indicate that
the suppression may originate from the loop suppression which can
be precisely evaluated in the framework of perturbative QCD. More
than 20 years ago, the OZI-suppressed radiative decays of
orthoquarkonia was investigated by K\"{o}rner et al. in
perturbative QCD\cite{Korner}, where reasonable approximations
were adopted. Since then, technique for calculating loop diagrams
has been greatly improved and knowledge on the wavefunctions of
light mesons is much enriched. Meanwhile more data have been
accumulated and the corresponding experimental measurements become
more precise\cite{exp1,exp2}, all the experimental progress indeed
provides us with a possibility to test our theoretical framework
where the perturbative and non-perturbative effects are factorized
and a convolution integral over them results in the physical
transition amplitude. Following their work, we have also
re-calculated the rates of $J/\psi(\Upsilon)\rightarrow
\gamma+\pi^0(\eta,\;\eta')$ which are respectively
isospin-violated, flavor-SU(3)-violated and flavor-SU(3)-favored
processes without any approximations at one-loop
level\cite{RevisitOZI}.

In fact, there may exist other possible mechanisms which also
contribute to the concerned processes of $J/\psi(\Upsilon)\to PP$
and $VP$ where $P$ and $V$ stand for pseudoscalar and vector
mesons respectively\cite{Close,HadronLoop,Chang}, therefore to
fully understand such reactions, a complete calculation on the
OZI-suppressed non-leptonic decay processes is obviously necessary
and should be possible with our present knowledge. Comparing with
the radiative decays, theoretical evaluation of the rate of the
non-leptonic decays is much more complicated. In the radiative
decays, a photon is emitted as a free particle escaping away from
the reaction and it does not participate in strong interaction.
For the non-leptonic decay, the two (at least) daughter hadrons
tangle together by exchanging gluons, therefore one not only needs
to carry out the complicated Feynman integrations of four-point
and five-point loop functions (i.e. D- and E-functions), but also
there are more Feynman diagrams than the radiative decays. In this
work we obtain the transition amplitude by carefully calculating
the loop integrations. Following the standard
procedure\cite{loop}, one can reduce the 5-point loop functions
into 4-point and 3-point loop functions which are then evaluated
in terms of the program "LoopTools"\cite{Dfunction,LoopTools}.
Moreover, one needs to carefully handle the color factors whereas
they are much simpler in the radiative decays.

In this work, we are going to make a full calculation on the
OZI-suppressed processes of $J/\psi(\Upsilon)\rightarrow \pi\pi$
and $J/\psi(\Upsilon)\rightarrow\rho\pi$ at the order of leading
twist.

The reason to only consider $\pi^{\pm,0}$ and $\rho^{\pm,0}$ as
the produced pseudoscalar and vector mesons is following. The
processes are non-leptonic decays, at least three hadrons are
involved and to theoretically evaluate the rates, one not only
needs to calculate the complicated loop integrations at
quark-gluon level, but also have to deal with the hadronic matrix
elements which are fully governed by the non-perturbative QCD
effects. However, at present, a completely reliable way to
calculate the non-perturbative QCD effects is lacking, so that
some phenomenological models must be invoked. In the decays of
$J/\psi(\Upsilon)\rightarrow\pi\pi,\rho\pi$, the product mesons
are light and can be nicely described in terms of the light-cone
distribution amplitudes. Since $\pi$ and $\rho$ are composed of
only $u,d$ and $\bar u,\bar d$ whose masses are approximately
equal, due to the obvious symmetry, the distribution functions are
more symmetric and reliable, at least for the leading twist order.
By contraries, for the distribution functions of $K(K^*)$, $\eta$
and $\eta'$, the produced mesons are composed of constituents
$u(d)$ and $s$ quarks which have a large difference in mass, thus
one would expect larger uncertainties in the evaluation of
hadronic matrix elements. Therefore, in this work, these final
states are not concerned.

A simple analysis indicates that $J/\psi(\Upsilon)\rightarrow
\pi\pi$ is an isospin-violating process. Namely the pions are
treated as identical particles once the isospin symmetry is
adopted in the analysis. Concretely, by the conservation of
angular momentum, the two pions are in the p-wave state, since
pions are identical bosons, the wave function of the two-pion
system must be totally symmetric, so that the isospin of the
system should be 1 as
$${1\over\sqrt 2}(|1,1\rangle|1,-1\rangle-|1,-1\rangle|1,1\rangle)\equiv {1\over \sqrt
2}(|\pi^+\rangle|\pi^-\rangle-|\pi^-\rangle|\pi^+\rangle).$$ That
requires that the process of $J/\psi\rightarrow \pi^0\pi^0$ is
strictly forbidden. The isospin violation effects are expressed in
the mass difference of $u$ and $d$ quarks which appears at the
loop calculations, and the factor $m_u-m_d$ will be explicitly
shown in the expressions derived at the quark level. Even though,
we only consider the leading twist contribution of the light-cone
wave functions which are independent of the quark masses, we still
count in the mass splitting which results in the isospin
violation. Turn to the processes
$J/\psi(\Upsilon)\rightarrow\rho\pi$, in contrast with the
$\pi\pi$ case, $\rho$ and $\pi$ are not identical particles,
therefore the anti-symmetry requirement which enforces the
two-pion system to be in isospin 1 state, is dismissed, so that
the $\rho\pi$ system can be in isospin 0 state and it guarantees
the iso-spin conservation for the decay process
$J/\psi(\Upsilon)\rightarrow \rho^0\pi^0$. This observation seems
to demand that the branching ratio of $J/\psi(\Upsilon)\to\rho\pi$
should be much larger than the isospin violating process
$J/\psi(\Upsilon)\rightarrow \pi\pi$ and the data indeed support
this statement. Moreover, in $J/\psi(\Upsilon)\rightarrow
\rho\pi$, the isospin 0 state of $\rho\pi$ is dominant, so that
the branching ratios of $J/\psi(\Upsilon)\rightarrow \rho^0\pi^0$
and $J/\psi(\Upsilon)\rightarrow \rho^+\pi^-+\rho^-\pi^+$ roughly
retain a relation of 1:2. However, as indicated in
Refs.\cite{Brodsky,HQP}, such processes violate the hadronic
helicity conservation because gluons and photon do not carry
hadronic helicities. A non-zero theoretical prediction on the rate
at the order of leading twist must come from a violation of the
hadronic helicity conservation. It is indicated that such a
violation is proportional to the light quark mass, therefore one
can expect that the directly calculated OZI suppressed amplitude
should be proportional to $m_q^2(q=u,d)$. Our calculation confirms
this mechanism (see the text for details).

The data\cite{PDG} tell us an opposite conclusion that the
helicity-violated process $J/\psi(\Upsilon)\rightarrow \rho\pi$
has sizable branching ratio and almost is one of the dominant
modes in $J/\psi$ and $\Upsilon$ decays. The discrepancy should be
explained, some suggestions that the next-to-leading twist
contribution, higher Fock states and other mechanisms such as the
hadronic loop and glueball intermediate states etc. are taken into
account, are proposed.

In this work, we only concern the OZI-suppressed processes for
$J/\psi(\Upsilon)\rightarrow \pi\pi, \; \rho\pi$, and will
explicitly demonstrate the isospin conserving and violating
effects and the helicity violation effects in the formulation.

As indicated above, to evaluate the hadronic matrix elements, one
has to deal with a convolution integrals over the distribution
amplitudes of the concerned hadrons. Because $J/\psi$ and $\Upsilon$
contain two heavy constituents, their bound-state effects can be
simply expressed in terms of the wave functions at origin which can
be easily obtained from the data of their leptonic decays. The
distribution functions of the two produced light hadrons might cause
uncertainties, even though as indicated above, for $\pi$ and $\rho$
mesons, they can be reduced to minimum. It is interesting to note
that the OZI-suppressed process for $J/\psi(\Upsilon)\rightarrow
\rho\pi$ is forbidden by the hadronic helicity conservation at the
leading twist, if the quark mass is neglected at the loop
calculations. However, it is not zero and the transition amplitudes
of such processes must be proportional to $m_q^2$. In this work we
only consider the contribution from the leading twist distribution
amplitudes of the mesons and show that as $m_q\to 0$, the amplitudes
would approaches zero, in other words, we confirm the statement that
the hadronic helicity conservation forbids the process
$J/\psi(\Upsilon)\to \rho\pi$ at the leading twist if quark mass is
neglected.

Following the literature, we can trust the calculations to a
relatively accurate level. A rough numerical estimate by changing
the input parameters in the distribution functions and its forms
given in literature, shows that the error can be of order of a few
tens percents.

After this introduction, we give all the formulas where we carry
out the four- and five-point Feynamn integrals to obtain the
hard-scattering amplitude at the quark level. The isospin
violation factor $m_u-m_d$ explicitly shows up for
$J/\psi(\Upsilon)\rightarrow \pi\pi$, and for the
helicity-violated $J/\psi(\Upsilon)\rightarrow \rho\pi$ the
amplitude is also proportional to the light-quark masses, then one
needs to convolute the hard kernel with the initial and final
states, and the convolution integration results in the physical
transition amplitude in Sec. II. In Sec. III, we carefully analyze
the infrared behavior in $J/\psi(\Upsilon) \to \pi\pi,\;\rho\pi$
and convince ourselves that all Feynman diagrams are infrared-safe
when the end-point behaviors of the wave functions are considered.
In Sec. IV, we make a numerical evaluation of the decay rates of
$J/\psi(\Upsilon)\rightarrow \pi^+\pi^-$ and $\rho^\pm\pi^\mp$ and
some necessary input parameters are explicitly given. The last
section is devoted to a simple discussion on the uncertainties in
our calculation and possible contributions to these processes from
other mechanisms and then draw our conclusion. Some tedious
details are collected in the appendices.

\section{Theoretical calculation on the rates of $J/\psi(\Upsilon) \to \pi \pi, \rho \pi$}

In this work, without invoking the so-called weak-binding
approximation which was adopted in literature\cite{Korner}, we
explicitly keep the masses of the heavy and light quarks at the
concerned propagators, when derive the transition amplitudes. The
amplitude is written as
\begin{eqnarray}
\mathcal{A}&=&
H\otimes\Phi_{J/\psi(\Upsilon)}\otimes\Phi_{P_1}\otimes\Phi_{P_2}\nonumber
\\
H&=&C\otimes \widetilde{H}
\end{eqnarray}
where the factors $C$, $\widetilde{H}$,
$\Phi_{J/\psi(\Upsilon),P_1,P_2}$ are the color factor, hard
kernel and distribution amplitudes of mesons, respectively. And
the labels $P_1, P_2$ denote the two produced mesons in the final
state. Indeed, here the perturbative and non-perturbative parts
are factorized and a convolution integral would associate them to
result in the physical amplitude. The detailed expressions of the
hard kernels are given in Appendix A.

Below, in Fig. 1, we present the relevant Feynman diagrams. In
these figures, we only explicitly draw the typical diagrams. There
exit their topologically deformed diagrams which are obtained by
exchanging the connections of the gluon-lines in the loop to the
light-quark-gluon vertices, namely the two gluon-lines cross with
each other. We do not explicitly show them in Fig. 1 just for
simplicity. But definitely, in our derivation the contributions
from those diagrams are included.
\begin{figure}[!htb]
\begin{center}
\begin{tabular}{cc}
\includegraphics[width=10cm]{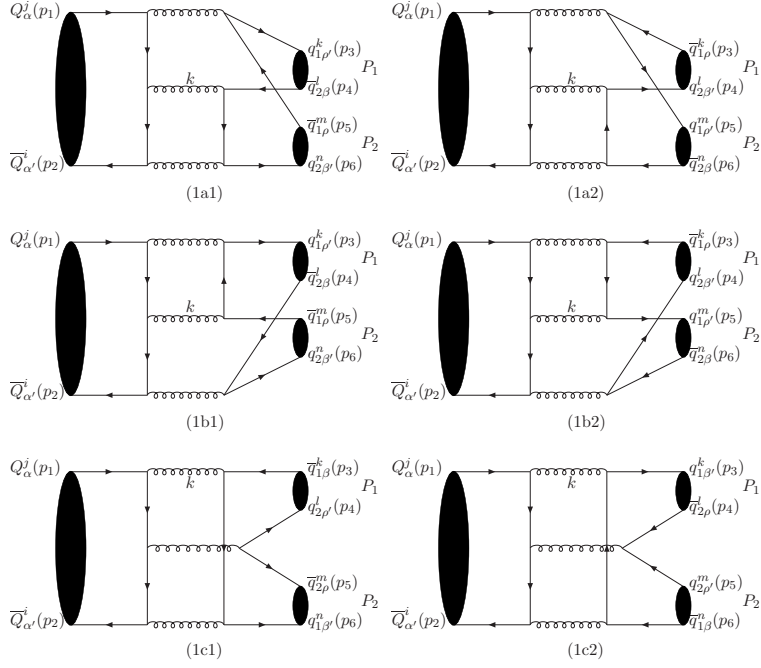}
\end{tabular}
\end{center}
\label{figure} \caption{The Feynman diagrams of $J/\psi(\Upsilon)
\to P_1 P_2$. Our calculation also includes such diagrams which
are topologically deformed from that shown above by exchanging the
connections of the gluon-lines in the loops to the
gluon-light-quark vertices, where the two gluon-lines cross with
each other.} \label{fig1}
\end{figure}

The amplitude of $J/\psi(\Upsilon) \to P_1 P_2$ can be divided
into three categories which correspond to Fig. \ref{fig1} (1a1)
and (1a2), (1b1) and (1b2), (1c1) and (1c2) respectively. To make
the text succinct we collect the detailed expressions of the
amplitudes in Appendix B, where the diagrams with gluon-lines in
the loops crossing each other are labelled as (2a1), (2a2), (2b1),
(2b2), (2c1) and (2c2) respectively.

Generally, the transition amplitude for $J/\psi(\Upsilon)\to
\pi^+\pi^-, \rho^+\pi^-, \rho^-\pi^+$ can be written as:
\begin{eqnarray}
\mathcal{A}^{J/\psi(\Upsilon)\to \pi^+\pi^-}&=&\sum_i
\mathcal{A}^{i}(P_1,P_2,m_q),\nonumber \\
\mathcal{A}^{J/\psi(\Upsilon)\to
\rho^+\pi^-(\rho^-\pi^+)}&=&\sum_i (\mathcal{A}^{ia}(P_1\to
\rho,P_2\to \pi,m_q)+\mathcal{A}^{ib}(P_1\to \pi,P_2\to
\rho,m_q)),
\end{eqnarray}
where summing over $i$ means including all diagrams listed above
and their topologically deformed diagrams which were depicted
above. For $\pi^+\pi^-$ final states one possible setting is that
$P_1$, $P_2$ correspond to $\pi^+$, $\pi^-$ respectively, and
another possibility of interchanging $\pi^+$ and $\pi^-$ is also
included in the sum. $m_q$ is either $m_u$ or $m_d$, which
respectively exist in different settings and their contributions
should be summed in the final amplitude. For $\rho^+\pi^-$ or
$\rho^-\pi^+$ final states interchanging $P_1$, $P_2$ would induce
obvious differences and therefore we use two new labels "a" and
"b" to distinguish the two different settings. Here we do not need
to calculate the rate of $J/\psi(\Upsilon)\rightarrow \rho^0\pi^0$
because as discussed above, it is an isospin conserving process
and the Clebsch-Gordan coefficients in I=0 state determines the
ratio of $\Gamma(J/\psi(\Upsilon)\rightarrow \rho^0\pi^0)/
\Gamma(J/\psi(\Upsilon)\rightarrow \rho\pi)$ and it should be
close to 1/3, both the data and the analysis according to the
topology of our diagrams shown in Fig. \ref{fig1} confirm it, even
though we only consider the contributions from the leading twist
distribution amplitudes of the mesons.

In the quark picture, hadrons are made of valence quarks whose
momenta-distributions are described by appropriate distribution
functions. The leading-twist distribution amplitudes of
$J/\psi(\Upsilon)$ is usually defined through the
correlator\cite{Ball}:
\begin{eqnarray}
&&\langle
0|\overline{c}_{\alpha}^i(y)c_{\beta}^{j}(x)|J/\psi(p)\rangle={\delta_{ij}\over
4N_c}\int^1_0due^{i\bar{u}p\cdot y+iup\cdot x}\times\nonumber \\
&&\left\{f_{J/\psi} m_{J/\psi} \rlap
/\varepsilon_{J/\psi}\phi_{J/\psi\parallel}(u)+{1\over
2}\sigma^{\mu'\nu'}if_{J/\psi}(\varepsilon_{J/\psi\mu'}p_{\nu'}-\varepsilon_{J/\psi\nu'}p_{\mu'})\phi_{J/\psi\perp}(u)\right\}_{\beta\alpha},
\end{eqnarray}
where $\varepsilon_{J/\psi}$ and $f_{J/\psi}$ are the polarization
vector and decay constant of $J/\psi$ respectively, and $\bar
u\equiv 1-u$. $\phi_{\parallel}$ and $\phi_{\perp}$ are the
leading-twist distribution functions corresponding to the
longitudinally and transversely polarized mesons, respectively, by
the definition given in literature\cite{Ball}.
For the case of $\Upsilon$ one only needs to replace all the
symbols of charm c into bottom b (of course as well as the
concerned parameters). The leading-twist distribution amplitude of
$\pi$ is usually defined through the correlator\cite{Beneke}:
\begin{eqnarray}
&&\langle
\pi(p')|\overline{q}_{1\alpha}^i(y)q_{2\beta}^{j}(x)|0\rangle=i{\delta_{ij}f_\pi\over
4N_c}\int^1_0due^{iup'\cdot y+i\bar{u}p'\cdot x}\left\{\rlap /p'
\gamma_5 \phi(u)\right\}_{\beta\alpha}\label{pi},
\end{eqnarray}
where $f_{\pi}$ is the decay constant of pion. And the
leading-twist distribution amplitude of $\rho$ is usually defined
through the correlator\cite{Ball}:
\begin{eqnarray}
&&\langle
\rho(p')|\overline{q}_{1\alpha}^i(y)q_{2\beta}^{j}(x)|0\rangle={\delta_{ij}\over
4N_c}\int^1_0due^{iup'\cdot y+i\bar{u}p'\cdot x}\times\nonumber \\
&&\left\{f_{\rho} m_{\rho} \rlap
/\varepsilon_{\rho}^\ast\phi_{\rho\parallel}(u)-{1\over
2}\sigma^{\mu'\nu'}if_{\rho}^T(\varepsilon_{\rho\mu'}^\ast
p'_{\nu'}-\varepsilon_{\rho\nu'}^\ast
p'_{\mu'})\phi_{\rho\perp}(u)\right\}_{\beta\alpha}\label{rho}
\end{eqnarray}
where $\varepsilon_{\rho}$ and $f_{\rho}$, $f_{\rho}^T$ are the
polarization vector and decay constant of $\rho$ respectively.

\section{Analysis on the infrared behaviors in $J/\psi(\Upsilon) \to P_1 P_2$}
It is definitely demanded that a reasonable theoretical prediction
on any practical process must be infrared safe, namely the
infrared divergence must be exactly cancels if it exists or
properly dealt with at the end of the calculation which
corresponds to real physical measurable quantities, such as the
decay width and cross section. In this work, we will explicitly
show that in the framework of perturbative QCD, the infrared
behavior of each individual Feynman diagram shown in Fig.
\ref{fig1} is benign, even though at first glimpse it seems to be
divergent.

There are several typical Feynman diagrams shown in Fig.
\ref{fig1}. We take the amplitude of Fig. \ref{fig1} (1a2) for
$J/\psi \to P_1 P_2$ as an example to analyze the infrared
behavior.

Its contribution to the transition amplitude reads
\begin{eqnarray}
&&\mathcal{A}^{1a2}(m_q,u,v)=H^{1a2}(m_q,u,v)\otimes\Phi_{P_1}(u)\otimes\Phi_{P_2}(v).\label{A1a2},
\end{eqnarray}
and $H^{1a2}(m_q,u,v)$ is given in Appendix A. The concerned
factor of the amplitude is
\begin{eqnarray}
{1\over
k^2(k+p_4+p_6)^2[(k+p_4)^2-m_q^2][(k+p_1-p_3-p_5)^2-m_Q^2](p_3+p_5)^2[(p_1-p_3-p_5)^2-m_Q^2]}.\label{deno}
\end{eqnarray}
Firstly, if there exists an infrared divergence in the loop
integration, it must come from the kinematic region $k\to 0$ and
the end-points of the distribution functions, therefore one only
needs to analyze two cases: (1) $k\to 0$, (2) at end-points.

To show the infrared behavior of the amplitude after integrating
out the loop function in the case (1), we may fix the external
momenta of the quarks and antiquarks in final states by a special
choice $p_3=p_4={1\over 2}p_{_{P_1}}, p_5=p_6={1\over
2}p_{_{P_2}}$ to avoid possible endpoint divergence. Looking at
the expression (\ref{deno}), as $m_q\neq 0$, the dangerous part is
proportional to $1/k^2$ which is finite after integrating over the
loop momentum $d^4k$, so that in this case there is no infrared
divergence coming from the loop integration.

Secondly, in case (2), when the momentum of one quark(antiquark)
in each of the final mesons is close to its endpoint, for example
$p_4,\; p_6\to 0$, while the other quark (antiquark) takes almost
all the momentum of the meson. It is observed that the factor
${1\over (p_3+p_5)^2}$ does not contribute a divergence. The
dangerous term comes from the factor $k^2(k+p_4+p_6)^2$ at the
denominator as $k\to 0$ and $p_4,\; p_6\to 0$, which seems to
cause a logarithmic divergence. However as we convolute the
amplitude with the distribution functions of the two mesons whose
distribution functions linearly approach to 0($\phi(u), \phi(v)\to
0$) at the end-points, because it turns to zero fasters than the
logarithmically divergent factor, i.e. $\lim_{u\to 0}u\ln u\to 0$,
the infrared behavior is safe.

Finally, when $p_3,\; p_5\to 0$, the loop integration does not
produce any divergence by the same sake of the first case. It is
noted, that the factor ${1\over (p_3+p_5)^2}\phi(u)\phi(v)$ is
finite, but there exists a subtlety. Namely in general the limit
depends on the ways how $x, y$ approach to 0 more or less. For
$J/\psi(\Upsilon)\rightarrow \pi\pi$ there are two possible
settings for each diagram, namely, an interchange
$\pi^{\pm}\leftrightarrow \pi^{\mp}$ brings one setting to
another, and their contributions have an opposite sign due to the
SU(2) symmetry and cancels each other (obviously, for the finite
term, their contributions cannot cancel each other due to the
SU(2) breaking i.e. $m_u\neq m_d$). Thus the dependence on the
order of limits disappears. By contraries, for
$J/\psi(\Upsilon)\to \rho\pi$, there is no such a cancellation, so
that even though infrared divergence does not exist, the final
numerical results somehow depend on the order of taking limits of
$u$ and $v$ approaching to zero. The strategy we adopt in this
work is to set the integration order as we integrate over $u$ and
then $v$ and it can be treated as a regularization scheme similar
to that we generally adopt for treating the ultraviolet
divergence.

\section{Numerical results}

The input parameters which we are going to use in the numerical
computations are \cite{RevisitOZI,PDG,Ball,data1,data2}:
$f_{J/\psi} = 551$ MeV, $f_\Upsilon= 710$ MeV, $f_\pi = 131$ MeV,
$f_\rho = 198$ MeV, $f_\rho^T = 160$ MeV, $m_{J/\psi} = 3096.87$
MeV, $m_\Upsilon=9460.3$ MeV, $m_{\pi^{\pm}} = 139.57$ MeV,
$m_{\rho^{\pm}} = 775.5$ MeV, $\alpha_s(m_c) = 0.32$,
$\alpha_s(m_b) = 0.21$, $m_c = 1300$ MeV, $m_b=4500$ MeV, and the
meson distribution functions respectively. For the numerical
evaluations, in Eqs.(\ref{pi}, \ref{rho}), we adopt three
different distribution functions for the light mesons in the
literatures\cite{Beneke,wave1,wave2,wave3} as $\phi_1(x) =
6x(1-x), \phi_2(x) = 30x^2(1-x)^2, \phi_3(x) = {15\over
2}(1-2x)^2[1-(1-2x)^2]$, and also let the current quark masses of
the u and d types vary within a reasonable range.

Below in Tables I, II, III and IV, we present our numerical
results on the decay rates of $J/\psi\rightarrow \pi^+\pi^-,\;
\rho^+\pi^-+\rho^-\pi^+$ and $\Upsilon \rightarrow \pi^+\pi^-,\;
\rho^+\pi^-+\rho^-\pi^+$ respectively.

\begin{table}[h]
\caption{Decay widths ($\Gamma$) of $J/\psi\to \pi^+ \pi^-$ based
on the three distribution functions, $\phi_1$, $\phi_2$ and
$\phi_3$, respectively}
\begin{center}
\begin{tabular}{|c|c|c|c|c|c|}
  \hline
  $m_u$(MeV) & $m_d$(MeV) & $\Gamma(\phi_1)$(MeV) & $\Gamma(\phi_2)$(MeV) & $\Gamma(\phi_3)$(MeV) & exp(MeV) \\
  \hline
  2 & 4 & $4.52\times 10^{-5}$ & $2.98\times 10^{-5}$ & $2.71\times 10^{-4}$ &  \\
  3 & 4 & $1.88\times 10^{-5}$ & $9.35\times 10^{-6}$ & $5.67\times 10^{-5}$ &  \\
  3 & 5 & $3.17\times 10^{-5}$ & $2.36\times 10^{-5}$ & $1.25\times 10^{-4}$ &  $(1.37\pm 0.21)\times 10^{-5}$\\
  4 & 5 & $1.03\times 10^{-5}$ & $8.12\times 10^{-6}$ & $4.26\times 10^{-5}$ &  \\
  4.5 & 6 & $2.29\times 10^{-5}$ & $1.43\times 10^{-5}$ & $8.85\times 10^{-5}$ &  \\
  \hline
\end{tabular}
\end{center}
\end{table}

\begin{table}[h]
\caption{Decay widths ($\Gamma$) of $\Upsilon\to \pi^+ \pi^-$
based on the three distribution functions, $\phi_1$, $\phi_2$ and
$\phi_3$, respectively}
\begin{center}
\begin{tabular}{|c|c|c|c|c|c|}
  \hline
  $m_u$(MeV) & $m_d$(MeV) & $\Gamma(\phi_1)$(MeV) & $\Gamma(\phi_2)$(MeV) & $\Gamma(\phi_3)$(MeV) & exp(MeV) \\
  \hline
  2 & 4 & $2.79\times 10^{-6}$ & $1.24\times 10^{-6}$ & $1.13\times 10^{-5}$ &  \\
  3 & 4 & $8.16\times 10^{-7}$ & $5.28\times 10^{-7}$ & $6.95\times 10^{-6}$ &  \\
  3 & 5 & $1.23\times 10^{-6}$ & $9.43\times 10^{-7}$ & $9.6\times 10^{-6}$ &  $<2.7\times 10^{-5}$\\
  4 & 5 & $7.39\times 10^{-7}$ & $2.72\times 10^{-7}$ & $5.11\times 10^{-6}$ &  \\
  4.5 & 6 & $9.78\times 10^{-7}$ & $7.5\times 10^{-7}$ & $8.93\times 10^{-6}$ &  \\
  \hline
\end{tabular}
\end{center}
\end{table}

\begin{table}[h]
\caption{Decay widths ($\Gamma$) of $J/\psi\to \pi^+ \rho^- +
\pi^- \rho^+$ based on the three distribution functions, $\phi_1$,
$\phi_2$ and $\phi_3$, respectively}
\begin{center}
\begin{tabular}{|c|c|c|c|c|c|}
  \hline
  $m_u$(MeV) & $m_d$(MeV) & $\Gamma(\phi_1)$(MeV) & $\Gamma(\phi_2)$(MeV) & $\Gamma(\phi_3)$(MeV) & exp(MeV) \\
  \hline
  2 & 2 & $1.04\times 10^{-4}$ & $7.21\times 10^{-5}$ & $5.11\times 10^{-4}$ &  \\
  3 & 3 & $2.36\times 10^{-4}$ & $1.6\times 10^{-4}$ & $1.17\times 10^{-3}$ &  \\
  4 & 4 & $4.12\times 10^{-4}$ & $2.9\times 10^{-4}$ & $2.08\times 10^{-3}$ &  $(1.06\pm 0.08)\times 10^{-3}$\\
  5 & 5 & $6.69\times 10^{-4}$ & $4.54\times 10^{-4}$ & $3.38\times 10^{-3}$ &  \\
  6 & 6 & $9.75\times 10^{-4}$ & $6.68\times 10^{-4}$ & $4.88\times 10^{-3}$ &  \\
  \hline
\end{tabular}
\end{center}
\end{table}

\begin{table}[h]
\caption{Decay widths ($\Gamma$) of $\Upsilon\to \pi^+ \rho^- +
\pi^- \rho^+$ based on the three distribution functions, $\phi_1$,
$\phi_2$ and $\phi_3$, respectively}
\begin{center}
\begin{tabular}{|c|c|c|c|c|c|}
  \hline
  $m_u$(MeV) & $m_d$(MeV) & $\Gamma(\phi_1)$(MeV) & $\Gamma(\phi_2)$(MeV) & $\Gamma(\phi_3)$(MeV) & exp(MeV) \\
  \hline
  2 & 2 & $2.23\times 10^{-6}$ & $1.56\times 10^{-6}$ & $8.54\times 10^{-6}$ &  \\
  3 & 3 & $5.04\times 10^{-6}$ & $3.67\times 10^{-6}$ & $1.84\times 10^{-5}$ &  \\
  4 & 4 & $8.95\times 10^{-6}$ & $6.25\times 10^{-6}$ & $3.4\times 10^{-5}$ &  $<1.08\times 10^{-5}$\\
  5 & 5 & $1.42\times 10^{-5}$ & $9.93\times 10^{-6}$ & $5.44\times 10^{-5}$ &  \\
  6 & 6 & $1.84\times 10^{-5}$ & $1.38\times 10^{-5}$ & $7.61\times 10^{-5}$ &  \\
  \hline
\end{tabular}
\end{center}
\end{table}

As discussed above the OZI-suppressed process $J/\psi(\Upsilon)\to
\pi^+ \pi^-$ is isospin-violated whereas $J/\psi(\Upsilon)\to
\pi^+ \rho^- + \pi^- \rho^+$ violates the hadronic helicity
conservation. It is noted that the theoretically evaluated values
on the OZI-suppressed processes $J/\psi\to \pi^+ \pi^-$ are
slightly larger than the experimental data depending on the
parameter choices such as $m_u,$ $m_d$ and types of the meson
distribution functions, whereas that on $J/\psi\to \pi^+ \rho^- +
\pi^- \rho^+$ are one order smaller than the data. It may imply
that some other mechanisms may also contribute to the decays and
we will remark on the results in next section.

\section{Discussion and Conclusion}

In this work, we calculate the contributions of the so-called OZI
forbidden processes to the decays $J/\psi(\Upsilon)\rightarrow
\pi\pi, \rho\pi$. As we discussed in the introduction, the process
$J/\psi(\Upsilon)\rightarrow \pi\pi$ is an isospin violating
reaction, whereas $J/\psi(\Upsilon)\rightarrow \rho\pi$ is an
isospin conserving one, on other aspect, the former one conserves
the hadronic helicity whereas the latter one  violates it. Our
numerical results on $J/\psi(\Upsilon)\to \pi^+ \pi^-$ are
reasonably consistent with the data at order of magnitude, but the
evaluated branching ratio of $J/\psi(\Upsilon)\rightarrow \rho\pi$
is obviously smaller than data by one order.

As shown in Tables I through IV, one can notice that the results
deviate from each other in a wider range as one adopts different
wave functions which all are suggested in literatures, as well as
the light-quark masses.

$J/\psi(\Upsilon)\rightarrow \pi\pi$ is an isospin violating
process and at the leading twist the OZI-suppressed process which
was supposed to be the main contribution to the mode of
$J/\psi(\Upsilon)\rightarrow \rho\pi$ violates the hadronic
helicity conservation. As well known, the source of isospin
violation can be either from a photon emission (absorption) and/or
quark mass difference, and for the helicity violating processes,
the decay width is proportional to $m_q^2$, so that both of the
processes are somehow sensitive to the light quark masses and
much suppressed. Our formulas explicitly show that as $m_q\to 0$,
the decay widths for both modes approach zero. This observation
confirms the above statements.

In this work, we only include the contributions from the
leading-twist distribution amplitude and our results confirm that
due to violation of helicity conservation, the theoretical
evaluated ratio is one order of magnitude smaller than the data.
It is also noted from our qualitative analysis that the rate of
isospin-violated process $J/\psi(\Upsilon)\to \pi\pi$ should be
proportional to the square of mass difference $(m_u-m_d)^2$,
whereas rate of the the hadronic helicity-violating process
$J/\psi(\Upsilon)\to \rho\pi$ is proportional to $(m_u+m_d)^2$,
i.e. it is natural to expect that $\Gamma(J/\psi(\Upsilon)\to
\rho\pi)$ which is theoretically estimated in this framework, is a
few times larger than $\Gamma(J/\psi(\Upsilon)\to \pi\pi)$, our
numerical results shown in Tables I through IV confirm this
statement, and if $m_u=m_d$, the estimated
$\Gamma(J/\psi(\Upsilon)\to \pi\pi)$ is zero, whereas
$\Gamma(J/\psi(\Upsilon)\to \rho\pi)$ is not. But this still does
not explain the largeness of the branching ratio of
$J/\psi(\Upsilon)\to \rho\pi$. It is indicated in
Refs.\cite{Brodsky,HQP}, the large branching ratio might be due to
higher twist contributions. Therefore it seems that to correctly
evaluate the branching ratio, in principle one needs to include
the contributions from higher twist distribution amplitudes in the
evaluation.

On other side, besides the contributions from higher twist
distribution amplitudes, there may exist other mechanisms which
may result in larger branching ratios for
$J/\psi(\Upsilon)\rightarrow \rho\pi$. As suggested by Suzuki
paper\cite{Suzuki} and our earlier work\cite{HadronLoop}, there
can be a contribution from the hadronic loops and by fitting data
(in the paper, the contributions from the OZI-forbidden processes
were not theoretically calculated as we do in this work, but
obtained by fitting data), we reached two conclusions that if only
the two mechanisms contributing, the hadronic loop contribution
would have the same order of magnitudes as that of the
OZI-forbidden processes (definitely including higher twist
contributions) and secondly the two contributions are destructive.

Of course, it may not be the end of the story that some authors
also suggested a glueball contribution which should be added to
that from the aforementioned mechanisms\cite{Close}, and then the
picture becomes more complicated, because we are unable to
reliably estimate the glueball mass and phenomenological behaviors
so far, unless we can borrow the lattice results. Therefore
further developments on theory are necessary.

Uncertainties in our theoretical evaluations come from the input
parameters, especially the light quark masses and the shapes of
the distribution functions while only the leading-twist
distribution amplitudes are accounted. One can note that the
shapes of the wave functions would cause order of magnitude
differences. So far, we still cannot really rule out any of them,
but wait for more accurate data to determine.

As indicated in the text, we only consider the processes of
$J/\psi(\Upsilon)\rightarrow \pi\pi,\;\rho\pi$ because as there no
strange flavor gets involved, the wave functions of the produced
mesons is simpler and more symmetric. For the processes involving
such as $K(K^*),\;\eta,\;\eta'$, the calculations become more
complicated and the results are not much reliable. Therefore we
postpone our study on such processes in our later works.

So far, the experimental data are not accurate yet, especially for
the measurements on $\Upsilon$ decays only upper limits are set.
However, we are inspired by the promises from the CLEO$_\text{c}$
and BES III collaborations, as they will provide a much larger
database on $J/\psi$ decays, and more data would be accumulated in
the B-meson factories, and then we will have concrete numbers
about the branching ratios of $\Upsilon\rightarrow
\pi\pi,\;\rho\pi$ instead of the upper limits set by the present
experimental measurements. Moreover, the LHC$_\text{b}$ and future
ILC can much enrich our knowledge on hadron structure. Conclusion
is definite that further work is necessary.

\noindent {\bf Acknowledgments}:

This work is supported by the National
Natural Science Foundation of China (NNSFC).\\

\noindent{\bf Appendix A: The hard-scattering amplitudes
$H^{i,\alpha'\beta'\gamma'\rho'\alpha\beta\gamma\rho}(m_q,u,v)$}\\

The hard-scattering amplitude corresponding to Fig. \ref{fig1}
(1a1) is:
\begin{eqnarray}
&&H^{1a1,\alpha\alpha'\beta\beta'\rho\rho'}(m_q,u,v)=\nonumber
\\
&&\int{d^4k\over
(2\pi)^4}\overline{Q}_{\alpha'}^i[(-ig_sT^a_{is}\gamma^\nu){i\over
\rlap /k+\rlap /p_1-\rlap /p_3-\rlap
/p_5-m_Q}(-ig_sT^b_{sr}\gamma^\mu){i\over \rlap /p_1-\rlap
/p_3-\rlap
/p_5-m_Q}(-ig_sT^c_{rj}\gamma^\lambda)]_{\alpha'\alpha}Q_\alpha^j\nonumber
\\
&& \overline{q}_{2\beta'}^n[(-ig_sT^a_{nw}\gamma_\nu){i\over
-\rlap /k-\rlap
/p_4-m_q}(-ig_sT^b_{wl}\gamma_\mu)]_{\beta'\beta}q_{2\beta}^l
\overline{q}_{1\rho'}^k[(-ig_sT^c_{km}\gamma_\lambda)]_{\rho'\rho}q_{1\rho}^m{-i\over
(p_3+p_5)^2}{-i\over k^2}{-i\over (k+p_4+p_6)^2}\nonumber
\\
\end{eqnarray}

The hard-scattering amplitude corresponding to Fig. \ref{fig1}
(1a2) is:
\begin{eqnarray}
&&H^{1a2,\alpha\alpha'\beta\beta'\rho\rho'}(m_q,u,v)=\nonumber
\\
&&\int{d^4k\over
(2\pi)^4}\overline{Q}_{\alpha'}^i[(-ig_sT^a_{is}\gamma^\nu){i\over
\rlap /k+\rlap /p_1-\rlap /p_3-\rlap
/p_5-m_Q}(-ig_sT^b_{sr}\gamma^\mu){i\over \rlap /p_1-\rlap
/p_3-\rlap
/p_5-m_Q}(-ig_sT^c_{rj}\gamma^\lambda)]_{\alpha'\alpha}Q_\alpha^j\nonumber
\\
&&\overline{q}_{2\beta'}^l[(-ig_sT^b_{lw}\gamma_\mu){i\over \rlap
/k+\rlap
/p_4-m_q}(-ig_sT^a_{wn}\gamma_\nu)]_{\beta'\beta}q_{2\beta}^n
\overline{q}_{1\rho'}^m[(-ig_sT^c_{mk}\gamma_\lambda)]_{\rho'\rho}q_{1\rho}^k{-i\over
(p_3+p_5)^2}{-i\over k^2}{-i\over (k+p_4+p_6)^2}\nonumber
\\
\end{eqnarray}

The hard-scattering amplitude corresponding to Fig. \ref{fig1}
(1b1) is:
\begin{eqnarray}
&&H^{1b1,\alpha\alpha'\beta\beta'\rho\rho'}(m_q,u,v)=\nonumber
\\
&&\int{d^4k\over
(2\pi)^4}\overline{Q}_{\alpha'}^i[(-ig_sT^a_{is}\gamma^\nu){i\over
\rlap /p_4+\rlap /p_6-\rlap
/p_2-m_Q}(-ig_sT^b_{sr}\gamma^\mu){i\over \rlap /k+\rlap
/p_4+\rlap /p_6-\rlap
/p_2-m_Q}(-ig_sT^c_{rj}\gamma^\lambda)]_{\alpha'\alpha}Q_\alpha^j\nonumber
\\
&&\overline{q}_{2\beta'}^n[(-ig_sT^a_{nl}\gamma_\nu)]_{\beta'\beta}q_{2\beta}^l
\overline{q}_{1\rho'}^k[(-ig_sT^c_{kw}\gamma_\lambda){i\over \rlap
/k-\rlap
/p_5-m_q}(-ig_sT^b_{wm}\gamma_\mu)]_{\rho'\rho}q_{1\rho}^m{-i\over
(p_4+p_6)^2}{-i\over k^2}{-i\over (k-p_3-p_5)^2}\nonumber
\\
\end{eqnarray}

The hard-scattering amplitude corresponding to Fig. \ref{fig1}
(1b2) is:
\begin{eqnarray}
&&H^{1b2,\alpha\alpha'\beta\beta'\rho\rho'}(m_q,u,v)=\nonumber
\\
&&\int{d^4k\over
(2\pi)^4}\overline{Q}_{\alpha'}^i[(-ig_sT^a_{is}\gamma^\nu){i\over
\rlap /p_4+\rlap /p_6-\rlap
/p_2-m_Q}(-ig_sT^b_{sr}\gamma^\mu){i\over \rlap /k+\rlap
/p_4+\rlap /p_6-\rlap
/p_2-m_Q}(-ig_sT^c_{rj}\gamma^\lambda)]_{\alpha'\alpha}Q_\alpha^j\nonumber
\\
&&\overline{q}_{2\beta'}^l[(-ig_sT^a_{ln}\gamma_\nu)]_{\beta'\beta}q_{2\beta}^n
\overline{q}_{1\rho'}^m[(-ig_sT^b_{mw}\gamma_\mu){i\over -\rlap
/k+\rlap
/p_5-m_q}(-ig_sT^c_{wk}\gamma_\lambda)]_{\rho'\rho}q_{1\rho}^k{-i\over
(p_4+p_6')^2}{-i\over k^2}{-i\over (k-p_3-p_5)^2}\nonumber
\\
\end{eqnarray}

The hard-scattering amplitude corresponding to Fig. \ref{fig1}
(1c1) is:
\begin{eqnarray}
&&H^{1c1,\alpha\alpha'\beta\beta'\rho\rho'}(m_q,u,v)=\nonumber
\\
&&\int{d^4k\over
(2\pi)^4}\overline{Q}_{\alpha'}^i[(-ig_sT^a_{is}\gamma^\nu){i\over
\rlap /k+\rlap /p_1-\rlap /p_4-\rlap
/p_5-m_Q}(-ig_sT^b_{sr}\gamma^\mu){i\over \rlap /k+\rlap
/p_1-m_Q}(-ig_sT^c_{rj}\gamma^\lambda)]_{\alpha'\alpha}Q_\alpha^j\nonumber
\\
&&\overline{q}_{2\beta'}^n[(-ig_sT^a_{nw}\gamma_\nu){i\over -\rlap
/k-\rlap
/p_3-m_q}(-ig_sT^c_{wk}\gamma_\lambda)]_{\beta'\beta}q_{2\beta}^k
\overline{q}_{1\rho'}^l[(-ig_sT^b_{lm}\gamma_\mu)]_{\rho'\rho}q_{1\rho}^m\nonumber
\\
&&{-i\over (p_4+p_5)^2}{-i\over k^2}{-i\over
(k+p_1+p_2-p_4-p_5)^2}
\end{eqnarray}

The hard scattering amplitude corresponding to Fig. \ref{fig1}
(1c2) is:
\begin{eqnarray}
&&H^{1c2,\alpha\alpha'\beta\beta'\rho\rho'}(m_q,u,v)=\nonumber
\\
&&\int{d^4k\over
(2\pi)^4}\overline{Q}_{\alpha'}^i[(-ig_sT^a_{is}\gamma^\nu){i\over
\rlap /k+\rlap /p_1-\rlap /p_4-\rlap
/p_5-m_Q}(-ig_sT^b_{sr}\gamma^\mu){i\over \rlap /k+\rlap
/p_1-m_Q}(-ig_sT^c_{rj}\gamma^\lambda)]_{\alpha'\alpha}Q_\alpha^j\nonumber
\\
&&\overline{q}_{2\beta'}^k[(-ig_sT^c_{kw}\gamma_\lambda){i\over
\rlap /k+\rlap
/p_3-m_q}(-ig_sT^a_{wn}\gamma_\nu)]_{\beta'\beta}q_{2\beta}^n
\overline{q}_{1\rho'}^m[(-ig_sT^b_{ml}\gamma_\mu)]_{\rho'\rho}q_{1\rho}^l\nonumber
\\
&&{-i\over (p_4+p_5)^2}{-i\over k^2}{-i\over
(k+p_1+p_2-p_4-p_5)^2}
\end{eqnarray}

The hard-scattering amplitude corresponding to the diagram which
is topologically deformed  from Fig. \ref{fig1} (1a1) by
exchanging the connection of the gluon-lines in the loop to the
gluon-light-quark vertices, is
\begin{eqnarray}
&&H^{2a1,\alpha\alpha'\beta\beta'\rho\rho'}(m_q,u,v)=\nonumber
\\
&&\int{d^4k\over
(2\pi)^4}\overline{Q}_{\alpha'}^i[(-ig_sT^a_{is}\gamma^\nu){i\over
\rlap /k+\rlap /p_1-\rlap /p_3-\rlap
/p_5-m_Q}(-ig_sT^b_{sr}\gamma^\mu){i\over \rlap /p_1-\rlap
/p_3-\rlap
/p_5-m_Q}(-ig_sT^c_{rj}\gamma^\lambda)]_{\alpha'\alpha}Q_\alpha^j\nonumber
\\
&& \overline{q}_{2\beta'}^n[(-ig_sT^b_{nw}\gamma_\mu){i\over \rlap
/k+\rlap
/p_6-m_q}(-ig_sT^a_{wl}\gamma_\nu)]_{\beta'\beta}q_{2\beta}^l
\overline{q}_{1\rho'}^k[(-ig_sT^c_{km}\gamma_\lambda)]_{\rho'\rho}q_{1\rho}^m{-i\over
(p_3+p_5)^2}{-i\over k^2}{-i\over (k+p_4+p_6)^2}\nonumber
\\
\end{eqnarray}

The hard-scattering amplitude corresponding to the topologically
deformed diagram from Fig. \ref{fig1} (1a2) is:
\begin{eqnarray}
&&H^{2a2,\alpha\alpha'\beta\beta'\rho\rho'}(m_q,u,v)=\nonumber
\\
&&\int{d^4k\over
(2\pi)^4}\overline{Q}_{\alpha'}^i[(-ig_sT^a_{is}\gamma^\nu){i\over
\rlap /k+\rlap /p_1-\rlap /p_3-\rlap
/p_5-m_Q}(-ig_sT^b_{sr}\gamma^\mu){i\over \rlap /p_1-\rlap
/p_3-\rlap
/p_5-m_Q}(-ig_sT^c_{rj}\gamma^\lambda)]_{\alpha'\alpha}Q_\alpha^j\nonumber
\\
&& \overline{q}_{2\beta'}^l[(-ig_sT^a_{lw}\gamma_\nu){i\over
-\rlap /k-\rlap
/p_6-m_q}(-ig_sT^b_{wn}\gamma_\mu)]_{\beta'\beta}q_{2\beta}^n
\overline{q}_{1\rho'}^m[(-ig_sT^c_{mk}\gamma_\lambda)]_{\rho'\rho}q_{1\rho}^k{-i\over
(p_3+p_5)^2}{-i\over k^2}{-i\over (k+p_4+p_6)^2}\nonumber
\\
\end{eqnarray}

The hard-scattering amplitude corresponding to the deformed
diagram from Fig. \ref{fig1} (1b1) is:
\begin{eqnarray}
&&H^{2b1,\alpha\alpha'\beta\beta'\rho\rho'}(m_q,u,v)=\nonumber
\\
&&\int{d^4k\over
(2\pi)^4}\overline{Q}_{\alpha'}^i[(-ig_sT^a_{is}\gamma^\nu){i\over
\rlap /p_4+\rlap /p_6-\rlap
/p_2-m_Q}(-ig_sT^b_{sr}\gamma^\mu){i\over \rlap /k+\rlap
/p_4+\rlap /p_6-\rlap
/p_2-m_Q}(-ig_sT^c_{rj}\gamma^\lambda)]_{\alpha'\alpha}Q_\alpha^j\nonumber
\\
&&\overline{q}_{2\beta'}^n[(-ig_sT^a_{nl}\gamma_\nu)]_{\beta'\beta}q_{2\beta}^l
\overline{q}_{1\rho'}^k[(-ig_sT^b_{kw}\gamma_\mu){i\over -\rlap
/k+\rlap
/p_3-m_q}(-ig_sT^c_{wm}\gamma_\lambda)]_{\rho'\rho}q_{1\rho}^m{-i\over
(p_4+p_6)^2}{-i\over k^2}{-i\over (k-p_3-p_5)^2}\nonumber
\\
\end{eqnarray}

The hard-scattering amplitude corresponding to the deformed
diagram from Fig. \ref{fig1} (1b2) is:
\begin{eqnarray}
&&H^{2b2,\alpha\alpha'\beta\beta'\rho\rho'}(m_q,u,v)=\nonumber
\\
&&\int{d^4k\over
(2\pi)^4}\overline{Q}_{\alpha'}^i[(-ig_sT^a_{is}\gamma^\nu){i\over
\rlap /p_4+\rlap /p_6-\rlap
/p_2-m_Q}(-ig_sT^b_{sr}\gamma^\mu){i\over \rlap /k+\rlap
/p_4+\rlap /p_6-\rlap
/p_2-m_Q}(-ig_sT^c_{rj}\gamma^\lambda)]_{\alpha'\alpha}Q_\alpha^j\nonumber
\\
&&\overline{q}_{2\beta'}^l[(-ig_sT^a_{ln}\gamma_\nu)]_{\beta'\beta}q_{2\beta}^n
\overline{q}_{1\rho'}^m[(-ig_sT^c_{mw}\gamma_\lambda){i\over \rlap
/k-\rlap
/p_3-m_q}(-ig_sT^b_{wk}\gamma_\mu)]_{\rho'\rho}q_{1\rho}^k{-i\over
(p_4+p_6)^2}{-i\over k^2}{-i\over (k-p_3-p_5)^2}\nonumber
\\
\end{eqnarray}

The hard-scattering amplitude corresponding to the deformed
diagram from Fig. \ref{fig1} (1c1) is:
\begin{eqnarray}
&&H^{2c1,\alpha\alpha'\beta\beta'\rho\rho'}(m_q,u,v)=\nonumber
\\
&&\int{d^4k\over
(2\pi)^4}\overline{Q}_{\alpha'}^i[(-ig_sT^a_{is}\gamma^\nu){i\over
\rlap /k+\rlap /p_1-\rlap /p_4-\rlap
/p_5-m_Q}(-ig_sT^b_{sr}\gamma^\mu){i\over \rlap /k+\rlap
/p_1-m_Q}(-ig_sT^c_{rj}\gamma^\lambda)]_{\alpha'\alpha}Q_\alpha^j\nonumber
\\
&&\overline{q}_{2\beta'}^n[(-ig_sT^c_{nw}\gamma_\lambda){i\over
\rlap /k+\rlap
/p_6-m_q}(-ig_sT^a_{wk}\gamma_\nu)]_{\beta'\beta}q_{2\beta}^k
\overline{q}_{1\rho'}^l[(-ig_sT^b_{lm}\gamma_\mu)]_{\rho'\rho}q_{1\rho}^m\nonumber
\\
&&{-i\over (p_4+p_5)^2}{-i\over k^2}{-i\over
(k+p_1+p_2-p_4-p_5)^2}
\end{eqnarray}

The hard-scattering amplitude corresponding to the deformed
diagram from Fig. \ref{fig1} (1c2) is:
\begin{eqnarray}
&&H^{2c2,\alpha\alpha'\beta\beta'\rho\rho'}(m_q,u,v)=\nonumber
\\
&&\int{d^4k\over
(2\pi)^4}\overline{Q}_{\alpha'}^i[(-ig_sT^a_{is}\gamma^\nu){i\over
\rlap /k+\rlap /p_1-\rlap /p_4-\rlap
/p_5-m_Q}(-ig_sT^b_{sr}\gamma^\mu){i\over \rlap /k+\rlap
/p_1-m_Q}(-ig_sT^c_{rj}\gamma^\lambda)]_{\alpha'\alpha}Q_\alpha^j\nonumber
\\
&&\overline{q}_{2\beta'}^k[(-ig_sT^a_{kw}\gamma_\nu){i\over -\rlap
/k-\rlap
/p_6'-m_q}(-ig_sT^c_{wn}\gamma_\lambda)]_{\beta'\beta}q_{2\beta}^n
\overline{q}_{1\rho'}^m[(-ig_sT^b_{ml}\gamma_\mu)]_{\rho'\rho}q_{1\rho}^l\nonumber
\\
&&{-i\over (p_4+p_5)^2}{-i\over k^2}{-i\over
(k+p_1+p_2-p_4-p_5)^2}
\end{eqnarray}

\noindent{\bf Appendix B: The amplitudes
$\mathcal{A}^i(m_q,u,v)$}\\

\textbf{1. For $\mathbf{J/\psi\to P P}$}

For amplitudes $\mathcal{A}^{1a1}$ and $\mathcal{A}^{1a2}$, we
have
\begin{eqnarray}
\mathcal{A}^{1a1}&=&C^{1a1}\widetilde{H}^{1a1}(m_q,u,v)\Phi_{J/\psi}\Phi_{P1}\Phi_{P2}\nonumber \\
\mathcal{A}^{1a2}&=&C^{1a2}\widetilde{H}^{1a2}(m_q,u,v)\Phi_{J/\psi}\Phi_{P1}\Phi_{P2}
\end{eqnarray}
with
\begin{eqnarray}
C^{1a1}&=&\text{Tr}(T^aT^bT^c)\text{Tr}(T^aT^bT^c)\nonumber \\
C^{1a2}&=&\text{Tr}(T^aT^bT^c)\text{Tr}(T^bT^aT^c)\nonumber \\
\widetilde{H}^{1a1}(m_q,u,v)&=&-\widetilde{H}^{1a2}(m_q,u,v)=-{i\pi^2\over
(2\pi)^4}g_s^6({1\over 4N_C})^3{1\over
(p_3+p_5)^2[(p_1-p_3-p_5)^2-m_Q^2]}\nonumber
\\
&&\{D_0(m_q,u,v)[-32m_{J/\psi}\varepsilon_{J/\psi}\cdot
p_{_{P1}}p_1'\cdot
p_{_{P2}}m_Q^2+96m_{J/\psi}\varepsilon_{J/\psi}\cdot
p_{_{P2}}p_1'\cdot p_{_{P1}}m_Q^2\nonumber \\
&&-32m_{J/\psi}\varepsilon_{J/\psi}\cdot p_1' p_{_{P1}}\cdot
p_{_{P2}}m_Q^2+32m_{J/\psi}\varepsilon_{J/\psi}\cdot
p_{_{P1}}p_1'\cdot
p_{_{P2}}p_3'^2-32m_{J/\psi}\varepsilon_{J/\psi}\cdot
p_{_{P2}}p_1'\cdot p_{_{P1}}p_3'^2\nonumber
\\
&&-64m_{J/\psi}\varepsilon_{J/\psi}\cdot p_{_{P1}}p_1'\cdot p_3'
p_3'\cdot p_{_{P2}}+64m_{J/\psi}\varepsilon_{J/\psi}\cdot p_3'
p_1'\cdot p_{_{P1}} p_3'\cdot p_{_{P2}}\nonumber
\\
&&-128m_{J/\psi}\varepsilon_{J/\psi}\cdot p_{_{P2}} p_1'\cdot p_3'
p_3'\cdot p_{_{P1}}-64m_{J/\psi}\varepsilon_{J/\psi}\cdot p_1'
p_3'\cdot p_{_{P2}} p_3'\cdot p_{_{P1}}\nonumber
\\
&&+32m_{J/\psi}\varepsilon_{J/\psi}\cdot p_1' p_3'^2
p_{_{P1}}\cdot p_{_{P2}}]\nonumber \\
&&+D_\mu(m_q,u,v)[-32m_{J/\psi}\varepsilon_{J/\psi}\cdot
p_{_{P1}}p^\mu_{_{P2}}m_Q^2+96m_{J/\psi}\varepsilon_{J/\psi}\cdot
p_{_{P2}}p^\mu_{_{P1}}m_Q^2\nonumber
\\
&&-32m_{J/\psi}\varepsilon^{\mu}_{J/\psi} p_{_{P1}}\cdot
p_{_{P2}}m_Q^2-32m_{J/\psi}\varepsilon_{J/\psi}\cdot p_{_{P1}}
p_{_{P2}}^\mu p_1'\cdot p_3'-32m_{J/\psi}\varepsilon_{J/\psi}\cdot
p_{_{P2}}
p_{_{P1}}^\mu p_1'\cdot p_3'\nonumber \\
&&+32m_{J/\psi}\varepsilon_{J/\psi}\cdot p_{_{P1}} p_3'^\mu
p_1'\cdot p_{_{P2}}+32m_{J/\psi}\varepsilon_{J/\psi}\cdot p_3'
p_{_{P1}}^\mu p_1'\cdot
p_{_{P2}}-32m_{J/\psi}\varepsilon_{J/\psi}\cdot p_{_{P2}} p_3'^\mu
p_1'\cdot p_{_{P1}}\nonumber
\\
&&+32m_{J/\psi}\varepsilon_{J/\psi}\cdot p_3' p_{_{P2}}^\mu
p_1'\cdot p_{_{P1}}+32m_{J/\psi}\varepsilon_{J/\psi}\cdot p_{_{P1}}
p_{_{P2}}^\mu p_3'^2-32m_{J/\psi}\varepsilon_{J/\psi}\cdot p_{_{P2}}
p_{_{P1}}^\mu p_3'^2\nonumber
\\
&&-32m_{J/\psi}\varepsilon_{J/\psi}\cdot p_{_{P1}} p_1'^\mu
p_3'\cdot p_{_{P2}}-64m_{J/\psi}\varepsilon_{J/\psi}\cdot p_{_{P1}}
p_3'^\mu p_3'\cdot p_{_{P2}}-32m_{J/\psi}\varepsilon_{J/\psi}\cdot
p_1'
p_{_{P1}}^\mu p_3'\cdot p_{_{P2}}\nonumber \\
&&+64m_{J/\psi}\varepsilon_{J/\psi}\cdot p_3' p_{_{P1}}^\mu
p_3'\cdot p_{_{P2}}+32m_{J/\psi}\varepsilon^{\mu}_{J/\psi} p_1'\cdot
p_{_{P1}} p_3'\cdot p_{_{P2}}-96m_{J/\psi}\varepsilon_{J/\psi}\cdot
p_{_{P2}}
p_1'^\mu p_3'\cdot p_{_{P1}}\nonumber \\
&&-128m_{J/\psi}\varepsilon_{J/\psi}\cdot p_{_{P2}} p_3'^\mu
p_3'\cdot p_{_{P1}}-32m_{J/\psi}\varepsilon_{J/\psi}\cdot p_1'
p_{_{P2}}^\mu p_3'\cdot
p_{_{P1}}-32m_{J/\psi}\varepsilon^{\mu}_{J/\psi} p_1'\cdot p_{_{P2}}
p_3'\cdot p_{_{P1}}\nonumber
\\
&&-64m_{J/\psi}\varepsilon^{\mu}_{J/\psi} p_3'\cdot p_{_{P2}}
p_3'\cdot p_{_{P1}}-32m_{J/\psi}\varepsilon_{J/\psi}\cdot p_3'
p_1'^\mu p_{_{P2}}\cdot
p_{_{P1}}+32m_{J/\psi}\varepsilon_{J/\psi}\cdot p_1' p_3'^\mu
p_{_{P2}}\cdot p_{_{P1}}\nonumber
\\
&&+32m_{J/\psi}\varepsilon^{\mu}_{J/\psi} p_1'\cdot p_3'
p_{_{P2}}\cdot p_{_{P1}}+32m_{J/\psi}\varepsilon^{\mu}_{J/\psi}
p_3'^2
p_{_{P2}}\cdot p_{_{P1}}]\nonumber \\
&&+D_{\mu\nu}(m_q,u,v)[-64m_{J/\psi}\varepsilon_{J/\psi}\cdot
p_{_{P2}}p_3'^\mu
p_{_{P1}}^\nu+64m_{J/\psi}\varepsilon_{J/\psi}\cdot p_3'
p_{_{P2}}^\mu p_{_{P1}}^\nu\nonumber \\
&&-32m_{J/\psi}\varepsilon_{J/\psi}\cdot p_{_{P1}} g^{\mu\nu}
p_3'\cdot p_{_{P2}}-96m_{J/\psi}\varepsilon_{J/\psi}\cdot p_{_{P2}}
g^{\mu\nu} p_3'\cdot
p_{_{P1}}-64m_{J/\psi}\varepsilon^{\mu}_{J/\psi} p_{_{P2}}^\nu
p_3'\cdot p_{_{P1}}\nonumber
\\
&&-32m_{J/\psi}\varepsilon_{J/\psi}\cdot p_3' g^{\mu\nu}
p_{_{P2}}\cdot p_{_{P1}}+64m_{J/\psi}\varepsilon^{\mu}_{J/\psi}
p_3'^\nu
p_{_{P2}}\cdot p_{_{P1}}]\}\nonumber \\
D_0(m_q,u,v)&=&{1\over i\pi^2}\int d^4k {1\over
k^2[(k+p_4)^2-m_q^2](k+p_4+p_6)^2[(k+p_1-p_3-p_5)^2-m_Q^2]}\nonumber \\
D_\mu(m_q,u,v)&=&{1\over i\pi^2}\int d^4k {k_\mu\over
k^2[(k+p_4)^2-m_q^2](k+p_4+p_6)^2[(k+p_1-p_3-p_5)^2-m_Q^2]}\nonumber \\
D_{\mu\nu}(m_q,u,v)&=&{1\over i\pi^2}\int d^4k {k_\mu k_\nu\over
k^2[(k+p_4)^2-m_q^2](k+p_4+p_6)^2[(k+p_1-p_3-p_5)^2-m_Q^2]}
\end{eqnarray}
where $p_1'=p_4, p_3'=p_1-p_3-p_5$.

For amplitudes $\mathcal{A}^{1b1}$ and $\mathcal{A}^{1b2}$, we
have
\begin{eqnarray}
\mathcal{A}^{1b1}&=&C^{1b1}\widetilde{H}^{1b1}(m_q,u,v)\Phi_{J/\psi}\Phi_{P1}\Phi_{P2}\nonumber \\
\mathcal{A}^{1b2}&=&C^{1b2}\widetilde{H}^{1b2}(m_q,u,v)\Phi_{J/\psi}\Phi_{P1}\Phi_{P2}
\end{eqnarray}
with
\begin{eqnarray}
C^{1b1}&=&\text{Tr}(T^aT^bT^c)\text{Tr}(T^bT^aT^c)\nonumber \\
C^{1b2}&=&\text{Tr}(T^aT^bT^c)\text{Tr}(T^aT^bT^c)\nonumber \\
\widetilde{H}^{1b1}(m_q,u,v)&=&-\widetilde{H}^{1b2}(m_q,u,v)={i\pi^2\over
(2\pi)^4}g_s^6({1\over 4N_C})^3{1\over
(p_4+p_6)^2[(p_4+p_6-p_2)^2-m_Q^2]}\nonumber
\\
&&\{D_0(m_q,u,v)[96m_{J/\psi}\varepsilon_{J/\psi}\cdot
p_{_{P1}}p_1'\cdot
p_{_{P2}}m_Q^2-32m_{J/\psi}\varepsilon_{J/\psi}\cdot
p_{_{P2}}p_1'\cdot p_{_{P1}}m_Q^2\nonumber
\\
&&-32m_{J/\psi}\varepsilon_{J/\psi}\cdot p_1' p_{_{P2}}\cdot
p_{_{P1}}m_Q^2-32m_{J/\psi}\varepsilon_{J/\psi}\cdot p_{_{P1}}
p_1'\cdot p_{_{P2}} p_3'^2+32m_{J/\psi}\varepsilon_{J/\psi}\cdot
p_{_{P2}} p_1'\cdot p_{_{P1}} p_3'^2\nonumber
\\
&&-128m_{J/\psi}\varepsilon_{J/\psi}\cdot p_{_{P1}} p_1'\cdot p_3'
p_3'\cdot p_{_{P2}}-64m_{J/\psi}\varepsilon_{J/\psi}\cdot p_{_{P2}}
p_1'\cdot p_3' p_3'\cdot p_{_{P1}}\nonumber
\\
&&+64m_{J/\psi}\varepsilon_{J/\psi}\cdot p_3' p_1'\cdot p_{_{P2}}
p_3'\cdot p_{_{P1}}-64m_{J/\psi}\varepsilon_{J/\psi}\cdot p_1'
p_3'\cdot p_{_{P2}} p_3'\cdot p_{_{P1}}\nonumber
\\
&&+32m_{J/\psi}\varepsilon_{J/\psi}\cdot p_1' p_3'^2
p_{_{P2}}\cdot p_{_{P1}}]\nonumber \\
&&+D_\mu(m_q,u,v)[96m_{J/\psi}\varepsilon_{J/\psi}\cdot
p_{_{P1}}p_{_{P2}}^\mu m_Q^2-32m_{J/\psi}\varepsilon_{J/\psi}\cdot
p_{_{P2}}p_{_{P1}}^\mu m_Q^2\nonumber
\\
&&-32m_{J/\psi}\varepsilon^{\mu}_{J/\psi}p_{_{P1}}\cdot
p_{_{P2}}m_Q^2-32m_{J/\psi}\varepsilon_{J/\psi}\cdot p_{_{P1}}
p_{_{P2}}^\mu p_1'\cdot p_3'-32m_{J/\psi}\varepsilon_{J/\psi}\cdot
p_{_{P2}} p_{_{P1}}^\mu p_1'\cdot p_3'\nonumber
\\
&&-32m_{J/\psi}\varepsilon_{J/\psi}\cdot p_{_{P1}} p_3'^\mu
p_1'\cdot p_{_{P2}}+32m_{J/\psi}\varepsilon_{J/\psi}\cdot p_3'
p_{_{P1}}^\mu p_1'\cdot
p_{_{P2}}+32m_{J/\psi}\varepsilon_{J/\psi}\cdot p_{_{P2}}
p_3'^\mu p_1'\cdot p_{_{P1}}\nonumber \\
&&+32m_{J/\psi}\varepsilon_{J/\psi}\cdot p_3' p_{_{P2}}^\mu
p_1'\cdot p_{_{P1}}-32m_{J/\psi}\varepsilon_{J/\psi}\cdot p_{_{P1}}
p_{_{P2}}^\mu p_3'^2+32m_{J/\psi}\varepsilon_{J/\psi}\cdot p_{_{P2}}
p_{_{P1}}^\mu p_3'^2\nonumber
\\
&&-96m_{J/\psi}\varepsilon_{J/\psi}\cdot p_{_{P1}} p_1'^\mu
p_3'\cdot p_{_{P2}}-128m_{J/\psi}\varepsilon_{J/\psi}\cdot p_{_{P1}}
p_3'^\mu p_3'\cdot p_{_{P2}}-32m_{J/\psi}\varepsilon_{J/\psi}\cdot
p_1' p_{_{P1}}^\mu p_3'\cdot p_{_{P2}}\nonumber
\\
&&-32m_{J/\psi}\varepsilon^{\mu}_{J/\psi} p_1'\cdot p_{_{P1}}
p_3'\cdot p_{_{P2}}-32m_{J/\psi}\varepsilon_{J/\psi}\cdot p_{_{P2}}
p_1'^\mu p_3'\cdot p_{_{P1}}-64m_{J/\psi}\varepsilon_{J/\psi}\cdot
p_{_{P2}} p_3'^\mu p_3'\cdot
p_{_{P1}}\nonumber \\
&&-32m_{J/\psi}\varepsilon_{J/\psi}\cdot p_1' p_{_{P2}}^\mu
p_3'\cdot p_{_{P1}}+64m_{J/\psi}\varepsilon_{J/\psi}\cdot p_3'
p_{_{P2}}^\mu p_3'\cdot
p_{_{P1}}+32m_{J/\psi}\varepsilon^{\mu}_{J/\psi} p_1'\cdot p_{_{P2}}
p_3'\cdot p_{_{P1}}\nonumber
\\
&&-64m_{J/\psi}\varepsilon^{\mu}_{J/\psi} p_3'\cdot p_{_{P2}}
p_3'\cdot p_{_{P1}}-32m_{J/\psi}\varepsilon_{J/\psi}\cdot p_3'
p_1'^\mu p_{_{P2}}\cdot
p_{_{P1}}+32m_{J/\psi}\varepsilon_{J/\psi}\cdot p_1' p_3'^\mu
p_{_{P2}}\cdot p_{_{P1}}\nonumber \\
&&+32m_{J/\psi}\varepsilon^{\mu}_{J/\psi} p_1'\cdot p_3'
p_{_{P2}}\cdot p_{_{P1}}+32m_{J/\psi}\varepsilon^{\mu}_{J/\psi}
p_3'^2
p_{_{P2}}\cdot p_{_{P1}}]\nonumber \\
&&+D_{\mu\nu}(m_q,u,v)[-64m_{J/\psi}\varepsilon_{J/\psi}\cdot
p_{_{P1}}p_3'^\mu
p_{_{P2}}^\nu+64m_{J/\psi}\varepsilon_{J/\psi}\cdot p_3'
p_{_{P2}}^\mu p_{_{P1}}^\nu\nonumber \\
&&-96m_{J/\psi}\varepsilon_{J/\psi}\cdot p_{_{P1}} g^{\mu\nu}
p_3'\cdot
p_{_{P2}}-64m_{J/\psi}\varepsilon^{\mu}_{J/\psi}p_{_{P1}}^\nu
p_3'\cdot p_{_{P2}}-32m_{J/\psi}\varepsilon_{J/\psi}\cdot p_{_{P2}}
g^{\mu\nu} p_3'\cdot p_{_{P1}}\nonumber
\\
&&-32m_{J/\psi}\varepsilon_{J/\psi}\cdot p_3' g^{\mu\nu}
p_{_{P2}}\cdot
p_{_{P1}}+64m_{J/\psi}\varepsilon^{\mu}_{J/\psi}p_3'^\nu
p_{_{P1}}\cdot p_{_{P2}}]\}\nonumber \\
D_0(m_q,u,v)&=&{1\over i\pi^2}\int d^4k {1\over
k^2[(k-p_5)^2-m_q^2](k-p_3-p_5)^2[(k+p_4+p_6-p_2)^2-m_Q^2]}\nonumber \\
D_\mu(m_q,u,v)&=&{1\over i\pi^2}\int d^4k {k_\mu\over
k^2[(k-p_5)^2-m_q^2](k-p_3-p_5)^2[(k+p_4+p_6-p_2)^2-m_Q^2]}\nonumber \\
D_{\mu\nu}(m_q,u,v)&=&{1\over i\pi^2}\int d^4k {k_\mu k_\nu\over
k^2[(k-p_5)^2-m_q^2](k-p_3-p_5)^2[(k+p_4+p_6-p_2)^2-m_Q^2]}
\end{eqnarray}
where $p_1'=-p_5, p_3'=p_4+p_6-p_2$.

For amplitudes $\mathcal{A}^{1c1}$ and $\mathcal{A}^{1c2}$, we
have
\begin{eqnarray}
\mathcal{A}^{1c1}&=&C^{1c1}\widetilde{H}^{1c1}(m_q,u,v)\Phi_{J/\psi}\Phi_{P1}\Phi_{P2}\nonumber \\
\mathcal{A}^{1c2}&=&C^{1c2}\widetilde{H}^{1c2}(m_q,u,v)\Phi_{J/\psi}\Phi_{P1}\Phi_{P2}
\end{eqnarray}
with
\begin{eqnarray}
&&C^{1c1}=\text{Tr}(T^aT^bT^c)\text{Tr}(T^bT^aT^c)\nonumber \\
&&C^{1c2}=\text{Tr}(T^aT^bT^c)\text{Tr}(T^aT^bT^c)\nonumber \\
&&\widetilde{H}^{1c1}(m_q,u,v)=-\widetilde{H}^{1c2}(m_q,u,v)=-{i\pi^2\over
(2\pi)^4}g_s^6({1\over 4N_C})^3{1\over (p_4+p_5)^2}\nonumber
\\
&&\{E_0(m_q,u,v)[-32m_{J/\psi}\varepsilon_{J/\psi}\cdot p_{_{P1}}
p_4'\cdot p_{_{P2}}m_Q^2-32m_{J/\psi}\varepsilon_{J/\psi}\cdot
p_{_{P2}} p_4'\cdot p_{_{P1}}m_Q^2\nonumber \\
&&+96m_{J/\psi}\varepsilon_{J/\psi}\cdot p_4' p_{_{P2}}\cdot
p_{_{P1}}m_Q^2-32m_{J/\psi}\varepsilon_{J/\psi}\cdot p_{_{P2}}
p_1'\cdot p_{_{P1}} p_2'\cdot p_4'\nonumber
\\
&&-32m_{J/\psi}\varepsilon_{J/\psi}\cdot p_{_{P1}} p_2'\cdot
p_{_{P2}} p_1'\cdot p_4'-96m_{J/\psi}\varepsilon_{J/\psi}\cdot p_4'
p_1'\cdot p_{_{P1}} p_2'\cdot p_{_{P2}}\nonumber
\\
&&-32m_{J/\psi}\varepsilon_{J/\psi}\cdot p_{_{P2}} p_1'\cdot p_4'
p_2'\cdot p_{_{P1}}-32m_{J/\psi}\varepsilon_{J/\psi}\cdot p_4'
p_1'\cdot p_{_{P2}} p_2'\cdot p_{_{P1}}\nonumber \\
&&+32m_{J/\psi}\varepsilon_{J/\psi}\cdot p_{_{P1}} p_1'\cdot p_2'
p_4'\cdot p_{_{P2}}-32m_{J/\psi}\varepsilon_{J/\psi}\cdot p_2'
p_1'\cdot p_{_{P1}} p_4'\cdot p_{_{P2}}\nonumber \\
&&+32m_{J/\psi}\varepsilon_{J/\psi}\cdot p_1' p_2'\cdot p_{_{P1}}
p_4'\cdot p_{_{P2}}+32m_{J/\psi}\varepsilon_{J/\psi}\cdot p_{_{P2}}
p_1'\cdot p_2' p_4'\cdot p_{_{P1}}\nonumber \\
&&+32m_{J/\psi}\varepsilon_{J/\psi}\cdot p_2' p_1'\cdot p_{_{P2}}
p_4'\cdot p_{_{P1}}-32m_{J/\psi}\varepsilon_{J/\psi}\cdot p_1'
p_2'\cdot p_{_{P2}} p_4'\cdot p_{_{P1}}\nonumber \\
&&-32m_{J/\psi}\varepsilon_{J/\psi}\cdot p_4' p_1'\cdot p_2'
p_{_{P2}}\cdot p_{_{P1}}+32m_{J/\psi}\varepsilon_{J/\psi}\cdot p_2'
p_1'\cdot p_4' p_{_{P2}}\cdot p_{_{P1}}\nonumber
\\
&&+32m_{J/\psi}\varepsilon_{J/\psi}\cdot p_1' p_2'\cdot p_4'
p_{_{P2}}\cdot p_{_{P1}}-32m_{J/\psi}\varepsilon_{J/\psi}\cdot
p_{_{P1}} p_1'\cdot p_{_{P2}} p_2'\cdot p_4']\nonumber \\
&&+E_\mu(m_q,u,v)[-32m_{J/\psi}\varepsilon_{J/\psi}\cdot p_{_{P1}}
p_{_{P2}}^\mu m_Q^2-32m_{J/\psi}\varepsilon_{J/\psi}\cdot p_{_{P2}}
p_{_{P1}}^\mu
m_Q^2\nonumber \\
&&+96m_{J/\psi}\varepsilon^{\mu}_{J/\psi}p_{_{P1}}\cdot
p_{_{P2}}m_Q^2+32m_{J/\psi}\varepsilon_{J/\psi}\cdot p_{_{P1}}
p_{_{P2}}^\mu p_1'\cdot p_2'+32m_{J/\psi}\varepsilon_{J/\psi}\cdot
p_{_{P2}}
p_{_{P1}}^\mu p_1'\cdot p_2'\nonumber \\
&&-32m_{J/\psi}\varepsilon_{J/\psi}\cdot p_{_{P1}} p_{_{P2}}^\mu
p_1'\cdot p_4'-32m_{J/\psi}\varepsilon_{J/\psi}\cdot p_{_{P2}}
p_{_{P1}}^\mu p_1'\cdot p_4'-32m_{J/\psi}\varepsilon_{J/\psi}\cdot
p_{_{P1}} p_2'^\mu
p_1'\cdot p_{_{P2}}\nonumber \\
&&-32m_{J/\psi}\varepsilon_{J/\psi}\cdot p_{_{P1}} p_4'^\mu
p_1'\cdot p_{_{P2}}+32m_{J/\psi}\varepsilon_{J/\psi}\cdot p_2'
p_{_{P1}}^\mu p_1'\cdot
p_{_{P2}}-32m_{J/\psi}\varepsilon_{J/\psi}\cdot p_4'
p_{_{P1}}^\mu p_1'\cdot p_{_{P2}}\nonumber \\
&&-32m_{J/\psi}\varepsilon_{J/\psi}\cdot p_{_{P2}} p_2'^\mu
p_1'\cdot p_{_{P1}}-32m_{J/\psi}\varepsilon_{J/\psi}\cdot p_{_{P2}}
p_4'^\mu p_1'\cdot p_{_{P1}}-32m_{J/\psi}\varepsilon_{J/\psi}\cdot
p_2'
p_{_{P2}}^\mu p_1'\cdot p_{_{P1}}\nonumber \\
&&-96m_{J/\psi}\varepsilon_{J/\psi}\cdot p_4' p_{_{P2}}^\mu
p_1'\cdot p_{_{P1}}-32m_{J/\psi}\varepsilon_{J/\psi}\cdot p_{_{P1}}
p_{_{P2}}^\mu p_2'\cdot p_4'-32m_{J/\psi}\varepsilon_{J/\psi}\cdot
p_{_{P2}}
p_{_{P1}}^\mu p_2'\cdot p_4'\nonumber \\
&&-32m_{J/\psi}\varepsilon_{J/\psi}\cdot p_{_{P1}} p_1'^\mu
p_2'\cdot p_{_{P2}}-32m_{J/\psi}\varepsilon_{J/\psi}\cdot p_{_{P1}}
p_4'^\mu p_2'\cdot p_{_{P2}}-32m_{J/\psi}\varepsilon_{J/\psi}\cdot
p_1'
p_{_{P1}}^\mu p_2'\cdot p_{_{P2}}\nonumber \\
&&-96m_{J/\psi}\varepsilon_{J/\psi}\cdot p_4' p_{_{P1}}^\mu
p_2'\cdot p_{_{P2}}-96m_{J/\psi}\varepsilon^{\mu}_{J/\psi} p_1'\cdot
p_{_{P1}} p_2'\cdot p_{_{P2}}-32m_{J/\psi}\varepsilon_{J/\psi}\cdot
p_{_{P2}} p_1'^\mu p_2'\cdot
p_{_{P1}}\nonumber \\
&&-32m_{J/\psi}\varepsilon_{J/\psi}\cdot p_{_{P2}} p_4'^\mu
p_2'\cdot p_{_{P1}}+32m_{J/\psi}\varepsilon_{J/\psi}\cdot p_1'
p_{_{P2}}^\mu p_2'\cdot
p_{_{P1}}-32m_{J/\psi}\varepsilon_{J/\psi}\cdot p_4'
p_{_{P2}}^\mu p_2'\cdot p_{_{P1}}\nonumber \\
&&-32m_{J/\psi}\varepsilon^{\mu}_{J/\psi} p_1'\cdot p_{_{P2}}
p_2'\cdot p_{_{P1}}+32m_{J/\psi}\varepsilon_{J/\psi}\cdot p_{_{P1}}
p_1'^\mu p_4'\cdot p_{_{P2}}+32m_{J/\psi}\varepsilon_{J/\psi}\cdot
p_{_{P1}}
p_2'^\mu p_4'\cdot p_{_{P2}}\nonumber \\
&&+32m_{J/\psi}\varepsilon_{J/\psi}\cdot p_1' p_{_{P1}}^\mu
p_4'\cdot p_{_{P2}}-32m_{J/\psi}\varepsilon_{J/\psi}\cdot p_2'
p_{_{P1}}^\mu p_4'\cdot
p_{_{P2}}-32m_{J/\psi}\varepsilon^{\mu}_{J/\psi} p_1'\cdot
p_{_{P1}} p_4'\cdot p_{_{P2}}\nonumber \\
&&+32m_{J/\psi}\varepsilon^{\mu}_{J/\psi} p_2'\cdot p_{_{P1}}
p_4'\cdot p_{_{P2}}+32m_{J/\psi}\varepsilon_{J/\psi}\cdot p_{_{P2}}
p_1'^\mu p_4'\cdot p_{_{P1}}+32m_{J/\psi}\varepsilon_{J/\psi}\cdot
p_{_{P2}}
p_2'^\mu p_4'\cdot p_{_{P1}}\nonumber \\
&&-32m_{J/\psi}\varepsilon_{J/\psi}\cdot p_1' p_{_{P2}}^\mu
p_4'\cdot p_{_{P1}}+32m_{J/\psi}\varepsilon_{J/\psi}\cdot p_2'
p_{_{P2}}^\mu p_4'\cdot
p_{_{P1}}+32m_{J/\psi}\varepsilon^{\mu}_{J/\psi} p_1'\cdot
p_{_{P2}} p_4'\cdot p_{_{P1}}\nonumber \\
&&-32m_{J/\psi}\varepsilon^{\mu}_{J/\psi} p_2'\cdot p_{_{P2}}
p_4'\cdot p_{_{P1}}+32m_{J/\psi}\varepsilon_{J/\psi}\cdot p_2'
p_1'^\mu p_{_{P2}}\cdot
p_{_{P1}}-32m_{J/\psi}\varepsilon_{J/\psi}\cdot p_4' p_1'^\mu
p_{_{P2}}\cdot p_{_{P1}}\nonumber \\
&&+32m_{J/\psi}\varepsilon_{J/\psi}\cdot p_1' p_2'^\mu
p_{_{P2}}\cdot p_{_{P1}}-32m_{J/\psi}\varepsilon_{J/\psi}\cdot p_4'
p_2'^\mu p_{_{P2}}\cdot
p_{_{P1}}+32m_{J/\psi}\varepsilon_{J/\psi}\cdot p_1' p_4'^\mu
p_{_{P2}}\cdot p_{_{P1}}\nonumber \\
&&+32m_{J/\psi}\varepsilon_{J/\psi}\cdot p_2' p_4'^\mu
p_{_{P2}}\cdot p_{_{P1}}-32m_{J/\psi}\varepsilon^{\mu}_{J/\psi}
p_2'\cdot p_1' p_{_{P2}}\cdot
p_{_{P1}}+32m_{J/\psi}\varepsilon^{\mu}_{J/\psi} p_4'\cdot p_1'
p_{_{P2}}\cdot p_{_{P1}}\nonumber \\
&&+32m_{J/\psi}\varepsilon^{\mu}_{J/\psi} p_4'\cdot p_2'
p_{_{P2}}\cdot p_{_{P1}}]\nonumber \\
&&+E_{\mu\nu}(m_q,u,v)[-64m_{J/\psi}\varepsilon_{J/\psi}\cdot
p_{_{P1}} p_4'^\mu
p_{_{P2}}^\nu-64m_{J/\psi}\varepsilon_{J/\psi}\cdot p_{_{P2}}
p_4'^\mu
p_{_{P1}}^\nu\nonumber \\
&&-128m_{J/\psi}\varepsilon_{J/\psi}\cdot p_4' p_{_{P2}}^\mu
p_{_{P1}}^\nu-32m_{J/\psi}\varepsilon_{J/\psi}\cdot
p_{_{P1}}g^{\mu\nu} p_1'\cdot
p_{_{P2}}-32m_{J/\psi}\varepsilon_{J/\psi}\cdot
p_{_{P2}}g^{\mu\nu} p_1'\cdot p_{_{P1}}\nonumber \\
&&-128m_{J/\psi}\varepsilon^{\mu}_{J/\psi}p_{_{P2}}^\nu p_1'\cdot
p_{_{P1}}-32m_{J/\psi}\varepsilon_{J/\psi}\cdot p_{_{P1}}g^{\mu\nu}
p_2'\cdot
p_{_{P2}}-128m_{J/\psi}\varepsilon^{\mu}_{J/\psi}p_{_{P1}}^\nu
p_2'\cdot p_{_{P2}}\nonumber \\
&&-32m_{J/\psi}\varepsilon_{J/\psi}\cdot p_{_{P2}}g^{\mu\nu}
p_2'\cdot p_{_{P1}}+32m_{J/\psi}\varepsilon_{J/\psi}\cdot
p_{_{P1}}g^{\mu\nu} p_4'\cdot
p_{_{P2}}+32m_{J/\psi}\varepsilon_{J/\psi}\cdot
p_{_{P2}}g^{\mu\nu} p_4'\cdot p_{_{P1}}\nonumber \\
&&+32m_{J/\psi}\varepsilon_{J/\psi}\cdot p_1'g^{\mu\nu}
p_{_{P2}}\cdot p_{_{P1}}+32m_{J/\psi}\varepsilon_{J/\psi}\cdot
p_2'g^{\mu\nu} p_{_{P2}}\cdot
p_{_{P1}}-32m_{J/\psi}\varepsilon_{J/\psi}\cdot
p_4'g^{\mu\nu} p_{_{P2}}\cdot p_{_{P1}}\nonumber \\
&&-64m_{J/\psi}\varepsilon^{\mu}_{J/\psi}p_4'^\nu
p_{_{P2}}\cdot p_{_{P1}}]\nonumber \\
&&+E_{\mu\nu\theta}(m_q,u,v)[-32m_{J/\psi}\varepsilon_{J/\psi}\cdot
p_{_{P1}}g^{\mu\nu}p_{_{P2}}^\theta-32m_{J/\psi}\varepsilon_{J/\psi}\cdot
p_{_{P2}}g^{\mu\nu}p_{_{P1}}^\theta\nonumber
\\
&&+32m_{J/\psi}\varepsilon^{\mu}_{J/\psi}g^{\nu\theta}p_{_{P1}}\cdot
p_{_{P2}}-128m_{J/\psi}\varepsilon^{\mu}_{J/\psi}p_{_{P1}}^\nu
p_{_{P2}}^\theta]\}
\end{eqnarray}
\begin{eqnarray}
&&E_0(m_q,u,v)=\nonumber \\
&&{1\over i\pi^2}\int d^4k {1\over
k^2[(k+p_1)^2-m_Q^2][(k+p_1-p_4-p_5)^2-m_Q^2][(k+p_3)^2-m_q^2](k+p_1+p_2-p_4-p_5)^2}\nonumber \\
&&E_\mu(m_q,u,v)=\nonumber \\
&&{1\over i\pi^2}\int d^4k {k_\mu\over
k^2[(k+p_1)^2-m_Q^2][(k+p_1-p_4-p_5)^2-m_Q^2][(k+p_3)^2-m_q^2](k+p_1+p_2-p_4-p_5)^2}\nonumber \\
&&E_{\mu\nu}(m_q,u,v)=\nonumber \\
&&{1\over i\pi^2}\int d^4k {k_\mu k_\nu\over
k^2[(k+p_1)^2-m_Q^2][(k+p_1-p_4-p_5)^2-m_Q^2][(k+p_3)^2-m_q^2](k+p_1+p_2-p_4-p_5)^2}\nonumber \\
&&E_{\mu\nu\theta}(m_q,u,v)=\nonumber \\
&&{1\over i\pi^2}\int d^4k {k_\mu k_\nu k_\theta\over
k^2[(k+p_1)^2-m_Q^2][(k+p_1-p_4-p_5)^2-m_Q^2][(k+p_3)^2-m_q^2](k+p_1+p_2-p_4-p_5)^2}\nonumber \\
\end{eqnarray}
where $p_1'=p_1, p_2'=p_1-p_4-p_5, p_4'=p_3$.

For amplitudes $\mathcal{A}^{2a1}$ and $\mathcal{A}^{2a2}$, we
have
\begin{eqnarray}
\mathcal{A}^{2a1}&=&C^{2a1}\widetilde{H}^{2a1}(m_q,u,v)\Phi_{J/\psi}\Phi_{P1}\Phi_{P2}\nonumber \\
\mathcal{A}^{2a2}&=&C^{2a2}\widetilde{H}^{2a2}(m_q,u,v)\Phi_{J/\psi}\Phi_{P1}\Phi_{P2}
\end{eqnarray}
with
\begin{eqnarray}
C^{2a1}&=&\text{Tr}(T^aT^bT^c)\text{Tr}(T^bT^aT^c)\nonumber \\
C^{2a2}&=&\text{Tr}(T^aT^bT^c)\text{Tr}(T^aT^bT^c)\nonumber \\
\widetilde{H}^{2a1}(m_q,u,v)&=&-\widetilde{H}^{2a2}(m_q,u,v)={i\pi^2\over
(2\pi)^4}g_s^6({1\over 4N_C})^3{1\over
(p_3+p_5)^2[(p_1-p_3-p_5)^2-m_Q^2]}\nonumber
\\
&&\{D_0(m_q,u,v)[96m_{J/\psi}\varepsilon_{J/\psi}\cdot
p_{_{P1}}p_1'\cdot
p_{_{P2}}m_Q^2-32m_{J/\psi}\varepsilon_{J/\psi}\cdot p_1'
p_{_{P1}}\cdot p_{_{P2}}m_Q^2\nonumber
\\
&&-32m_{J/\psi}\varepsilon_{J/\psi}\cdot p_{_{P2}}p_1'\cdot
p_{_{P1}}m_Q^2-32m_{J/\psi}\varepsilon_{J/\psi}\cdot
p_{_{P1}}p_1'\cdot p_{_{P2}}
p_3'^2+32m_{J/\psi}\varepsilon_{J/\psi}\cdot p_{_{P2}}p_1'\cdot
p_{_{P1}} p_3'^2\nonumber
\\
&&-128m_{J/\psi}\varepsilon_{J/\psi}\cdot p_{_{P1}}p_1'\cdot p_3'
p_3'\cdot p_{_{P2}}-64m_{J/\psi}\varepsilon_{J/\psi}\cdot
p_{_{P2}}p_1'\cdot p_3' p_3'\cdot p_{_{P1}}\nonumber
\\
&&+64m_{J/\psi}\varepsilon_{J/\psi}\cdot p_3' p_1'\cdot p_{_{P2}}
p_3'\cdot p_{_{P1}}-64m_{J/\psi}\varepsilon_{J/\psi}\cdot p_1'
p_3'\cdot p_{_{P2}} p_3'\cdot p_{_{P1}}\nonumber
\\
&&+32m_{J/\psi}\varepsilon_{J/\psi}\cdot p_1' p_3'^2 p_{_{P2}}\cdot
p_{_{P1}}]\nonumber
\\
&&D_\mu(m_q,u,v)[-32m_{J/\psi}\varepsilon_{J/\psi}\cdot
p_{_{P2}}p^\mu_{_{P1}}m_Q^2+96m_{J/\psi}\varepsilon_{J/\psi}\cdot
p_{_{P1}}p^\mu_{_{P2}}m_Q^2\nonumber
\\
&&-32m_{J/\psi}\varepsilon^{\mu}_{J/\psi} p_{_{P1}}\cdot
p_{_{P2}}m_Q^2-32m_{J/\psi}\varepsilon_{J/\psi}\cdot p_{_{P1}}
p_{_{P2}}^\mu p_1'\cdot p_3'-32m_{J/\psi}\varepsilon_{J/\psi}\cdot
p_{_{P2}}
p_{_{P1}}^\mu p_1'\cdot p_3'\nonumber \\
&&-32m_{J/\psi}\varepsilon_{J/\psi}\cdot p_{_{P1}} p_3'^\mu
p_1'\cdot p_{_{P2}}+32m_{J/\psi}\varepsilon_{J/\psi}\cdot p_3'
p_{_{P1}}^\mu p_1'\cdot
p_{_{P2}}+32m_{J/\psi}\varepsilon_{J/\psi}\cdot p_{_{P2}} p_3'^\mu
p_1'\cdot p_{_{P1}}\nonumber
\\
&&+32m_{J/\psi}\varepsilon_{J/\psi}\cdot p_3' p_{_{P2}}^\mu
p_1'\cdot p_{_{P1}}-32m_{J/\psi}\varepsilon_{J/\psi}\cdot p_{_{P1}}
p_{_{P2}}^\mu p_3'^2+32m_{J/\psi}\varepsilon_{J/\psi}\cdot p_{_{P2}}
p_{_{P1}}^\mu p_3'^2\nonumber
\\
&&-96m_{J/\psi}\varepsilon_{J/\psi}\cdot p_{_{P1}} p_1'^\mu
p_3'\cdot p_{_{P2}}-128m_{J/\psi}\varepsilon_{J/\psi}\cdot p_{_{P1}}
p_3'^\mu p_3'\cdot p_{_{P2}}-32m_{J/\psi}\varepsilon_{J/\psi}\cdot
p_1'
p_{_{P1}}^\mu p_3'\cdot p_{_{P2}}\nonumber \\
&&-32m_{J/\psi}\varepsilon^{\mu}_{J/\psi} p_1'\cdot p_{_{P1}}
p_3'\cdot p_{_{P2}}-32m_{J/\psi}\varepsilon_{J/\psi}\cdot p_{_{P2}}
p_1'^\mu p_3'\cdot p_{_{P1}}-64m_{J/\psi}\varepsilon_{J/\psi}\cdot
p_{_{P2}}
p_3'^\mu p_3'\cdot p_{_{P1}}\nonumber \\
&&-32m_{J/\psi}\varepsilon_{J/\psi}\cdot p_1' p_{_{P2}}^\mu
p_3'\cdot p_{_{P1}}+64m_{J/\psi}\varepsilon_{J/\psi}\cdot p_3'
p_{_{P2}}^\mu p_3'\cdot
p_{_{P1}}+32m_{J/\psi}\varepsilon^{\mu}_{J/\psi} p_1'\cdot p_{_{P2}}
p_3'\cdot p_{_{P1}}\nonumber
\\
&&-64m_{J/\psi}\varepsilon^{\mu}_{J/\psi} p_3'\cdot p_{_{P2}}
p_3'\cdot p_{_{P1}}-32m_{J/\psi}\varepsilon_{J/\psi}\cdot p_3'
p_1'^\mu p_{_{P2}}\cdot
p_{_{P1}}+32m_{J/\psi}\varepsilon_{J/\psi}\cdot p_1' p_3'^\mu
p_{_{P2}}\cdot p_{_{P1}}\nonumber
\\
&&+32m_{J/\psi}\varepsilon^{\mu}_{J/\psi} p_1'\cdot p_3'
p_{_{P2}}\cdot p_{_{P1}}+32m_{J/\psi}\varepsilon^{\mu}_{J/\psi}
p_3'^2
p_{_{P2}}\cdot p_{_{P1}}]\nonumber \\
&&+D_{\mu\nu}(m_q,u,v)[-64m_{J/\psi}\varepsilon_{J/\psi}\cdot
p_{_{P1}}p_3'^\mu
p_{_{P2}}^\nu+64m_{J/\psi}\varepsilon_{J/\psi}\cdot p_3'
p_{_{P1}}^\mu p_{_{P2}}^\nu\nonumber \\
&&-96m_{J/\psi}\varepsilon_{J/\psi}\cdot p_{_{P1}} g^{\mu\nu}
p_3'\cdot p_{_{P2}}-64m_{J/\psi}\varepsilon^{\mu}_{J/\psi}\cdot
p_{_{P1}}^\nu p_3'\cdot
p_{_{P2}}-32m_{J/\psi}\varepsilon_{J/\psi}\cdot p_{_{P2}}g^{\mu\nu}
p_3'\cdot p_{_{P1}}\nonumber
\\
&&-32m_{J/\psi}\varepsilon_{J/\psi}\cdot p_3' g^{\mu\nu}
p_{_{P2}}\cdot p_{_{P1}}+64m_{J/\psi}\varepsilon^{\mu}_{J/\psi}
p_3'^\nu
p_{_{P2}}\cdot p_{_{P1}}]\}\nonumber \\
D_0(m_q,u,v)&=&{1\over i\pi^2}\int d^4k {1\over
k^2[(k+p_6)^2-m_q^2](k+p_4+p_6)^2[(k+p_1-p_3-p_5)^2-m_Q^2]}\nonumber \\
D_\mu(m_q,u,v)&=&{1\over i\pi^2}\int d^4k {k_\mu\over
k^2[(k+p_6)^2-m_q^2](k+p_4+p_6)^2[(k+p_1-p_3-p_5)^2-m_Q^2]}\nonumber \\
D_{\mu\nu}(m_q,u,v)&=&{1\over i\pi^2}\int d^4k {k_\mu k_\nu\over
k^2[(k+p_6)^2-m_q^2](k+p_4+p_6)^2[(k+p_1-p_3-p_5)^2-m_Q^2]}
\end{eqnarray}
where $p_1'=p_6, p_3'=p_1-p_3-p_5$.

For amplitudes $\mathcal{A}^{2b1}$ and $\mathcal{A}^{2b2}$, we
have
\begin{eqnarray}
\mathcal{A}^{2b1}&=&C^{2b1}\widetilde{H}^{2b1}(m_q,u,v)\Phi_{J/\psi}\Phi_{P1}\Phi_{P2}\nonumber \\
\mathcal{A}^{2b2}&=&C^{2b2}\widetilde{H}^{2b2}(m_q,u,v)\Phi_{J/\psi}\Phi_{P1}\Phi_{P2}
\end{eqnarray}
with
\begin{eqnarray}
C^{2b1}&=&\text{Tr}(T^aT^bT^c)\text{Tr}(T^aT^bT^c)\nonumber \\
C^{2b2}&=&\text{Tr}(T^aT^bT^c)\text{Tr}(T^bT^aT^c)\nonumber \\
\widetilde{H}^{2b1}(m_q,u,v)&=&-\widetilde{H}^{2b2}(m_q,u,v)=-{i\pi^2\over
(2\pi)^4}g_s^6({1\over 4N_C})^3{1\over
(p_4+p_6)^2[(p_4+p_6-p_2)^2-m_Q^2]}\nonumber \\
&&\{D_0(m_q,u,v)[96m_{J/\psi}\varepsilon_{J/\psi}\cdot
p_{_{P2}}p_1'\cdot
p_{_{P1}}m_Q^2-32m_{J/\psi}\varepsilon_{J/\psi}\cdot
p_{_{P1}}p_1'\cdot p_{_{P2}}m_Q^2\nonumber
\\
&&+32m_{J/\psi}\varepsilon_{J/\psi}\cdot p_1' p_{_{P2}}\cdot
p_{_{P1}}m_Q^2+32m_{J/\psi}\varepsilon_{J/\psi}\cdot p_{_{P1}}
p_1'\cdot p_{_{P2}} p_3'^2-32m_{J/\psi}\varepsilon_{J/\psi}\cdot
p_{_{P2}} p_1'\cdot p_{_{P1}} p_3'^2\nonumber
\\
&&-64m_{J/\psi}\varepsilon_{J/\psi}\cdot p_{_{P1}} p_1'\cdot p_3'
p_3'\cdot p_{_{P2}}+64m_{J/\psi}\varepsilon_{J/\psi}\cdot p_3'
p_1'\cdot p_{_{P2}} p_3'\cdot p_{_{P2}}\nonumber
\\
&&-128m_{J/\psi}\varepsilon_{J/\psi}\cdot p_{_{P2}} p_1'\cdot p_3'
p_3'\cdot p_{_{P1}}-64m_{J/\psi}\varepsilon_{J/\psi}\cdot p_1'
p_3'\cdot p_{_{P2}} p_3'\cdot p_{_{P1}}\nonumber
\\
&&+32m_{J/\psi}\varepsilon_{J/\psi}\cdot p_1' p_3'^2
p_{_{P2}}\cdot p_{_{P1}}]\nonumber \\
&&+D_\mu(m_q,u,v)[96m_{J/\psi}\varepsilon_{J/\psi}\cdot
p_{_{P2}}p_{_{P1}}^\mu m_Q^2-32m_{J/\psi}\varepsilon_{J/\psi}\cdot
p_{_{P1}}p_{_{P2}}^\mu m_Q^2\nonumber
\\
&&-32m_{J/\psi}\varepsilon^{\mu}_{J/\psi}p_{_{P1}}\cdot
p_{_{P2}}m_Q^2-32m_{J/\psi}\varepsilon_{J/\psi}\cdot p_{_{P1}}
p_{_{P2}}^\mu p_1'\cdot p_3'-32m_{J/\psi}\varepsilon_{J/\psi}\cdot
p_{_{P2}} p_{_{P1}}^\mu p_1'\cdot p_3'\nonumber
\\
&&+32m_{J/\psi}\varepsilon_{J/\psi}\cdot p_{_{P1}} p_3'^\mu
p_1'\cdot p_{_{P2}}+32m_{J/\psi}\varepsilon_{J/\psi}\cdot p_3'
p_{_{P1}}^\mu p_1'\cdot
p_{_{P2}}-32m_{J/\psi}\varepsilon_{J/\psi}\cdot p_{_{P2}}
p_3'^\mu p_1'\cdot p_{_{P1}}\nonumber \\
&&+32m_{J/\psi}\varepsilon_{J/\psi}\cdot p_3' p_{_{P2}}^\mu
p_1'\cdot p_{_{P1}}+32m_{J/\psi}\varepsilon_{J/\psi}\cdot p_{_{P1}}
p_{_{P2}}^\mu p_3'^2-32m_{J/\psi}\varepsilon_{J/\psi}\cdot p_{_{P2}}
p_{_{P1}}^\mu p_3'^2\nonumber
\\
&&-32m_{J/\psi}\varepsilon_{J/\psi}\cdot p_{_{P1}} p_1'^\mu
p_3'\cdot p_{_{P2}}-64m_{J/\psi}\varepsilon_{J/\psi}\cdot p_{_{P1}}
p_3'^\mu p_3'\cdot p_{_{P2}}-32m_{J/\psi}\varepsilon_{J/\psi}\cdot
p_1' p_{_{P1}}^\mu p_3'\cdot p_{_{P2}}\nonumber
\\
&&+64m_{J/\psi}\varepsilon_{J/\psi}\cdot p_3' p_{_{P1}}^\mu
p_3'\cdot p_{_{P2}}+32m_{J/\psi}\varepsilon^{\mu}_{J/\psi} p_1'\cdot
p_{_{P1}} p_3'\cdot p_{_{P2}}-96m_{J/\psi}\varepsilon_{J/\psi}\cdot
p_{_{P2}} p_1'^\mu p_3'\cdot
p_{_{P1}}\nonumber \\
&&-128m_{J/\psi}\varepsilon_{J/\psi}\cdot p_{_{P2}} p_3'^\mu
p_3'\cdot p_{_{P1}}-32m_{J/\psi}\varepsilon_{J/\psi}\cdot p_1'
p_{_{P2}}^\mu p_3'\cdot
p_{_{P1}}-32m_{J/\psi}\varepsilon^{\mu}_{J/\psi} p_1'\cdot p_{_{P2}}
p_3'\cdot p_{_{P1}}\nonumber
\\
&&-64m_{J/\psi}\varepsilon^{\mu}_{J/\psi} p_3'\cdot p_{_{P2}}
p_3'\cdot p_{_{P1}}-32m_{J/\psi}\varepsilon_{J/\psi}\cdot p_3'
p_1'^\mu p_{_{P2}}\cdot
p_{_{P1}}+32m_{J/\psi}\varepsilon_{J/\psi}\cdot p_1' p_3'^\mu
p_{_{P2}}\cdot p_{_{P1}}\nonumber \\
&&+32m_{J/\psi}\varepsilon^{\mu}_{J/\psi} p_1'\cdot p_3'
p_{_{P2}}\cdot p_{_{P1}}+32m_{J/\psi}\varepsilon^{\mu}_{J/\psi}
p_3'^2
p_{_{P2}}\cdot p_{_{P1}}]\nonumber \\
&&+D_{\mu\nu}(m_q,u,v)[-64m_{J/\psi}\varepsilon_{J/\psi}\cdot
p_{_{P2}}p_{_{P1}}^\mu
p_3'^\nu+64m_{J/\psi}\varepsilon_{J/\psi}\cdot p_3'
p_{_{P2}}^\mu p_{_{P1}}^\nu\nonumber \\
&&-32m_{J/\psi}\varepsilon_{J/\psi}\cdot p_{_{P1}} g^{\mu\nu}
p_3'\cdot p_{_{P2}}-96m_{J/\psi}\varepsilon_{J/\psi}\cdot p_{_{P2}}
g^{\mu\nu} p_3'\cdot
p_{_{P1}}-64m_{J/\psi}\varepsilon^{\mu}_{J/\psi} p_{_{P2}}^\nu
p_3'\cdot p_{_{P1}}\nonumber
\\
&&-32m_{J/\psi}\varepsilon_{J/\psi}\cdot p_3' g^{\mu\nu}
p_{_{P2}}\cdot
p_{_{P1}}+64m_{J/\psi}\varepsilon^{\mu}_{J/\psi}p_3'^\nu
p_{_{P1}}\cdot p_{_{P2}}]\}\nonumber \\
D_0(m_q,u,v)&=&{1\over i\pi^2}\int d^4k {1\over
k^2[(k-p_3)^2-m_q^2](k-p_3-p_5)^2[(k+p_4+p_6-p_2)^2-m_Q^2]}\nonumber \\
D_\mu(m_q,u,v)&=&{1\over i\pi^2}\int d^4k {k_\mu\over
k^2[(k-p_3)^2-m_q^2](k-p_3-p_5)^2[(k+p_4+p_6-p_2)^2-m_Q^2]}\nonumber \\
D_{\mu\nu}(m_q,u,v)&=&{1\over i\pi^2}\int d^4k {k_\mu k_\nu\over
k^2[(k-p_3)^2-m_q^2](k-p_3-p_5)^2[(k+p_4+p_6-p_2)^2-m_Q^2]}
\end{eqnarray}
where $p_1'=-p_3, p_3'=p_4+p_6-p_2$.

For amplitudes $\mathcal{A}^{2c1}$ and $\mathcal{A}^{2c2}$, we
have
\begin{eqnarray}
\mathcal{A}^{2c1}&=&C^{2c1}\widetilde{H}^{2c1}(m_q,u,v)\Phi_{J/\psi}\Phi_{P1}\Phi_{P2}\nonumber \\
\mathcal{A}^{2c2}&=&C^{2c2}\widetilde{H}^{2c2}(m_q,u,v)\Phi_{J/\psi}\Phi_{P1}\Phi_{P2}
\end{eqnarray}
with
\begin{eqnarray}
&&C^{2c1}=\text{Tr}(T^aT^bT^c)\text{Tr}(T^aT^bT^c)\nonumber \\
&&C^{2c2}=\text{Tr}(T^aT^bT^c)\text{Tr}(T^bT^aT^c)\nonumber \\
&&\widetilde{H}^{2c1}(m_q,u,v)=-\widetilde{H}^{2c2}(m_q,u,v)={i\pi^2\over
(2\pi)^4}g_s^6({1\over 4N_C})^3{1\over (p_4+p_5)^2}\nonumber
\\
&&\{E_0(m_q,u,v)[-32m_{J/\psi}\varepsilon_{J/\psi}\cdot p_{_{P1}}
p_4'\cdot p_{_{P2}}m_Q^2-32m_{J/\psi}\varepsilon_{J/\psi}\cdot
p_{_{P2}} p_4'\cdot p_{_{P1}}m_Q^2\nonumber \\
&&+96m_{J/\psi}\varepsilon_{J/\psi}\cdot p_4' p_{_{P2}}\cdot
p_{_{P1}}m_Q^2-32m_{J/\psi}\varepsilon_{J/\psi}\cdot p_{_{P1}}
p_1'\cdot p_{_{P2}} p_2'\cdot p_4'\nonumber
\\
&&-32m_{J/\psi}\varepsilon_{J/\psi}\cdot p_{_{P2}} p_1'\cdot
p_{_{P1}} p_2'\cdot p_4'-32m_{J/\psi}\varepsilon_{J/\psi}\cdot
p_{_{P1}} p_1'\cdot p_4' p_2'\cdot p_{_{P2}}\nonumber
\\
&&-96m_{J/\psi}\varepsilon_{J/\psi}\cdot p_4' p_1'\cdot p_{_{P1}}
p_2'\cdot p_{_{P2}}-32m_{J/\psi}\varepsilon_{J/\psi}\cdot p_{_{P2}}
p_1'\cdot p_4' p_2'\cdot p_{_{P1}}\nonumber \\
&&-96m_{J/\psi}\varepsilon_{J/\psi}\cdot p_4' p_1'\cdot p_{_{P2}}
p_2'\cdot p_{_{P1}}+32m_{J/\psi}\varepsilon_{J/\psi}\cdot p_{_{P1}}
p_1'\cdot p_2' p_4'\cdot p_{_{P2}}\nonumber \\
&&+32m_{J/\psi}\varepsilon_{J/\psi}\cdot p_2' p_1'\cdot p_{_{P1}}
p_4'\cdot p_{_{P2}}-32m_{J/\psi}\varepsilon_{J/\psi}\cdot p_1'
p_2'\cdot p_{_{P1}} p_4'\cdot p_{_{P2}}\nonumber \\
&&+32m_{J/\psi}\varepsilon_{J/\psi}\cdot p_{_{P2}} p_1'\cdot p_2'
p_4'\cdot p_{_{P1}}-32m_{J/\psi}\varepsilon_{J/\psi}\cdot p_2'
p_1'\cdot p_{_{P2}} p_4'\cdot p_{_{P1}}\nonumber \\
&&+32m_{J/\psi}\varepsilon_{J/\psi}\cdot p_1' p_2'\cdot p_{_{P2}}
p_4'\cdot p_{_{P1}}-32m_{J/\psi}\varepsilon_{J/\psi}\cdot p_4'
p_1'\cdot p_2' p_{_{P2}}\cdot p_{_{P1}}\nonumber
\\
&&+32m_{J/\psi}\varepsilon_{J/\psi}\cdot p_2' p_1'\cdot p_4'
p_{_{P2}}\cdot p_{_{P1}}+32m_{J/\psi}\varepsilon_{J/\psi}\cdot p_1'
p_2'\cdot p_4'
p_{_{P2}}\cdot p_{_{P1}}]\nonumber \\
&&+E_\mu(m_q,u,v)[-32m_{J/\psi}\varepsilon_{J/\psi}\cdot p_{_{P1}}
p_{_{P2}}^\mu m_Q^2-32m_{J/\psi}\varepsilon_{J/\psi}\cdot p_{_{P2}}
p_{_{P1}}^\mu
m_Q^2\nonumber \\
&&+96m_{J/\psi}\varepsilon^{\mu}_{J/\psi}p_{_{P1}}\cdot
p_{_{P2}}m_Q^2+32m_{J/\psi}\varepsilon_{J/\psi}\cdot p_{_{P1}}
p_{_{P2}}^\mu p_1'\cdot p_2'+32m_{J/\psi}\varepsilon_{J/\psi}\cdot
p_{_{P2}}
p_{_{P1}}^\mu p_1'\cdot p_2'\nonumber \\
&&-32m_{J/\psi}\varepsilon_{J/\psi}\cdot p_{_{P1}} p_{_{P2}}^\mu
p_1'\cdot p_4'-32m_{J/\psi}\varepsilon_{J/\psi}\cdot p_{_{P2}}
p_{_{P1}}^\mu p_1'\cdot p_4'-32m_{J/\psi}\varepsilon_{J/\psi}\cdot
p_{_{P1}} p_2'^\mu
p_1'\cdot p_{_{P2}}\nonumber \\
&&-32m_{J/\psi}\varepsilon_{J/\psi}\cdot p_{_{P1}} p_4'^\mu
p_1'\cdot p_{_{P2}}-32m_{J/\psi}\varepsilon_{J/\psi}\cdot p_2'
p_{_{P1}}^\mu p_1'\cdot
p_{_{P2}}-96m_{J/\psi}\varepsilon_{J/\psi}\cdot p_4'
p_{_{P1}}^\mu p_1'\cdot p_{_{P2}}\nonumber \\
&&-32m_{J/\psi}\varepsilon_{J/\psi}\cdot p_{_{P2}} p_2'^\mu
p_1'\cdot p_{_{P1}}-32m_{J/\psi}\varepsilon_{J/\psi}\cdot p_{_{P2}}
p_4'^\mu p_1'\cdot p_{_{P1}}+32m_{J/\psi}\varepsilon_{J/\psi}\cdot
p_2'
p_{_{P2}}^\mu p_1'\cdot p_{_{P1}}\nonumber \\
&&-32m_{J/\psi}\varepsilon_{J/\psi}\cdot p_4' p_{_{P2}}^\mu
p_1'\cdot p_{_{P1}}-32m_{J/\psi}\varepsilon_{J/\psi}\cdot p_{_{P1}}
p_{_{P2}}^\mu p_2'\cdot p_4'-32m_{J/\psi}\varepsilon_{J/\psi}\cdot
p_{_{P2}}
p_{_{P1}}^\mu p_2'\cdot p_4'\nonumber \\
&&-32m_{J/\psi}\varepsilon_{J/\psi}\cdot p_{_{P1}} p_1'^\mu
p_2'\cdot p_{_{P2}}-32m_{J/\psi}\varepsilon_{J/\psi}\cdot p_{_{P1}}
p_4'^\mu p_2'\cdot p_{_{P2}}+32m_{J/\psi}\varepsilon_{J/\psi}\cdot
p_1'
p_{_{P1}}^\mu p_2'\cdot p_{_{P2}}\nonumber \\
&&-32m_{J/\psi}\varepsilon_{J/\psi}\cdot p_4' p_{_{P1}}^\mu
p_2'\cdot p_{_{P2}}-32m_{J/\psi}\varepsilon^{\mu}_{J/\psi} p_1'\cdot
p_{_{P1}} p_2'\cdot p_{_{P2}}-32m_{J/\psi}\varepsilon_{J/\psi}\cdot
p_{_{P2}} p_1'^\mu p_2'\cdot
p_{_{P1}}\nonumber \\
&&-32m_{J/\psi}\varepsilon_{J/\psi}\cdot p_{_{P2}} p_4'^\mu
p_2'\cdot p_{_{P1}}-32m_{J/\psi}\varepsilon_{J/\psi}\cdot p_1'
p_{_{P2}}^\mu p_2'\cdot
p_{_{P1}}-96m_{J/\psi}\varepsilon_{J/\psi}\cdot p_4'
p_{_{P2}}^\mu p_2'\cdot p_{_{P1}}\nonumber \\
&&-96m_{J/\psi}\varepsilon^{\mu}_{J/\psi} p_1'\cdot p_{_{P2}}
p_2'\cdot p_{_{P1}}+32m_{J/\psi}\varepsilon_{J/\psi}\cdot p_{_{P1}}
p_1'^\mu p_4'\cdot p_{_{P2}}+32m_{J/\psi}\varepsilon_{J/\psi}\cdot
p_{_{P1}}
p_2'^\mu p_4'\cdot p_{_{P2}}\nonumber \\
&&-32m_{J/\psi}\varepsilon_{J/\psi}\cdot p_1' p_{_{P1}}^\mu
p_4'\cdot p_{_{P2}}+32m_{J/\psi}\varepsilon_{J/\psi}\cdot p_2'
p_{_{P1}}^\mu p_4'\cdot
p_{_{P2}}+32m_{J/\psi}\varepsilon^{\mu}_{J/\psi} p_1'\cdot
p_{_{P1}} p_4'\cdot p_{_{P2}}\nonumber \\
&&-32m_{J/\psi}\varepsilon^{\mu}_{J/\psi} p_2'\cdot p_{_{P1}}
p_4'\cdot p_{_{P2}}+32m_{J/\psi}\varepsilon_{J/\psi}\cdot p_{_{P2}}
p_1'^\mu p_4'\cdot p_{_{P1}}+32m_{J/\psi}\varepsilon_{J/\psi}\cdot
p_{_{P2}}
p_2'^\mu p_4'\cdot p_{_{P1}}\nonumber \\
&&+32m_{J/\psi}\varepsilon_{J/\psi}\cdot p_1' p_{_{P2}}^\mu
p_4'\cdot p_{_{P1}}-32m_{J/\psi}\varepsilon_{J/\psi}\cdot p_2'
p_{_{P2}}^\mu p_4'\cdot
p_{_{P1}}-32m_{J/\psi}\varepsilon^{\mu}_{J/\psi} p_1'\cdot
p_{_{P2}} p_4'\cdot p_{_{P1}}\nonumber \\
&&+32m_{J/\psi}\varepsilon^{\mu}_{J/\psi} p_2'\cdot p_{_{P2}}
p_4'\cdot p_{_{P1}}+32m_{J/\psi}\varepsilon_{J/\psi}\cdot p_2'
p_1'^\mu p_{_{P2}}\cdot
p_{_{P1}}-32m_{J/\psi}\varepsilon_{J/\psi}\cdot p_4' p_1'^\mu
p_{_{P2}}\cdot p_{_{P1}}\nonumber \\
&&+32m_{J/\psi}\varepsilon_{J/\psi}\cdot p_1' p_2'^\mu
p_{_{P2}}\cdot p_{_{P1}}-32m_{J/\psi}\varepsilon_{J/\psi}\cdot p_4'
p_2'^\mu p_{_{P2}}\cdot
p_{_{P1}}+32m_{J/\psi}\varepsilon_{J/\psi}\cdot p_1' p_4'^\mu
p_{_{P2}}\cdot p_{_{P1}}\nonumber \\
&&+32m_{J/\psi}\varepsilon_{J/\psi}\cdot p_2' p_4'^\mu
p_{_{P2}}\cdot p_{_{P1}}-32m_{J/\psi}\varepsilon^{\mu}_{J/\psi}
p_2'\cdot p_1' p_{_{P2}}\cdot
p_{_{P1}}+32m_{J/\psi}\varepsilon^{\mu}_{J/\psi} p_4'\cdot p_1'
p_{_{P2}}\cdot p_{_{P1}}\nonumber \\
&&+32m_{J/\psi}\varepsilon^{\mu}_{J/\psi} p_4'\cdot p_2'
p_{_{P2}}\cdot p_{_{P1}}]\nonumber \\
&&+E_{\mu\nu}(m_q,u,v)[-64m_{J/\psi}\varepsilon_{J/\psi}\cdot
p_{_{P1}} p_4'^\mu
p_{_{P2}}^\nu-64m_{J/\psi}\varepsilon_{J/\psi}\cdot p_{_{P2}}
p_4'^\mu
p_{_{P1}}^\nu\nonumber \\
&&-128m_{J/\psi}\varepsilon_{J/\psi}\cdot p_4' p_{_{P2}}^\mu
p_{_{P1}}^\nu-32m_{J/\psi}\varepsilon_{J/\psi}\cdot
p_{_{P1}}g^{\mu\nu} p_1'\cdot
p_{_{P2}}-32m_{J/\psi}\varepsilon_{J/\psi}\cdot
p_{_{P2}}g^{\mu\nu} p_1'\cdot p_{_{P1}}\nonumber \\
&&-128m_{J/\psi}\varepsilon^{\mu}_{J/\psi}p_{_{P1}}^\nu p_1'\cdot
p_{_{P2}}-32m_{J/\psi}\varepsilon_{J/\psi}\cdot p_{_{P1}}g^{\mu\nu}
p_2'\cdot
p_{_{P2}}-128m_{J/\psi}\varepsilon^{\mu}_{J/\psi}p_{_{P2}}^\nu
p_2'\cdot p_{_{P1}}\nonumber \\
&&-32m_{J/\psi}\varepsilon_{J/\psi}\cdot p_{_{P2}}g^{\mu\nu}
p_2'\cdot p_{_{P1}}+32m_{J/\psi}\varepsilon_{J/\psi}\cdot
p_{_{P1}}g^{\mu\nu} p_4'\cdot
p_{_{P2}}+32m_{J/\psi}\varepsilon_{J/\psi}\cdot
p_{_{P2}}g^{\mu\nu} p_4'\cdot p_{_{P1}}\nonumber \\
&&+32m_{J/\psi}\varepsilon_{J/\psi}\cdot p_1'g^{\mu\nu}
p_{_{P2}}\cdot p_{_{P1}}+32m_{J/\psi}\varepsilon_{J/\psi}\cdot
p_2'g^{\mu\nu} p_{_{P2}}\cdot
p_{_{P1}}-32m_{J/\psi}\varepsilon_{J/\psi}\cdot
p_4'g^{\mu\nu} p_{_{P2}}\cdot p_{_{P1}}\nonumber \\
&&-64m_{J/\psi}\varepsilon^{\mu}_{J/\psi}p_4'^\nu
p_{_{P2}}\cdot p_{_{P1}}]\nonumber \\
&&+E_{\mu\nu\theta}(m_q,u,v)[-32m_{J/\psi}\varepsilon_{J/\psi}\cdot
p_{_{P1}}g^{\mu\nu}p_{_{P2}}^\theta-32m_{J/\psi}\varepsilon_{J/\psi}\cdot
p_{_{P2}}g^{\mu\nu}p_{_{P1}}^\theta\nonumber
\\
&&+32m_{J/\psi}\varepsilon^{\mu}_{J/\psi}g^{\nu\theta}p_{_{P1}}\cdot
p_{_{P2}}-128m_{J/\psi}\varepsilon^{\mu}_{J/\psi}p_{_{P1}}^\nu
p_{_{P2}}^\theta]\}
\end{eqnarray}
\begin{eqnarray}
&&E_0(m_q,u,v)=\nonumber \\
&&{1\over i\pi^2}\int d^4k {1\over
k^2[(k+p_1)^2-m_Q^2][(k+p_1-p_4-p_5)^2-m_Q^2][(k+p_6)^2-m_q^2](k+p_1+p_2-p_4-p_5)^2}\nonumber \\
&&E_\mu(m_q,u,v)=\nonumber \\
&&{1\over i\pi^2}\int d^4k {k_\mu\over
k^2[(k+p_1)^2-m_Q^2][(k+p_1-p_4-p_5)^2-m_Q^2][(k+p_6)^2-m_q^2](k+p_1+p_2-p_4-p_5)^2}\nonumber \\
&&E_{\mu\nu}(m_q,u,v)=\nonumber \\
&&{1\over i\pi^2}\int d^4k {k_\mu k_\nu\over
k^2[(k+p_1)^2-m_Q^2][(k+p_1-p_4-p_5)^2-m_Q^2][(k+p_6)^2-m_q^2](k+p_1+p_2-p_4-p_5)^2}\nonumber \\
&&E_{\mu\nu\theta}(m_q,u,v)=\nonumber \\
&&{1\over i\pi^2}\int d^4k {k_\mu k_\nu k_\theta\over
k^2[(k+p_1)^2-m_Q^2][(k+p_1-p_4-p_5)^2-m_Q^2][(k+p_6)^2-m_q^2](k+p_1+p_2-p_4-p_5)^2}\nonumber \\
\end{eqnarray}
where $p_1'=p_1, p_2'=p_1-p_4-p_5, p_4'=p_6$.
\\
\\
\\
\textbf{2. The case of $\mathbf{J/\psi\to V P}$}

For amplitude $\mathcal{A}^{1a1a}$, we have
\begin{eqnarray}
\mathcal{A}^{1a1a}&=&C^{1a1a}\widetilde{H}^{1a1a}(m_q,u,v)\Phi_{J/\psi}\Phi_{V}\Phi_{P}
\end{eqnarray}
with
\begin{eqnarray}
C^{1a1a}&=&\text{Tr}(T^aT^bT^c)\text{Tr}(T^aT^bT^c)\nonumber \\
\widetilde{H}^{1a1a}(m_q,u,v)&=&-{i\pi^2\over
(2\pi)^4}g_s^6({1\over 4N_C})^3{1\over
(p_3+p_5)^2[(p_1-p_3-p_5)^2-m_Q^2]}\nonumber
\\
&&m_Qm_q\{D_0(m_q,u,v)\varepsilon_{\mu\nu\alpha\beta}
[-64\varepsilon_{J/\psi}^{\mu}\varepsilon_{_V}^{\ast\nu}p_{_{P}}^\alpha
p_{_{V}}^\beta p_3'\cdot
p_{J/\psi}-64\varepsilon_{J/\psi}^{\mu}\varepsilon_{_V}^{\ast\alpha}p_{J/\psi}^\nu
p_{_{V}}^\beta p_3'\cdot p_{_{P}}\nonumber \\
&&+64\varepsilon_{J/\psi}^{\mu}\varepsilon_{_V}^{\ast\alpha}p_{J/\psi}^\nu
p_{_{P}}^\beta p_3'\cdot
p_{_{V}}+64\varepsilon_{J/\psi}^{\nu}\varepsilon_{_V}^{\ast\alpha}p_3'^\mu
p_{_{P}}^\beta p_{J/\psi}\cdot
p_{_{V}}-64\varepsilon_{J/\psi}^{\mu}\varepsilon_{_V}^{\ast\beta}p_{J/\psi}^\alpha
p_3'^\nu p_{_{V}}\cdot p_{_{P}}]\nonumber \\
&&+[D_\theta(m_q,u,v)\varepsilon_{\mu\nu\alpha\beta}
(32\varepsilon_{J/\psi}^\mu\varepsilon_{_V}^{\ast\nu}p_{_P}^\alpha
p_{_V}^\beta
p_{J/\psi}^\theta+32\varepsilon_{J/\psi}^\mu\varepsilon_{_V}^{\ast\alpha}p_{J/\psi}^\nu
p_{_V}^\beta
p_{_P}^\theta+32\varepsilon_{J/\psi}^\mu\varepsilon_{_V}^{\ast\alpha}p_{J/\psi}^\nu
p_{_P}^\beta p_{_V}^\theta)\nonumber
\\
&&+D^\nu(m_q,u,v)\varepsilon_{\mu\nu\alpha\beta}(32\varepsilon_{J/\psi}^\mu\varepsilon_{_V}^{\ast\alpha}
p_{_V}^\beta p_{J/\psi}\cdot
p_{_P}-32\varepsilon_{J/\psi}^\mu\varepsilon_{_V}^{\ast\alpha}
p_{_P}^\beta p_{J/\psi}\cdot p_{_V}\nonumber
\\
&&-32\varepsilon_{J/\psi}^\mu\varepsilon_{_V}^{\ast\beta}
p_{J/\psi}^\alpha p_{_V}\cdot p_{_P})]\}
\end{eqnarray}
where $p_1'=p_4, p_3'=p_1-p_3-p_5$.

For amplitude $\mathcal{A}^{1a1b}$, we have
\begin{eqnarray}
\mathcal{A}^{1a1b}&=&C^{1a1b}\widetilde{H}^{1a1b}(m_q,u,v)\Phi_{J/\psi}\Phi_{V}\Phi_{P}
\end{eqnarray}
with
\begin{eqnarray}
C^{1a1b}&=&\text{Tr}(T^aT^bT^c)\text{Tr}(T^aT^bT^c)\nonumber \\
\widetilde{H}^{1a1b}(m_q,u,v)&=&-{i\pi^2\over
(2\pi)^4}g_s^6({1\over 4N_C})^3{1\over
(p_3+p_5)^2[(p_1-p_3-p_5)^2-m_Q^2]}\nonumber
\\
&&m_Qm_q\{D_0(m_q,u,v)\varepsilon_{\mu\nu\alpha\beta}
[-64\varepsilon_{J/\psi}^{\mu}\varepsilon_{_V}^{\ast\nu}p_{_{P}}^\alpha
p_{_{V}}^\beta p_3'\cdot
p_{J/\psi}-64\varepsilon_{J/\psi}^{\mu}\varepsilon_{_V}^{\ast\alpha}p_{J/\psi}^\nu
p_{_{V}}^\beta p_3'\cdot p_{_{P}}\nonumber \\
&&+64\varepsilon_{J/\psi}^{\mu}\varepsilon_{_V}^{\ast\alpha}p_{J/\psi}^\nu
p_{_{P}}^\beta p_3'\cdot
p_{_{V}}-64\varepsilon_{J/\psi}^{\nu}\varepsilon_{_V}^{\ast\alpha}p_3'^\mu
p_{_{V}}^\beta p_{J/\psi}\cdot
p_{_{P}}+64\varepsilon_{J/\psi}^{\mu}\varepsilon_{_V}^{\ast\beta}p_{J/\psi}^\alpha
p_3'^\nu p_{_{V}}\cdot p_{_{P}}]\nonumber \\
&&+[D_\theta(m_q,u,v)\varepsilon_{\mu\nu\alpha\beta}
(96\varepsilon_{J/\psi}^\mu\varepsilon_{_V}^{\ast\nu}p_{_P}^\alpha
p_{_V}^\beta
p_{J/\psi}^\theta-32\varepsilon_{J/\psi}^\mu\varepsilon_{_V}^{\ast\alpha}p_{J/\psi}^\nu
p_{_V}^\beta
p_{_P}^\theta-32\varepsilon_{J/\psi}^\mu\varepsilon_{_V}^{\ast\alpha}p_{J/\psi}^\nu
p_{_P}^\beta p_{_V}^\theta)\nonumber
\\
&&+D^\nu(m_q,u,v)\varepsilon_{\mu\nu\alpha\beta}(32\varepsilon_{J/\psi}^\mu\varepsilon_{_V}^{\ast\alpha}
p_{_V}^\beta p_{J/\psi}\cdot
p_{_P}-32\varepsilon_{J/\psi}^\mu\varepsilon_{_V}^{\ast\alpha}
p_{_P}^\beta p_{J/\psi}\cdot p_{_V}\nonumber
\\
&&+32\varepsilon_{J/\psi}^\mu\varepsilon_{_V}^{\ast\beta}
p_{J/\psi}^\alpha p_{_V}\cdot p_{_P})]\}
\end{eqnarray}
where $p_1'=p_4, p_3'=p_1-p_3-p_5$.

For amplitude $\mathcal{A}^{1a2a}$, we have
\begin{eqnarray}
\mathcal{A}^{1a2a}&=&C^{1a2a}\widetilde{H}^{1a2a}(m_q,u,v)\Phi_{J/\psi}\Phi_{V}\Phi_{P}
\end{eqnarray}
with
\begin{eqnarray}
C^{1a2a}&=&\text{Tr}(T^aT^bT^c)\text{Tr}(T^bT^aT^c)\nonumber \\
\widetilde{H}^{1a2a}(m_q,u,v)&=&{i\pi^2\over
(2\pi)^4}g_s^6({1\over 4N_C})^3{1\over
(p_3+p_5)^2[(p_1-p_3-p_5)^2-m_Q^2]}\nonumber
\\
&&m_Qm_q\{D_0(m_q,u,v)\varepsilon_{\mu\nu\alpha\beta}
[64\varepsilon_{J/\psi}^{\mu}\varepsilon_{_V}^{\ast\nu}p_{_{P}}^\alpha
p_{_{V}}^\beta p_3'\cdot
p_{J/\psi}+64\varepsilon_{J/\psi}^{\mu}\varepsilon_{_V}^{\ast\alpha}p_{J/\psi}^\nu
p_{_{V}}^\beta p_3'\cdot p_{_{P}}\nonumber \\
&&+64\varepsilon_{J/\psi}^{\mu}\varepsilon_{_V}^{\ast\alpha}p_{J/\psi}^\nu
p_{_{P}}^\beta p_3'\cdot
p_{_{V}}+64\varepsilon_{J/\psi}^{\nu}\varepsilon_{_V}^{\ast\alpha}p_3'^\mu
p_{_{P}}^\beta p_{J/\psi}\cdot
p_{_{V}}-64\varepsilon_{J/\psi}^{\mu}\varepsilon_{_V}^{\ast\beta}p_{J/\psi}^\alpha
p_3'^\nu p_{_{V}}\cdot p_{_{P}}]\nonumber \\
&&+[D_\theta(m_q,u,v)\varepsilon_{\mu\nu\alpha\beta}
(-32\varepsilon_{J/\psi}^\mu\varepsilon_{_V}^{\ast\nu}p_{_P}^\alpha
p_{_V}^\beta
p_{J/\psi}^\theta-32\varepsilon_{J/\psi}^\mu\varepsilon_{_V}^{\ast\alpha}p_{J/\psi}^\nu
p_{_V}^\beta
p_{_P}^\theta+32\varepsilon_{J/\psi}^\mu\varepsilon_{_V}^{\ast\alpha}p_{J/\psi}^\nu
p_{_P}^\beta p_{_V}^\theta)\nonumber
\\
&&+D^\nu(m_q,u,v)\varepsilon_{\mu\nu\alpha\beta}(-32\varepsilon_{J/\psi}^\mu\varepsilon_{_V}^{\ast\alpha}
p_{_V}^\beta p_{J/\psi}\cdot
p_{_P}-32\varepsilon_{J/\psi}^\mu\varepsilon_{_V}^{\ast\alpha}
p_{_P}^\beta p_{J/\psi}\cdot p_{_V}\nonumber
\\
&&-32\varepsilon_{J/\psi}^\mu\varepsilon_{_V}^{\ast\beta}
p_{J/\psi}^\alpha p_{_V}\cdot p_{_P})]\}
\end{eqnarray}
where $p_1'=p_4, p_3'=p_1-p_3-p_5$.

For amplitude $\mathcal{A}^{1a2b}$, we have
\begin{eqnarray}
\mathcal{A}^{1a2b}&=&C^{1a2b}\widetilde{H}^{1a2b}(m_q,u,v)\Phi_{J/\psi}\Phi_{V}\Phi_{P}
\end{eqnarray}
with
\begin{eqnarray}
C^{1a2b}&=&\text{Tr}(T^aT^bT^c)\text{Tr}(T^bT^aT^c)\nonumber \\
\widetilde{H}^{1a2b}(m_q,u,v)&=&{i\pi^2\over
(2\pi)^4}g_s^6({1\over 4N_C})^3{1\over
(p_3+p_5)^2[(p_1-p_3-p_5)^2-m_Q^2]}\nonumber
\\
&&m_Qm_q\{D_0(m_q,u,v)\varepsilon_{\mu\nu\alpha\beta}
[64\varepsilon_{J/\psi}^{\mu}\varepsilon_{_V}^{\ast\nu}p_{_{P}}^\alpha
p_{_{V}}^\beta p_3'\cdot
p_{J/\psi}-64\varepsilon_{J/\psi}^{\mu}\varepsilon_{_V}^{\ast\alpha}p_{J/\psi}^\nu
p_{_{V}}^\beta p_3'\cdot p_{_{P}}\nonumber \\
&&-64\varepsilon_{J/\psi}^{\mu}\varepsilon_{_V}^{\ast\alpha}p_{J/\psi}^\nu
p_{_{P}}^\beta p_3'\cdot
p_{_{V}}-64\varepsilon_{J/\psi}^{\nu}\varepsilon_{_V}^{\ast\alpha}p_3'^\mu
p_{_{V}}^\beta p_{J/\psi}\cdot
p_{_{P}}+64\varepsilon_{J/\psi}^{\mu}\varepsilon_{_V}^{\ast\beta}p_{J/\psi}^\alpha
p_3'^\nu p_{_{V}}\cdot p_{_{P}}]\nonumber \\
&&+[D_\theta(m_q,u,v)\varepsilon_{\mu\nu\alpha\beta}
(-64\varepsilon_{J/\psi}^\mu\varepsilon_{_V}^{\ast\nu}p_{_P}^\alpha
p_{_V}^\beta
p_{J/\psi}^\theta-32\varepsilon_{J/\psi}^\mu\varepsilon_{_V}^{\ast\alpha}p_{J/\psi}^\nu
p_{_V}^\beta
p_{_P}^\theta+32\varepsilon_{J/\psi}^\mu\varepsilon_{_V}^{\ast\alpha}p_{J/\psi}^\nu
p_{_P}^\beta p_{_V}^\theta)\nonumber
\\
&&+D^\nu(m_q,u,v)\varepsilon_{\mu\nu\alpha\beta}(32\varepsilon_{J/\psi}^\mu\varepsilon_{_V}^{\ast\alpha}
p_{_V}^\beta p_{J/\psi}\cdot
p_{_P}+32\varepsilon_{J/\psi}^\mu\varepsilon_{_V}^{\ast\alpha}
p_{_P}^\beta p_{J/\psi}\cdot p_{_V}\nonumber
\\
&&+32\varepsilon_{J/\psi}^\mu\varepsilon_{_V}^{\ast\beta}
p_{J/\psi}^\alpha p_{_V}\cdot p_{_P})]\}
\end{eqnarray}
where $p_1'=p_4, p_3'=p_1-p_3-p_5$.

For amplitude $\mathcal{A}^{1b1a}$, we have
\begin{eqnarray}
\mathcal{A}^{1b1a}&=&C^{1b1a}\widetilde{H}^{1b1a}(m_q,u,v)\Phi_{J/\psi}\Phi_{V}\Phi_{P}
\end{eqnarray}
with
\begin{eqnarray}
C^{1b1a}&=&\text{Tr}(T^aT^bT^c)\text{Tr}(T^bT^aT^c)\nonumber \\
\widetilde{H}^{1b1a}(m_q,u,v)&=&{i\pi^2\over
(2\pi)^4}g_s^6({1\over 4N_C})^3{1\over
(p_4+p_6)^2[(p_4+p_6-p_2)^2-m_Q^2]}\nonumber
\\
&&m_Qm_q\{D_0(m_q,u,v)\varepsilon_{\mu\nu\alpha\beta}
[64\varepsilon_{J/\psi}^{\mu}\varepsilon_{_V}^{\ast\alpha}p_{J/\psi}^\nu
p_{_{V}}^\beta p_3'\cdot
p_{_P}+64\varepsilon_{J/\psi}^{\mu}\varepsilon_{_V}^{\ast\alpha}p_{J/\psi}^\nu
p_{_{P}}^\beta p_3'\cdot p_{_{V}}\nonumber \\
&&+64\varepsilon_{J/\psi}^{\nu}\varepsilon_{_V}^{\ast\alpha}p_3'^\mu
p_{_{P}}^\beta p_{J/\psi}\cdot
p_{_{V}}-64\varepsilon_{J/\psi}^{\mu}\varepsilon_{_V}^{\ast\beta}p_3'^\nu
p_{J/\psi}^\alpha p_{_V}\cdot
p_{_{P}}]\nonumber \\
&&+[D_\theta(m_q,u,v)\varepsilon_{\mu\nu\alpha\beta}
(32\varepsilon_{J/\psi}^\mu\varepsilon_{_V}^{\ast\nu}p_{_P}^\alpha
p_{_V}^\beta
p_{J/\psi}^\theta-32\varepsilon_{J/\psi}^\mu\varepsilon_{_V}^{\ast\alpha}p_{J/\psi}^\nu
p_{_V}^\beta
p_{_P}^\theta+32\varepsilon_{J/\psi}^\mu\varepsilon_{_V}^{\ast\alpha}p_{J/\psi}^\nu
p_{_P}^\beta p_{_V}^\theta)\nonumber
\\
&&+D^\nu(m_q,u,v)\varepsilon_{\mu\nu\alpha\beta}(-32\varepsilon_{J/\psi}^\mu\varepsilon_{_V}^{\ast\alpha}
p_{_V}^\beta p_{J/\psi}\cdot
p_{_P}-32\varepsilon_{J/\psi}^\mu\varepsilon_{_V}^{\ast\alpha}
p_{_P}^\beta p_{J/\psi}\cdot p_{_V}\nonumber
\\
&&-32\varepsilon_{J/\psi}^\mu\varepsilon_{_V}^{\ast\beta}
p_{J/\psi}^\alpha p_{_V}\cdot p_{_P})]\}
\end{eqnarray}
where $p_1'=-p_5, p_3'=p_4+p_6-p_2$.

For amplitude $\mathcal{A}^{1b1b}$, we have
\begin{eqnarray}
\mathcal{A}^{1b1b}&=&C^{1b1b}\widetilde{H}^{1b1b}(m_q,u,v)\Phi_{J/\psi}\Phi_{V}\Phi_{P}
\end{eqnarray}
with
\begin{eqnarray}
C^{1b1b}&=&\text{Tr}(T^aT^bT^c)\text{Tr}(T^bT^aT^c)\nonumber \\
\widetilde{H}^{1b1b}(m_q,u,v)&=&{i\pi^2\over
(2\pi)^4}g_s^6({1\over 4N_C})^3{1\over
(p_4+p_6)^2[(p_4+p_6-p_2)^2-m_Q^2]}\nonumber
\\
&&m_Qm_q\{D_0(m_q,u,v)\varepsilon_{\mu\nu\alpha\beta}
[-64\varepsilon_{J/\psi}^{\mu}\varepsilon_{_V}^{\ast\alpha}p_{J/\psi}^\nu
p_{_{V}}^\beta p_3'\cdot
p_{_P}-64\varepsilon_{J/\psi}^{\mu}\varepsilon_{_V}^{\ast\alpha}p_{J/\psi}^\nu
p_{_{P}}^\beta p_3'\cdot p_{_{V}}\nonumber \\
&&-64\varepsilon_{J/\psi}^{\nu}\varepsilon_{_V}^{\ast\alpha}p_3'^\mu
p_{_{V}}^\beta p_{J/\psi}\cdot
p_{_{P}}+64\varepsilon_{J/\psi}^{\mu}\varepsilon_{_V}^{\ast\beta}p_3'^\nu
p_{J/\psi}^\alpha p_{_V}\cdot
p_{_{P}}]\nonumber \\
&&+[D_\theta(m_q,u,v)\varepsilon_{\mu\nu\alpha\beta}
(-32\varepsilon_{J/\psi}^\mu\varepsilon_{_V}^{\ast\nu}p_{_P}^\alpha
p_{_V}^\beta
p_{J/\psi}^\theta-32\varepsilon_{J/\psi}^\mu\varepsilon_{_V}^{\ast\alpha}p_{J/\psi}^\nu
p_{_V}^\beta
p_{_P}^\theta+32\varepsilon_{J/\psi}^\mu\varepsilon_{_V}^{\ast\alpha}p_{J/\psi}^\nu
p_{_P}^\beta p_{_V}^\theta)\nonumber
\\
&&+D^\nu(m_q,u,v)\varepsilon_{\mu\nu\alpha\beta}(32\varepsilon_{J/\psi}^\mu\varepsilon_{_V}^{\ast\alpha}
p_{_V}^\beta p_{J/\psi}\cdot
p_{_P}+32\varepsilon_{J/\psi}^\mu\varepsilon_{_V}^{\ast\alpha}
p_{_P}^\beta p_{J/\psi}\cdot p_{_V}\nonumber
\\
&&+32\varepsilon_{J/\psi}^\mu\varepsilon_{_V}^{\ast\beta}
p_{J/\psi}^\alpha p_{_V}\cdot p_{_P})]\}
\end{eqnarray}
where $p_1'=-p_5, p_3'=p_4+p_6-p_2$.

For amplitude $\mathcal{A}^{1b2a}$, we have
\begin{eqnarray}
\mathcal{A}^{1b2a}&=&C^{1b2a}\widetilde{H}^{1b2a}(m_q,u,v)\Phi_{J/\psi}\Phi_{V}\Phi_{P}
\end{eqnarray}
with
\begin{eqnarray}
C^{1b2a}&=&\text{Tr}(T^aT^bT^c)\text{Tr}(T^aT^bT^c)\nonumber \\
\widetilde{H}^{1b2a}(m_q,u,v)&=&-{i\pi^2\over
(2\pi)^4}g_s^6({1\over 4N_C})^3{1\over
(p_4+p_6)^2[(p_4+p_6-p_2)^2-m_Q^2]}\nonumber
\\
&&m_Qm_q\{D_0(m_q,u,v)\varepsilon_{\mu\nu\alpha\beta}
[-64\varepsilon_{J/\psi}^{\mu}\varepsilon_{_V}^{\ast\alpha}p_{J/\psi}^\nu
p_{_{V}}^\beta p_3'\cdot
p_{_P}+64\varepsilon_{J/\psi}^{\mu}\varepsilon_{_V}^{\ast\alpha}p_{J/\psi}^\nu
p_{_{P}}^\beta p_3'\cdot p_{_{V}}\nonumber \\
&&+64\varepsilon_{J/\psi}^{\nu}\varepsilon_{_V}^{\ast\alpha}p_3'^\mu
p_{_{P}}^\beta p_{J/\psi}\cdot
p_{_{V}}-64\varepsilon_{J/\psi}^{\mu}\varepsilon_{_V}^{\ast\beta}p_3'^\nu
p_{J/\psi}^\alpha p_{_V}\cdot
p_{_{P}}]\nonumber \\
&&+[D_\theta(m_q,u,v)\varepsilon_{\mu\nu\alpha\beta}
(-32\varepsilon_{J/\psi}^\mu\varepsilon_{_V}^{\ast\nu}p_{_P}^\alpha
p_{_V}^\beta
p_{J/\psi}^\theta+32\varepsilon_{J/\psi}^\mu\varepsilon_{_V}^{\ast\alpha}p_{J/\psi}^\nu
p_{_V}^\beta
p_{_P}^\theta+32\varepsilon_{J/\psi}^\mu\varepsilon_{_V}^{\ast\alpha}p_{J/\psi}^\nu
p_{_P}^\beta p_{_V}^\theta)\nonumber
\\
&&+D^\nu(m_q,u,v)\varepsilon_{\mu\nu\alpha\beta}(32\varepsilon_{J/\psi}^\mu\varepsilon_{_V}^{\ast\alpha}
p_{_V}^\beta p_{J/\psi}\cdot
p_{_P}-32\varepsilon_{J/\psi}^\mu\varepsilon_{_V}^{\ast\alpha}
p_{_P}^\beta p_{J/\psi}\cdot p_{_V}\nonumber
\\
&&-32\varepsilon_{J/\psi}^\mu\varepsilon_{_V}^{\ast\beta}
p_{J/\psi}^\alpha p_{_V}\cdot p_{_P})]\}
\end{eqnarray}
where $p_1'=-p_5, p_3'=p_4+p_6-p_2$.

For amplitude $\mathcal{A}^{1b2b}$, we have
\begin{eqnarray}
\mathcal{A}^{1b2b}&=&C^{1b2b}\widetilde{H}^{1b2b}(m_q,u,v)\Phi_{J/\psi}\Phi_{V}\Phi_{P}
\end{eqnarray}
with
\begin{eqnarray}
C^{1b2b}&=&\text{Tr}(T^aT^bT^c)\text{Tr}(T^aT^bT^c)\nonumber \\
\widetilde{H}^{1b2b}(m_q,u,v)&=&-{i\pi^2\over
(2\pi)^4}g_s^6({1\over 4N_C})^3{1\over
(p_4+p_6)^2[(p_4+p_6-p_2)^2-m_Q^2]}\nonumber
\\
&&m_Qm_q\{D_0(m_q,u,v)\varepsilon_{\mu\nu\alpha\beta}
[-64\varepsilon_{J/\psi}^{\mu}\varepsilon_{_V}^{\ast\alpha}p_{J/\psi}^\nu
p_{_{V}}^\beta p_3'\cdot
p_{_P}+64\varepsilon_{J/\psi}^{\mu}\varepsilon_{_V}^{\ast\alpha}p_{J/\psi}^\nu
p_{_{P}}^\beta p_3'\cdot p_{_{V}}\nonumber \\
&&-64\varepsilon_{J/\psi}^{\nu}\varepsilon_{_V}^{\ast\alpha}p_3'^\mu
p_{_{V}}^\beta p_{J/\psi}\cdot
p_{_{P}}+64\varepsilon_{J/\psi}^{\mu}\varepsilon_{_V}^{\ast\beta}p_3'^\nu
p_{J/\psi}^\alpha p_{_V}\cdot
p_{_{P}}]\nonumber \\
&&+[D_\theta(m_q,u,v)\varepsilon_{\mu\nu\alpha\beta}
(32\varepsilon_{J/\psi}^\mu\varepsilon_{_V}^{\ast\nu}p_{_P}^\alpha
p_{_V}^\beta
p_{J/\psi}^\theta-32\varepsilon_{J/\psi}^\mu\varepsilon_{_V}^{\ast\alpha}p_{J/\psi}^\nu
p_{_V}^\beta
p_{_P}^\theta-32\varepsilon_{J/\psi}^\mu\varepsilon_{_V}^{\ast\alpha}p_{J/\psi}^\nu
p_{_P}^\beta p_{_V}^\theta)\nonumber
\\
&&+D^\nu(m_q,u,v)\varepsilon_{\mu\nu\alpha\beta}(32\varepsilon_{J/\psi}^\mu\varepsilon_{_V}^{\ast\alpha}
p_{_V}^\beta p_{J/\psi}\cdot
p_{_P}-32\varepsilon_{J/\psi}^\mu\varepsilon_{_V}^{\ast\alpha}
p_{_P}^\beta p_{J/\psi}\cdot p_{_V}\nonumber
\\
&&+32\varepsilon_{J/\psi}^\mu\varepsilon_{_V}^{\ast\beta}
p_{J/\psi}^\alpha p_{_V}\cdot p_{_P})]\}
\end{eqnarray}
where $p_1'=-p_5, p_3'=p_4+p_6-p_2$.

For amplitude $\mathcal{A}^{1c1a}$, we have
\begin{eqnarray}
\mathcal{A}^{1c1a}&=&C^{1c1a}\widetilde{H}^{1c1a}(m_q,u,v)\Phi_{J/\psi}\Phi_{V}\Phi_{P}
\end{eqnarray}
with
\begin{eqnarray}
&&C^{1c1a}=\text{Tr}(T^aT^bT^c)\text{Tr}(T^bT^aT^c)\nonumber \\
&&\widetilde{H}^{1c1a}(m_q,u,v)=-{i\pi^2\over
(2\pi)^4}g_s^6({1\over 4N_C})^3{1\over (p_4+p_5)^2}\nonumber
\\
&&m_Qm_q\{E_0(m_q,u,v)\varepsilon_{\mu\nu\alpha\beta}[-32\varepsilon_{J/\psi}^\mu
\varepsilon_{_V}^{\ast\nu}p_{_P}^\alpha p_{_V}^\beta p_2'\cdot
p_{J/\psi}-32\varepsilon_{J/\psi}^\mu
\varepsilon_{_V}^{\ast\alpha}p_{J/\psi}^\nu p_{_V}^\beta p_2'\cdot
p_{_P}-32\varepsilon_{J/\psi}^\mu
\varepsilon_{_V}^{\ast\alpha}p_{J/\psi}^\nu p_{_P}^\beta p_2'\cdot
p_{_V}\nonumber \\
&&-32\varepsilon_{J/\psi}^\mu
\varepsilon_{_V}^{\ast\nu}p_{_P}^\alpha p_{_V}^\beta p_1'\cdot
p_{J/\psi}-32\varepsilon_{J/\psi}^\mu
\varepsilon_{_V}^{\ast\alpha}p_{J/\psi}^\nu p_{_V}^\beta p_1'\cdot
p_{_P}-32\varepsilon_{J/\psi}^\mu
\varepsilon_{_V}^{\ast\alpha}p_{J/\psi}^\nu p_{_P}^\beta p_1'\cdot
p_{_V}\nonumber \\
&&+32\varepsilon_{J/\psi}^\nu
\varepsilon_{_V}^{\ast\alpha}p_2'^\mu p_{_V}^\beta p_{J/\psi}\cdot
p_{_P} +32\varepsilon_{J/\psi}^\nu
\varepsilon_{_V}^{\ast\alpha}p_1'^\mu p_{_V}^\beta p_{J/\psi}\cdot
p_{_P}-32\varepsilon_{J/\psi}^\nu
\varepsilon_{_V}^{\ast\alpha}p_2'^\mu p_{_P}^\beta p_{J/\psi}\cdot
p_{_V}\nonumber \\
&&-32\varepsilon_{J/\psi}^\nu
\varepsilon_{_V}^{\ast\alpha}p_1'^\mu p_{_P}^\beta p_{J/\psi}\cdot
p_{_V}-32\varepsilon_{J/\psi}^\nu
\varepsilon_{_V}^{\ast\beta}p_2'^\mu p_{J/\psi}^\alpha p_{_P}\cdot
p_{_V}-32\varepsilon_{J/\psi}^\nu
\varepsilon_{_V}^{\ast\beta}p_1'^\mu p_{J/\psi}^\alpha p_{_P}\cdot
p_{_V}]\nonumber \\
&&+[E_{1\theta}(m_q,u,v)\varepsilon_{\mu\nu\alpha\beta}
(-64\varepsilon_{J/\psi}^\mu\varepsilon_{_V}^{\ast\nu}p_{_P}^\alpha
p_{_V}^\beta
p_{J/\psi}^\theta-64\varepsilon_{J/\psi}^\mu\varepsilon_{_V}^{\ast\alpha}p_{J/\psi}^\nu
p_{_V}^\beta
p_{_P}^\theta-64\varepsilon_{J/\psi}^\mu\varepsilon_{_V}^{\ast\alpha}p_{J/\psi}^\nu
p_{_P}^\beta p_{_V}^\theta)\nonumber
\\
&&+E^\nu_1(m_q,u,v)\varepsilon_{\mu\nu\alpha\beta}(-64\varepsilon_{J/\psi}^\mu\varepsilon_{_V}^{\ast\alpha}
p_{_V}^\beta p_{J/\psi}\cdot
p_{_P}+64\varepsilon_{J/\psi}^\mu\varepsilon_{_V}^{\ast\alpha}
p_{_P}^\beta p_{J/\psi}\cdot
p_{_V}+64\varepsilon_{J/\psi}^\mu\varepsilon_{_V}^{\ast\beta}
p_{J/\psi}^\alpha p_{_V}\cdot p_{_P})]\}\nonumber \\
\end{eqnarray}
where $p_1'=p_1, p_2'=p_1-p_4-p_5, p_4'=p_3$.

For amplitude $\mathcal{A}^{1c1b}$, we have
\begin{eqnarray}
\mathcal{A}^{1c1b}&=&C^{1c1b}\widetilde{H}^{1c1b}(m_q,u,v)\Phi_{J/\psi}\Phi_{V}\Phi_{P}
\end{eqnarray}
with
\begin{eqnarray}
&&C^{1c1b}=\text{Tr}(T^aT^bT^c)\text{Tr}(T^bT^aT^c)\nonumber \\
&&\widetilde{H}^{1c1b}(m_q,u,v)=-{i\pi^2\over
(2\pi)^4}g_s^6({1\over 4N_C})^3{1\over (p_4+p_5)^2}\nonumber
\\
&&m_Qm_q\{E_0(m_q,u,v)\varepsilon_{\mu\nu\alpha\beta}[-32\varepsilon_{J/\psi}^\mu
\varepsilon_{_V}^{\ast\nu}p_{_P}^\alpha p_{_V}^\beta p_2'\cdot
p_{J/\psi}+32\varepsilon_{J/\psi}^\mu
\varepsilon_{_V}^{\ast\alpha}p_{J/\psi}^\nu p_{_V}^\beta p_2'\cdot
p_{_P}+32\varepsilon_{J/\psi}^\mu
\varepsilon_{_V}^{\ast\alpha}p_{J/\psi}^\nu p_{_P}^\beta p_2'\cdot
p_{_V}\nonumber \\
&&-32\varepsilon_{J/\psi}^\mu
\varepsilon_{_V}^{\ast\nu}p_{_P}^\alpha p_{_V}^\beta p_1'\cdot
p_{J/\psi}+32\varepsilon_{J/\psi}^\mu
\varepsilon_{_V}^{\ast\alpha}p_{J/\psi}^\nu p_{_V}^\beta p_1'\cdot
p_{_P}+32\varepsilon_{J/\psi}^\mu
\varepsilon_{_V}^{\ast\alpha}p_{J/\psi}^\nu p_{_P}^\beta p_1'\cdot
p_{_V}\nonumber \\
&&+32\varepsilon_{J/\psi}^\nu
\varepsilon_{_V}^{\ast\alpha}p_2'^\mu p_{_V}^\beta p_{J/\psi}\cdot
p_{_P} +32\varepsilon_{J/\psi}^\nu
\varepsilon_{_V}^{\ast\alpha}p_1'^\mu p_{_V}^\beta p_{J/\psi}\cdot
p_{_P}-32\varepsilon_{J/\psi}^\nu
\varepsilon_{_V}^{\ast\alpha}p_2'^\mu p_{_P}^\beta p_{J/\psi}\cdot
p_{_V}\nonumber \\
&&-32\varepsilon_{J/\psi}^\nu
\varepsilon_{_V}^{\ast\alpha}p_1'^\mu p_{_P}^\beta p_{J/\psi}\cdot
p_{_V}+32\varepsilon_{J/\psi}^\nu
\varepsilon_{_V}^{\ast\beta}p_2'^\mu p_{J/\psi}^\alpha p_{_P}\cdot
p_{_V}+32\varepsilon_{J/\psi}^\nu
\varepsilon_{_V}^{\ast\beta}p_1'^\mu p_{J/\psi}^\alpha p_{_P}\cdot
p_{_V}]\nonumber \\
&&+[E_{1\theta}(m_q,u,v)\varepsilon_{\mu\nu\alpha\beta}
(-64\varepsilon_{J/\psi}^\mu\varepsilon_{_V}^{\ast\nu}p_{_P}^\alpha
p_{_V}^\beta
p_{J/\psi}^\theta+64\varepsilon_{J/\psi}^\mu\varepsilon_{_V}^{\ast\alpha}p_{J/\psi}^\nu
p_{_V}^\beta
p_{_P}^\theta+64\varepsilon_{J/\psi}^\mu\varepsilon_{_V}^{\ast\alpha}p_{J/\psi}^\nu
p_{_P}^\beta p_{_V}^\theta)\nonumber
\\
&&+E^\nu_1(m_q,u,v)\varepsilon_{\mu\nu\alpha\beta}(-64\varepsilon_{J/\psi}^\mu\varepsilon_{_V}^{\ast\alpha}
p_{_V}^\beta p_{J/\psi}\cdot
p_{_P}+64\varepsilon_{J/\psi}^\mu\varepsilon_{_V}^{\ast\alpha}
p_{_P}^\beta p_{J/\psi}\cdot
p_{_V}-64\varepsilon_{J/\psi}^\mu\varepsilon_{_V}^{\ast\beta}
p_{J/\psi}^\alpha p_{_V}\cdot p_{_P})]\}\nonumber \\
\end{eqnarray}
where $p_1'=p_1, p_2'=p_1-p_4-p_5, p_4'=p_3$.

For amplitude $\mathcal{A}^{1c2a}$, we have
\begin{eqnarray}
\mathcal{A}^{1c2a}&=&C^{1c2a}\widetilde{H}^{1c2a}(m_q,u,v)\Phi_{J/\psi}\Phi_{V}\Phi_{P}
\end{eqnarray}
with
\begin{eqnarray}
&&C^{1c2a}=\text{Tr}(T^aT^bT^c)\text{Tr}(T^aT^bT^c)\nonumber \\
&&\widetilde{H}^{1c2a}(m_q,u,v)={i\pi^2\over
(2\pi)^4}g_s^6({1\over 4N_C})^3{1\over (p_4+p_5)^2}\nonumber
\\
&&m_Qm_q\{E_0(m_q,u,v)\varepsilon_{\mu\nu\alpha\beta}[32\varepsilon_{J/\psi}^\mu
\varepsilon_{_V}^{\ast\nu}p_{_P}^\alpha p_{_V}^\beta p_2'\cdot
p_{J/\psi}+32\varepsilon_{J/\psi}^\mu
\varepsilon_{_V}^{\ast\alpha}p_{J/\psi}^\nu p_{_V}^\beta p_2'\cdot
p_{_P}-32\varepsilon_{J/\psi}^\mu
\varepsilon_{_V}^{\ast\alpha}p_{J/\psi}^\nu p_{_P}^\beta p_2'\cdot
p_{_V}\nonumber \\
&&+32\varepsilon_{J/\psi}^\mu
\varepsilon_{_V}^{\ast\nu}p_{_P}^\alpha p_{_V}^\beta p_1'\cdot
p_{J/\psi}+32\varepsilon_{J/\psi}^\mu
\varepsilon_{_V}^{\ast\alpha}p_{J/\psi}^\nu p_{_V}^\beta p_1'\cdot
p_{_P}-32\varepsilon_{J/\psi}^\mu
\varepsilon_{_V}^{\ast\alpha}p_{J/\psi}^\nu p_{_P}^\beta p_1'\cdot
p_{_V}\nonumber \\
&&-32\varepsilon_{J/\psi}^\nu
\varepsilon_{_V}^{\ast\alpha}p_2'^\mu p_{_V}^\beta p_{J/\psi}\cdot
p_{_P}-32\varepsilon_{J/\psi}^\nu
\varepsilon_{_V}^{\ast\alpha}p_1'^\mu p_{_V}^\beta p_{J/\psi}\cdot
p_{_P}-32\varepsilon_{J/\psi}^\nu
\varepsilon_{_V}^{\ast\alpha}p_2'^\mu p_{_P}^\beta p_{J/\psi}\cdot
p_{_V}\nonumber \\
&&-32\varepsilon_{J/\psi}^\nu
\varepsilon_{_V}^{\ast\alpha}p_1'^\mu p_{_P}^\beta p_{J/\psi}\cdot
p_{_V}-32\varepsilon_{J/\psi}^\nu
\varepsilon_{_V}^{\ast\beta}p_2'^\mu p_{J/\psi}^\alpha p_{_P}\cdot
p_{_V}-32\varepsilon_{J/\psi}^\nu
\varepsilon_{_V}^{\ast\beta}p_1'^\mu p_{J/\psi}^\alpha p_{_P}\cdot
p_{_V}]\nonumber \\
&&+[E_{1\theta}(m_q,u,v)\varepsilon_{\mu\nu\alpha\beta}
(64\varepsilon_{J/\psi}^\mu\varepsilon_{_V}^{\ast\nu}p_{_P}^\alpha
p_{_V}^\beta
p_{J/\psi}^\theta+64\varepsilon_{J/\psi}^\mu\varepsilon_{_V}^{\ast\alpha}p_{J/\psi}^\nu
p_{_V}^\beta
p_{_P}^\theta-64\varepsilon_{J/\psi}^\mu\varepsilon_{_V}^{\ast\alpha}p_{J/\psi}^\nu
p_{_P}^\beta p_{_V}^\theta)\nonumber
\\
&&+E^\nu_1(m_q,u,v)\varepsilon_{\mu\nu\alpha\beta}(64\varepsilon_{J/\psi}^\mu\varepsilon_{_V}^{\ast\alpha}
p_{_V}^\beta p_{J/\psi}\cdot
p_{_P}+64\varepsilon_{J/\psi}^\mu\varepsilon_{_V}^{\ast\alpha}
p_{_P}^\beta p_{J/\psi}\cdot
p_{_V}+64\varepsilon_{J/\psi}^\mu\varepsilon_{_V}^{\ast\beta}
p_{J/\psi}^\alpha p_{_V}\cdot p_{_P})]\}\nonumber \\
\end{eqnarray}
where $p_1'=p_1, p_2'=p_1-p_4-p_5, p_4'=p_3$.

For amplitude $\mathcal{A}^{1c2b}$, we have
\begin{eqnarray}
\mathcal{A}^{1c2b}&=&C^{1c2b}\widetilde{H}^{1c2b}(m_q,u,v)\Phi_{J/\psi}\Phi_{V}\Phi_{P}
\end{eqnarray}
with
\begin{eqnarray}
&&C^{1c2b}=\text{Tr}(T^aT^bT^c)\text{Tr}(T^aT^bT^c)\nonumber \\
&&\widetilde{H}^{1c2b}(m_q,u,v)={i\pi^2\over
(2\pi)^4}g_s^6({1\over 4N_C})^3{1\over (p_4+p_5)^2}\nonumber
\\
&&m_Qm_q\{E_0(m_q,u,v)\varepsilon_{\mu\nu\alpha\beta}[32\varepsilon_{J/\psi}^\mu
\varepsilon_{_V}^{\ast\nu}p_{_P}^\alpha p_{_V}^\beta p_2'\cdot
p_{J/\psi}+32\varepsilon_{J/\psi}^\mu
\varepsilon_{_V}^{\ast\alpha}p_{J/\psi}^\nu p_{_V}^\beta p_2'\cdot
p_{_P}-32\varepsilon_{J/\psi}^\mu
\varepsilon_{_V}^{\ast\alpha}p_{J/\psi}^\nu p_{_P}^\beta p_2'\cdot
p_{_V}\nonumber \\
&&+32\varepsilon_{J/\psi}^\mu
\varepsilon_{_V}^{\ast\nu}p_{_P}^\alpha p_{_V}^\beta p_1'\cdot
p_{J/\psi}+32\varepsilon_{J/\psi}^\mu
\varepsilon_{_V}^{\ast\alpha}p_{J/\psi}^\nu p_{_V}^\beta p_1'\cdot
p_{_P}-32\varepsilon_{J/\psi}^\mu
\varepsilon_{_V}^{\ast\alpha}p_{J/\psi}^\nu p_{_P}^\beta p_1'\cdot
p_{_V}\nonumber \\
&&+32\varepsilon_{J/\psi}^\nu
\varepsilon_{_V}^{\ast\alpha}p_2'^\mu p_{_V}^\beta p_{J/\psi}\cdot
p_{_P}-32\varepsilon_{J/\psi}^\nu
\varepsilon_{_V}^{\ast\alpha}p_1'^\mu p_{_V}^\beta p_{J/\psi}\cdot
p_{_P}+32\varepsilon_{J/\psi}^\nu
\varepsilon_{_V}^{\ast\alpha}p_2'^\mu p_{_P}^\beta p_{J/\psi}\cdot
p_{_V}\nonumber \\
&&+32\varepsilon_{J/\psi}^\nu
\varepsilon_{_V}^{\ast\alpha}p_1'^\mu p_{_P}^\beta p_{J/\psi}\cdot
p_{_V}+32\varepsilon_{J/\psi}^\nu
\varepsilon_{_V}^{\ast\beta}p_2'^\mu p_{J/\psi}^\alpha p_{_P}\cdot
p_{_V}+32\varepsilon_{J/\psi}^\nu
\varepsilon_{_V}^{\ast\beta}p_1'^\mu p_{J/\psi}^\alpha p_{_P}\cdot
p_{_V}]\nonumber \\
&&+[E_{1\theta}(m_q,u,v)\varepsilon_{\mu\nu\alpha\beta}
(64\varepsilon_{J/\psi}^\mu\varepsilon_{_V}^{\ast\nu}p_{_P}^\alpha
p_{_V}^\beta
p_{J/\psi}^\theta+64\varepsilon_{J/\psi}^\mu\varepsilon_{_V}^{\ast\alpha}p_{J/\psi}^\nu
p_{_V}^\beta
p_{_P}^\theta-64\varepsilon_{J/\psi}^\mu\varepsilon_{_V}^{\ast\alpha}p_{J/\psi}^\nu
p_{_P}^\beta p_{_V}^\theta)\nonumber
\\
&&+E^\nu_1(m_q,u,v)\varepsilon_{\mu\nu\alpha\beta}(-64\varepsilon_{J/\psi}^\mu\varepsilon_{_V}^{\ast\alpha}
p_{_V}^\beta p_{J/\psi}\cdot
p_{_P}-64\varepsilon_{J/\psi}^\mu\varepsilon_{_V}^{\ast\alpha}
p_{_P}^\beta p_{J/\psi}\cdot
p_{_V}-64\varepsilon_{J/\psi}^\mu\varepsilon_{_V}^{\ast\beta}
p_{J/\psi}^\alpha p_{_V}\cdot p_{_P})]\}\nonumber \\
\end{eqnarray}
where $p_1'=p_1, p_2'=p_1-p_4-p_5, p_4'=p_3$.

For amplitude $\mathcal{A}^{2a1a}$, we have
\begin{eqnarray}
\mathcal{A}^{2a1a}&=&C^{2a1a}\widetilde{H}^{2a1a}(m_q,u,v)\Phi_{J/\psi}\Phi_{V}\Phi_{P}
\end{eqnarray}
with
\begin{eqnarray}
C^{2a1a}&=&\text{Tr}(T^aT^bT^c)\text{Tr}(T^bT^aT^c)\nonumber \\
\widetilde{H}^{2a1a}(m_q,u,v)&=&{i\pi^2\over
(2\pi)^4}g_s^6({1\over 4N_C})^3{1\over
(p_3+p_5)^2[(p_1-p_3-p_5)^2-m_Q^2]}\nonumber
\\
&&m_Qm_q\{D_0(m_q,u,v)\varepsilon_{\mu\nu\alpha\beta}
[-64\varepsilon_{J/\psi}^{\mu}\varepsilon_{_V}^{\ast\alpha}p_{J/\psi}^\nu
p_{_{V}}^\beta p_3'\cdot
p_{_P}+64\varepsilon_{J/\psi}^{\mu}\varepsilon_{_V}^{\ast\alpha}p_{J/\psi}^\nu
p_{_{P}}^\beta p_3'\cdot p_{_{V}}\nonumber \\
&&+64\varepsilon_{J/\psi}^{\nu}\varepsilon_{_V}^{\ast\alpha}p_3'^\mu
p_{_{P}}^\beta p_{J/\psi}\cdot
p_{_{V}}-64\varepsilon_{J/\psi}^{\mu}\varepsilon_{_V}^{\ast\beta}p_3'^\nu
p_{J/\psi}^\alpha p_{_P}\cdot
p_{_{V}}]\nonumber \\
&&+[D_\theta(m_q,u,v)\varepsilon_{\mu\nu\alpha\beta}
(-32\varepsilon_{J/\psi}^\mu\varepsilon_{_V}^{\ast\nu}p_{_P}^\alpha
p_{_V}^\beta
p_{J/\psi}^\theta+32\varepsilon_{J/\psi}^\mu\varepsilon_{_V}^{\ast\alpha}p_{J/\psi}^\nu
p_{_V}^\beta
p_{_P}^\theta+32\varepsilon_{J/\psi}^\mu\varepsilon_{_V}^{\ast\alpha}p_{J/\psi}^\nu
p_{_P}^\beta p_{_V}^\theta)\nonumber
\\
&&+D^\nu(m_q,u,v)\varepsilon_{\mu\nu\alpha\beta}(32\varepsilon_{J/\psi}^\mu\varepsilon_{_V}^{\ast\alpha}
p_{_V}^\beta p_{J/\psi}\cdot
p_{_P}-32\varepsilon_{J/\psi}^\mu\varepsilon_{_V}^{\ast\alpha}
p_{_P}^\beta p_{J/\psi}\cdot p_{_V}\nonumber
\\
&&-32\varepsilon_{J/\psi}^\mu\varepsilon_{_V}^{\ast\beta}
p_{J/\psi}^\alpha p_{_V}\cdot p_{_P})]\}
\end{eqnarray}
where $p_1'=p_6, p_3'=p_1-p_3-p_5$.

For amplitude $\mathcal{A}^{2a1b}$, we have
\begin{eqnarray}
\mathcal{A}^{2a1b}&=&C^{2a1b}\widetilde{H}^{2a1b}(m_q,u,v)\Phi_{J/\psi}\Phi_{V}\Phi_{P}
\end{eqnarray}
with
\begin{eqnarray}
C^{2a1b}&=&\text{Tr}(T^aT^bT^c)\text{Tr}(T^bT^aT^c)\nonumber \\
\widetilde{H}^{2a1b}(m_q,u,v)&=&{i\pi^2\over
(2\pi)^4}g_s^6({1\over 4N_C})^3{1\over
(p_3+p_5)^2[(p_1-p_3-p_5)^2-m_Q^2]}\nonumber
\\
&&m_Qm_q\{D_0(m_q,u,v)\varepsilon_{\mu\nu\alpha\beta}
[-64\varepsilon_{J/\psi}^{\mu}\varepsilon_{_V}^{\ast\alpha}p_{J/\psi}^\nu
p_{_{V}}^\beta p_3'\cdot
p_{_P}+64\varepsilon_{J/\psi}^{\mu}\varepsilon_{_V}^{\ast\alpha}p_{J/\psi}^\nu
p_{_{P}}^\beta p_3'\cdot p_{_{V}}\nonumber \\
&&-64\varepsilon_{J/\psi}^{\nu}\varepsilon_{_V}^{\ast\alpha}p_3'^\mu
p_{_{P}}^\beta p_{J/\psi}\cdot
p_{_{V}}+64\varepsilon_{J/\psi}^{\mu}\varepsilon_{_V}^{\ast\beta}p_3'^\nu
p_{J/\psi}^\alpha p_{_P}\cdot
p_{_{V}}]\nonumber \\
&&+[D_\theta(m_q,u,v)\varepsilon_{\mu\nu\alpha\beta}
(32\varepsilon_{J/\psi}^\mu\varepsilon_{_V}^{\ast\nu}p_{_P}^\alpha
p_{_V}^\beta
p_{J/\psi}^\theta-32\varepsilon_{J/\psi}^\mu\varepsilon_{_V}^{\ast\alpha}p_{J/\psi}^\nu
p_{_V}^\beta
p_{_P}^\theta-32\varepsilon_{J/\psi}^\mu\varepsilon_{_V}^{\ast\alpha}p_{J/\psi}^\nu
p_{_P}^\beta p_{_V}^\theta)\nonumber
\\
&&+D^\nu(m_q,u,v)\varepsilon_{\mu\nu\alpha\beta}(32\varepsilon_{J/\psi}^\mu\varepsilon_{_V}^{\ast\alpha}
p_{_V}^\beta p_{J/\psi}\cdot
p_{_P}-32\varepsilon_{J/\psi}^\mu\varepsilon_{_V}^{\ast\alpha}
p_{_P}^\beta p_{J/\psi}\cdot p_{_V}\nonumber
\\
&&+32\varepsilon_{J/\psi}^\mu\varepsilon_{_V}^{\ast\beta}
p_{J/\psi}^\alpha p_{_V}\cdot p_{_P})]\}
\end{eqnarray}
where $p_1'=p_6, p_3'=p_1-p_3-p_5$.

For amplitude $\mathcal{A}^{2a2a}$, we have
\begin{eqnarray}
\mathcal{A}^{2a2a}&=&C^{2a2a}\widetilde{H}^{2a2a}(m_q,u,v)\Phi_{J/\psi}\Phi_{V}\Phi_{P}
\end{eqnarray}
with
\begin{eqnarray}
C^{2a2a}&=&\text{Tr}(T^aT^bT^c)\text{Tr}(T^aT^bT^c)\nonumber \\
\widetilde{H}^{2a2a}(m_q,u,v)&=&-{i\pi^2\over
(2\pi)^4}g_s^6({1\over 4N_C})^3{1\over
(p_3+p_5)^2[(p_1-p_3-p_5)^2-m_Q^2]}\nonumber
\\
&&m_Qm_q\{D_0(m_q,u,v)\varepsilon_{\mu\nu\alpha\beta}
[64\varepsilon_{J/\psi}^{\mu}\varepsilon_{_V}^{\ast\alpha}p_{J/\psi}^\nu
p_{_{V}}^\beta p_3'\cdot
p_{_P}+64\varepsilon_{J/\psi}^{\mu}\varepsilon_{_V}^{\ast\alpha}p_{J/\psi}^\nu
p_{_{P}}^\beta p_3'\cdot p_{_{V}}\nonumber \\
&&+64\varepsilon_{J/\psi}^{\nu}\varepsilon_{_V}^{\ast\alpha}p_3'^\mu
p_{_{P}}^\beta p_{J/\psi}\cdot
p_{_{V}}-64\varepsilon_{J/\psi}^{\mu}\varepsilon_{_V}^{\ast\beta}p_3'^\nu
p_{J/\psi}^\alpha p_{_P}\cdot
p_{_{V}}]\nonumber \\
&&+[D_\theta(m_q,u,v)\varepsilon_{\mu\nu\alpha\beta}
(32\varepsilon_{J/\psi}^\mu\varepsilon_{_V}^{\ast\nu}p_{_P}^\alpha
p_{_V}^\beta
p_{J/\psi}^\theta-32\varepsilon_{J/\psi}^\mu\varepsilon_{_V}^{\ast\alpha}p_{J/\psi}^\nu
p_{_V}^\beta
p_{_P}^\theta+32\varepsilon_{J/\psi}^\mu\varepsilon_{_V}^{\ast\alpha}p_{J/\psi}^\nu
p_{_P}^\beta p_{_V}^\theta)\nonumber
\\
&&+D^\nu(m_q,u,v)\varepsilon_{\mu\nu\alpha\beta}(-32\varepsilon_{J/\psi}^\mu\varepsilon_{_V}^{\ast\alpha}
p_{_V}^\beta p_{J/\psi}\cdot
p_{_P}-32\varepsilon_{J/\psi}^\mu\varepsilon_{_V}^{\ast\alpha}
p_{_P}^\beta p_{J/\psi}\cdot p_{_V}\nonumber
\\
&&-32\varepsilon_{J/\psi}^\mu\varepsilon_{_V}^{\ast\beta}
p_{J/\psi}^\alpha p_{_V}\cdot p_{_P})]\}
\end{eqnarray}
where $p_1'=p_6, p_3'=p_1-p_3-p_5$.

For amplitude $\mathcal{A}^{2a2b}$, we have
\begin{eqnarray}
\mathcal{A}^{2a2b}&=&C^{2a2b}\widetilde{H}^{2a2b}(m_q,u,v)\Phi_{J/\psi}\Phi_{V}\Phi_{P}
\end{eqnarray}
with
\begin{eqnarray}
C^{2a2b}&=&\text{Tr}(T^aT^bT^c)\text{Tr}(T^aT^bT^c)\nonumber \\
\widetilde{H}^{2a2b}(m_q,u,v)&=&-{i\pi^2\over
(2\pi)^4}g_s^6({1\over 4N_C})^3{1\over
(p_3+p_5)^2[(p_1-p_3-p_5)^2-m_Q^2]}\nonumber
\\
&&m_Qm_q\{D_0(m_q,u,v)\varepsilon_{\mu\nu\alpha\beta}
[-64\varepsilon_{J/\psi}^{\mu}\varepsilon_{_V}^{\ast\alpha}p_{J/\psi}^\nu
p_{_{V}}^\beta p_3'\cdot
p_{_P}-64\varepsilon_{J/\psi}^{\mu}\varepsilon_{_V}^{\ast\alpha}p_{J/\psi}^\nu
p_{_{P}}^\beta p_3'\cdot p_{_{V}}\nonumber \\
&&-64\varepsilon_{J/\psi}^{\nu}\varepsilon_{_V}^{\ast\alpha}p_3'^\mu
p_{_{P}}^\beta p_{J/\psi}\cdot
p_{_{V}}+64\varepsilon_{J/\psi}^{\mu}\varepsilon_{_V}^{\ast\beta}p_3'^\nu
p_{J/\psi}^\alpha p_{_P}\cdot
p_{_{V}}]\nonumber \\
&&+[D_\theta(m_q,u,v)\varepsilon_{\mu\nu\alpha\beta}
(-32\varepsilon_{J/\psi}^\mu\varepsilon_{_V}^{\ast\nu}p_{_P}^\alpha
p_{_V}^\beta
p_{J/\psi}^\theta-32\varepsilon_{J/\psi}^\mu\varepsilon_{_V}^{\ast\alpha}p_{J/\psi}^\nu
p_{_V}^\beta
p_{_P}^\theta+32\varepsilon_{J/\psi}^\mu\varepsilon_{_V}^{\ast\alpha}p_{J/\psi}^\nu
p_{_P}^\beta p_{_V}^\theta)\nonumber
\\
&&+D^\nu(m_q,u,v)\varepsilon_{\mu\nu\alpha\beta}(32\varepsilon_{J/\psi}^\mu\varepsilon_{_V}^{\ast\alpha}
p_{_V}^\beta p_{J/\psi}\cdot
p_{_P}+32\varepsilon_{J/\psi}^\mu\varepsilon_{_V}^{\ast\alpha}
p_{_P}^\beta p_{J/\psi}\cdot p_{_V}\nonumber
\\
&&+32\varepsilon_{J/\psi}^\mu\varepsilon_{_V}^{\ast\beta}
p_{J/\psi}^\alpha p_{_V}\cdot p_{_P})]\}
\end{eqnarray}
where $p_1'=p_6, p_3'=p_1-p_3-p_5$.

For amplitude $\mathcal{A}^{2b1a}$, we have
\begin{eqnarray}
\mathcal{A}^{2b1a}&=&C^{2b1a}\widetilde{H}^{2b1a}(m_q,u,v)\Phi_{J/\psi}\Phi_{V}\Phi_{P}
\end{eqnarray}
with
\begin{eqnarray}
C^{2b1a}&=&\text{Tr}(T^aT^bT^c)\text{Tr}(T^aT^bT^c)\nonumber \\
\widetilde{H}^{2b1a}(m_q,u,v)&=&-{i\pi^2\over
(2\pi)^4}g_s^6({1\over 4N_C})^3{1\over
(p_4+p_6)^2[(p_4+p_6-p_2)^2-m_Q^2]}\nonumber
\\
&&m_Qm_q\{D_0(m_q,u,v)\varepsilon_{\mu\nu\alpha\beta}
[64\varepsilon_{J/\psi}^{\mu}\varepsilon_{_V}^{\ast\nu}p_{_{P}}^\alpha
p_{_{V}}^\beta p_3'\cdot
p_{J/\psi}+64\varepsilon_{J/\psi}^{\mu}\varepsilon_{_V}^{\ast\alpha}p_{J/\psi}^\nu
p_{_{V}}^\beta p_3'\cdot p_{_{P}}\nonumber \\
&&+64\varepsilon_{J/\psi}^{\mu}\varepsilon_{_V}^{\ast\alpha}p_{J/\psi}^\nu
p_{_{P}}^\beta p_3'\cdot
p_{_{V}}+64\varepsilon_{J/\psi}^{\nu}\varepsilon_{_V}^{\ast\alpha}p_3'^\mu
p_{_{P}}^\beta p_{J/\psi}\cdot
p_{_{V}}-64\varepsilon_{J/\psi}^{\mu}\varepsilon_{_V}^{\ast\beta}p_{J/\psi}^\alpha
p_3'^\nu p_{_{V}}\cdot p_{_{P}}]\nonumber \\
&&+[D_\theta(m_q,u,v)\varepsilon_{\mu\nu\alpha\beta}
(-32\varepsilon_{J/\psi}^\mu\varepsilon_{_V}^{\ast\nu}p_{_P}^\alpha
p_{_V}^\beta
p_{J/\psi}^\theta-32\varepsilon_{J/\psi}^\mu\varepsilon_{_V}^{\ast\alpha}p_{J/\psi}^\nu
p_{_V}^\beta
p_{_P}^\theta+32\varepsilon_{J/\psi}^\mu\varepsilon_{_V}^{\ast\alpha}p_{J/\psi}^\nu
p_{_P}^\beta p_{_V}^\theta)\nonumber
\\
&&+D^\nu(m_q,u,v)\varepsilon_{\mu\nu\alpha\beta}(-32\varepsilon_{J/\psi}^\mu\varepsilon_{_V}^{\ast\alpha}
p_{_V}^\beta p_{J/\psi}\cdot
p_{_P}-32\varepsilon_{J/\psi}^\mu\varepsilon_{_V}^{\ast\alpha}
p_{_P}^\beta p_{J/\psi}\cdot p_{_V}\nonumber
\\
&&-32\varepsilon_{J/\psi}^\mu\varepsilon_{_V}^{\ast\beta}
p_{J/\psi}^\alpha p_{_V}\cdot p_{_P})]\}
\end{eqnarray}
where $p_1'=-p_3, p_3'=p_4+p_6-p_2$.

For amplitude $\mathcal{A}^{2b1b}$, we have
\begin{eqnarray}
\mathcal{A}^{2b1b}&=&C^{2b1b}\widetilde{H}^{2b1b}(m_q,u,v)\Phi_{J/\psi}\Phi_{V}\Phi_{P}
\end{eqnarray}
with
\begin{eqnarray}
C^{2b1b}&=&\text{Tr}(T^aT^bT^c)\text{Tr}(T^aT^bT^c)\nonumber \\
\widetilde{H}^{2b1b}(m_q,u,v)&=&-{i\pi^2\over
(2\pi)^4}g_s^6({1\over 4N_C})^3{1\over
(p_4+p_6)^2[(p_4+p_6-p_2)^2-m_Q^2]}\nonumber
\\
&&m_Qm_q\{D_0(m_q,u,v)\varepsilon_{\mu\nu\alpha\beta}
[64\varepsilon_{J/\psi}^{\mu}\varepsilon_{_V}^{\ast\nu}p_{_{P}}^\alpha
p_{_{V}}^\beta p_3'\cdot
p_{J/\psi}-64\varepsilon_{J/\psi}^{\mu}\varepsilon_{_V}^{\ast\alpha}p_{J/\psi}^\nu
p_{_{V}}^\beta p_3'\cdot p_{_{P}}\nonumber \\
&&-64\varepsilon_{J/\psi}^{\mu}\varepsilon_{_V}^{\ast\alpha}p_{J/\psi}^\nu
p_{_{P}}^\beta p_3'\cdot
p_{_{V}}-64\varepsilon_{J/\psi}^{\nu}\varepsilon_{_V}^{\ast\alpha}p_3'^\mu
p_{_{P}}^\beta p_{J/\psi}\cdot
p_{_{V}}+64\varepsilon_{J/\psi}^{\mu}\varepsilon_{_V}^{\ast\beta}p_{J/\psi}^\alpha
p_3'^\nu p_{_{V}}\cdot p_{_{P}}]\nonumber \\
&&+[D_\theta(m_q,u,v)\varepsilon_{\mu\nu\alpha\beta}
(-96\varepsilon_{J/\psi}^\mu\varepsilon_{_V}^{\ast\nu}p_{_P}^\alpha
p_{_V}^\beta
p_{J/\psi}^\theta-32\varepsilon_{J/\psi}^\mu\varepsilon_{_V}^{\ast\alpha}p_{J/\psi}^\nu
p_{_V}^\beta
p_{_P}^\theta+32\varepsilon_{J/\psi}^\mu\varepsilon_{_V}^{\ast\alpha}p_{J/\psi}^\nu
p_{_P}^\beta p_{_V}^\theta)\nonumber
\\
&&+D^\nu(m_q,u,v)\varepsilon_{\mu\nu\alpha\beta}(32\varepsilon_{J/\psi}^\mu\varepsilon_{_V}^{\ast\alpha}
p_{_V}^\beta p_{J/\psi}\cdot
p_{_P}+32\varepsilon_{J/\psi}^\mu\varepsilon_{_V}^{\ast\alpha}
p_{_P}^\beta p_{J/\psi}\cdot p_{_V}\nonumber
\\
&&+32\varepsilon_{J/\psi}^\mu\varepsilon_{_V}^{\ast\beta}
p_{J/\psi}^\alpha p_{_V}\cdot p_{_P})]\}
\end{eqnarray}
where $p_1'=-p_3, p_3'=p_4+p_6-p_2$.

For amplitude $\mathcal{A}^{2b2a}$, we have
\begin{eqnarray}
\mathcal{A}^{2b2a}&=&C^{2b2a}\widetilde{H}^{2b2a}(m_q,u,v)\Phi_{J/\psi}\Phi_{V}\Phi_{P}
\end{eqnarray}
with
\begin{eqnarray}
C^{2b2a}&=&\text{Tr}(T^aT^bT^c)\text{Tr}(T^bT^aT^c)\nonumber \\
\widetilde{H}^{2b2a}(m_q,u,v)&=&{i\pi^2\over
(2\pi)^4}g_s^6({1\over 4N_C})^3{1\over
(p_4+p_6)^2[(p_4+p_6-p_2)^2-m_Q^2]}\nonumber
\\
&&m_Qm_q\{D_0(m_q,u,v)\varepsilon_{\mu\nu\alpha\beta}
[-64\varepsilon_{J/\psi}^{\mu}\varepsilon_{_V}^{\ast\nu}p_{_{P}}^\alpha
p_{_{V}}^\beta p_3'\cdot
p_{J/\psi}-64\varepsilon_{J/\psi}^{\mu}\varepsilon_{_V}^{\ast\alpha}p_{J/\psi}^\nu
p_{_{V}}^\beta p_3'\cdot p_{_{P}}\nonumber \\
&&+64\varepsilon_{J/\psi}^{\mu}\varepsilon_{_V}^{\ast\alpha}p_{J/\psi}^\nu
p_{_{P}}^\beta p_3'\cdot
p_{_{V}}+64\varepsilon_{J/\psi}^{\nu}\varepsilon_{_V}^{\ast\alpha}p_3'^\mu
p_{_{P}}^\beta p_{J/\psi}\cdot
p_{_{V}}-64\varepsilon_{J/\psi}^{\mu}\varepsilon_{_V}^{\ast\beta}p_{J/\psi}^\alpha
p_3'^\nu p_{_{V}}\cdot p_{_{P}}]\nonumber \\
&&+[D_\theta(m_q,u,v)\varepsilon_{\mu\nu\alpha\beta}
(32\varepsilon_{J/\psi}^\mu\varepsilon_{_V}^{\ast\nu}p_{_P}^\alpha
p_{_V}^\beta
p_{J/\psi}^\theta+32\varepsilon_{J/\psi}^\mu\varepsilon_{_V}^{\ast\alpha}p_{J/\psi}^\nu
p_{_V}^\beta
p_{_P}^\theta+32\varepsilon_{J/\psi}^\mu\varepsilon_{_V}^{\ast\alpha}p_{J/\psi}^\nu
p_{_P}^\beta p_{_V}^\theta)\nonumber
\\
&&+D^\nu(m_q,u,v)\varepsilon_{\mu\nu\alpha\beta}(32\varepsilon_{J/\psi}^\mu\varepsilon_{_V}^{\ast\alpha}
p_{_V}^\beta p_{J/\psi}\cdot
p_{_P}-32\varepsilon_{J/\psi}^\mu\varepsilon_{_V}^{\ast\alpha}
p_{_P}^\beta p_{J/\psi}\cdot p_{_V}\nonumber
\\
&&-32\varepsilon_{J/\psi}^\mu\varepsilon_{_V}^{\ast\beta}
p_{J/\psi}^\alpha p_{_V}\cdot p_{_P})]\}
\end{eqnarray}
where $p_1'=-p_3, p_3'=p_4+p_6-p_2$.

For amplitude $\mathcal{A}^{2b2b}$, we have
\begin{eqnarray}
\mathcal{A}^{2b2b}&=&C^{2b2b}\widetilde{H}^{2b2b}(m_q,u,v)\Phi_{J/\psi}\Phi_{V}\Phi_{P}
\end{eqnarray}
with
\begin{eqnarray}
C^{2b2b}&=&\text{Tr}(T^aT^bT^c)\text{Tr}(T^bT^aT^c)\nonumber \\
\widetilde{H}^{2b2b}(m_q,u,v)&=&{i\pi^2\over
(2\pi)^4}g_s^6({1\over 4N_C})^3{1\over
(p_4+p_6)^2[(p_4+p_6-p_2)^2-m_Q^2]}\nonumber
\\
&&m_Qm_q\{D_0(m_q,u,v)\varepsilon_{\mu\nu\alpha\beta}
[-64\varepsilon_{J/\psi}^{\mu}\varepsilon_{_V}^{\ast\nu}p_{_{P}}^\alpha
p_{_{V}}^\beta p_3'\cdot
p_{J/\psi}-64\varepsilon_{J/\psi}^{\mu}\varepsilon_{_V}^{\ast\alpha}p_{J/\psi}^\nu
p_{_{V}}^\beta p_3'\cdot p_{_{P}}\nonumber \\
&&+64\varepsilon_{J/\psi}^{\mu}\varepsilon_{_V}^{\ast\alpha}p_{J/\psi}^\nu
p_{_{P}}^\beta p_3'\cdot
p_{_{V}}-64\varepsilon_{J/\psi}^{\nu}\varepsilon_{_V}^{\ast\alpha}p_3'^\mu
p_{_{P}}^\beta p_{J/\psi}\cdot
p_{_{V}}+64\varepsilon_{J/\psi}^{\mu}\varepsilon_{_V}^{\ast\beta}p_{J/\psi}^\alpha
p_3'^\nu p_{_{V}}\cdot p_{_{P}}]\nonumber \\
&&+[D_\theta(m_q,u,v)\varepsilon_{\mu\nu\alpha\beta}
(96\varepsilon_{J/\psi}^\mu\varepsilon_{_V}^{\ast\nu}p_{_P}^\alpha
p_{_V}^\beta
p_{J/\psi}^\theta-32\varepsilon_{J/\psi}^\mu\varepsilon_{_V}^{\ast\alpha}p_{J/\psi}^\nu
p_{_V}^\beta
p_{_P}^\theta-32\varepsilon_{J/\psi}^\mu\varepsilon_{_V}^{\ast\alpha}p_{J/\psi}^\nu
p_{_P}^\beta p_{_V}^\theta)\nonumber
\\
&&+D^\nu(m_q,u,v)\varepsilon_{\mu\nu\alpha\beta}(32\varepsilon_{J/\psi}^\mu\varepsilon_{_V}^{\ast\alpha}
p_{_V}^\beta p_{J/\psi}\cdot
p_{_P}-32\varepsilon_{J/\psi}^\mu\varepsilon_{_V}^{\ast\alpha}
p_{_P}^\beta p_{J/\psi}\cdot p_{_V}\nonumber
\\
&&+32\varepsilon_{J/\psi}^\mu\varepsilon_{_V}^{\ast\beta}
p_{J/\psi}^\alpha p_{_V}\cdot p_{_P})]\}
\end{eqnarray}
where $p_1'=-p_3, p_3'=p_4+p_6-p_2$.

For amplitude $\mathcal{A}^{2c1a}$, we have
\begin{eqnarray}
\mathcal{A}^{2c1a}&=&C^{2c1a}\widetilde{H}^{2c1a}(m_q,u,v)\Phi_{J/\psi}\Phi_{V}\Phi_{P}
\end{eqnarray}
with
\begin{eqnarray}
&&C^{2c1a}=\text{Tr}(T^aT^bT^c)\text{Tr}(T^aT^bT^c)\nonumber \\
&&\widetilde{H}^{2c1a}(m_q,u,v)={i\pi^2\over
(2\pi)^4}g_s^6({1\over 4N_C})^3{1\over (p_4+p_5)^2}\nonumber
\\
&&m_Qm_q\{E_0(m_q,u,v)\varepsilon_{\mu\nu\alpha\beta}[-32\varepsilon_{J/\psi}^\mu
\varepsilon_{_V}^{\ast\nu}p_{_P}^\alpha p_{_V}^\beta p_2'\cdot
p_{J/\psi}-32\varepsilon_{J/\psi}^\mu
\varepsilon_{_V}^{\ast\alpha}p_{J/\psi}^\nu p_{_V}^\beta p_2'\cdot
p_{_P}-32\varepsilon_{J/\psi}^\mu
\varepsilon_{_V}^{\ast\alpha}p_{J/\psi}^\nu p_{_P}^\beta p_2'\cdot
p_{_V}\nonumber \\
&&-32\varepsilon_{J/\psi}^\mu
\varepsilon_{_V}^{\ast\nu}p_{_P}^\alpha p_{_V}^\beta p_1'\cdot
p_{J/\psi}-32\varepsilon_{J/\psi}^\mu
\varepsilon_{_V}^{\ast\alpha}p_{J/\psi}^\nu p_{_V}^\beta p_1'\cdot
p_{_P}-32\varepsilon_{J/\psi}^\mu
\varepsilon_{_V}^{\ast\alpha}p_{J/\psi}^\nu p_{_P}^\beta p_1'\cdot
p_{_V}\nonumber \\
&&+32\varepsilon_{J/\psi}^\nu
\varepsilon_{_V}^{\ast\alpha}p_2'^\mu p_{_V}^\beta p_{J/\psi}\cdot
p_{_P} +32\varepsilon_{J/\psi}^\nu
\varepsilon_{_V}^{\ast\alpha}p_1'^\mu p_{_V}^\beta p_{J/\psi}\cdot
p_{_P}-32\varepsilon_{J/\psi}^\nu
\varepsilon_{_V}^{\ast\alpha}p_2'^\mu p_{_P}^\beta p_{J/\psi}\cdot
p_{_V}\nonumber \\
&&-32\varepsilon_{J/\psi}^\nu
\varepsilon_{_V}^{\ast\alpha}p_1'^\mu p_{_P}^\beta p_{J/\psi}\cdot
p_{_V}-32\varepsilon_{J/\psi}^\nu
\varepsilon_{_V}^{\ast\beta}p_2'^\mu p_{J/\psi}^\alpha p_{_P}\cdot
p_{_V}-32\varepsilon_{J/\psi}^\nu
\varepsilon_{_V}^{\ast\beta}p_1'^\mu p_{J/\psi}^\alpha p_{_P}\cdot
p_{_V}]\nonumber \\
&&+[E_{1\theta}(m_q,u,v)\varepsilon_{\mu\nu\alpha\beta}
(-64\varepsilon_{J/\psi}^\mu\varepsilon_{_V}^{\ast\nu}p_{_P}^\alpha
p_{_V}^\beta
p_{J/\psi}^\theta-64\varepsilon_{J/\psi}^\mu\varepsilon_{_V}^{\ast\alpha}p_{J/\psi}^\nu
p_{_V}^\beta
p_{_P}^\theta-64\varepsilon_{J/\psi}^\mu\varepsilon_{_V}^{\ast\alpha}p_{J/\psi}^\nu
p_{_P}^\beta p_{_V}^\theta)\nonumber
\\
&&+E^\nu_1(m_q,u,v)\varepsilon_{\mu\nu\alpha\beta}(-64\varepsilon_{J/\psi}^\mu\varepsilon_{_V}^{\ast\alpha}
p_{_V}^\beta p_{J/\psi}\cdot
p_{_P}+64\varepsilon_{J/\psi}^\mu\varepsilon_{_V}^{\ast\alpha}
p_{_P}^\beta p_{J/\psi}\cdot
p_{_V}+64\varepsilon_{J/\psi}^\mu\varepsilon_{_V}^{\ast\beta}
p_{J/\psi}^\alpha p_{_V}\cdot p_{_P})]\}\nonumber \\
\end{eqnarray}
where $p_1'=p_1, p_2'=p_1-p_4-p_5, p_4'=p_6$.

For amplitude $\mathcal{A}^{2c1b}$, we have
\begin{eqnarray}
\mathcal{A}^{2c1b}&=&C^{2c1b}\widetilde{H}^{2c1b}(m_q,u,v)\Phi_{J/\psi}\Phi_{V}\Phi_{P}
\end{eqnarray}
with
\begin{eqnarray}
&&C^{2c1b}=\text{Tr}(T^aT^bT^c)\text{Tr}(T^aT^bT^c)\nonumber \\
&&\widetilde{H}^{2c1b}(m_q,u,v)={i\pi^2\over
(2\pi)^4}g_s^6({1\over 4N_C})^3{1\over (p_4+p_5)^2}\nonumber
\\
&&m_Qm_q\{E_0(m_q,u,v)\varepsilon_{\mu\nu\alpha\beta}[-32\varepsilon_{J/\psi}^\mu
\varepsilon_{_V}^{\ast\nu}p_{_P}^\alpha p_{_V}^\beta p_2'\cdot
p_{J/\psi}+32\varepsilon_{J/\psi}^\mu
\varepsilon_{_V}^{\ast\alpha}p_{J/\psi}^\nu p_{_V}^\beta p_2'\cdot
p_{_P}+32\varepsilon_{J/\psi}^\mu
\varepsilon_{_V}^{\ast\alpha}p_{J/\psi}^\nu p_{_P}^\beta p_2'\cdot
p_{_V}\nonumber \\
&&-32\varepsilon_{J/\psi}^\mu
\varepsilon_{_V}^{\ast\nu}p_{_P}^\alpha p_{_V}^\beta p_1'\cdot
p_{J/\psi}+32\varepsilon_{J/\psi}^\mu
\varepsilon_{_V}^{\ast\alpha}p_{J/\psi}^\nu p_{_V}^\beta p_1'\cdot
p_{_P}+32\varepsilon_{J/\psi}^\mu
\varepsilon_{_V}^{\ast\alpha}p_{J/\psi}^\nu p_{_P}^\beta p_1'\cdot
p_{_V}\nonumber \\
&&+32\varepsilon_{J/\psi}^\nu
\varepsilon_{_V}^{\ast\alpha}p_2'^\mu p_{_V}^\beta p_{J/\psi}\cdot
p_{_P} +32\varepsilon_{J/\psi}^\nu
\varepsilon_{_V}^{\ast\alpha}p_1'^\mu p_{_V}^\beta p_{J/\psi}\cdot
p_{_P}-32\varepsilon_{J/\psi}^\nu
\varepsilon_{_V}^{\ast\alpha}p_2'^\mu p_{_P}^\beta p_{J/\psi}\cdot
p_{_V}\nonumber \\
&&-32\varepsilon_{J/\psi}^\nu
\varepsilon_{_V}^{\ast\alpha}p_1'^\mu p_{_P}^\beta p_{J/\psi}\cdot
p_{_V}+32\varepsilon_{J/\psi}^\nu
\varepsilon_{_V}^{\ast\beta}p_2'^\mu p_{J/\psi}^\alpha p_{_P}\cdot
p_{_V}+32\varepsilon_{J/\psi}^\nu
\varepsilon_{_V}^{\ast\beta}p_1'^\mu p_{J/\psi}^\alpha p_{_P}\cdot
p_{_V}]\nonumber \\
&&+[E_{1\theta}(m_q,u,v)\varepsilon_{\mu\nu\alpha\beta}
(-64\varepsilon_{J/\psi}^\mu\varepsilon_{_V}^{\ast\nu}p_{_P}^\alpha
p_{_V}^\beta
p_{J/\psi}^\theta+64\varepsilon_{J/\psi}^\mu\varepsilon_{_V}^{\ast\alpha}p_{J/\psi}^\nu
p_{_V}^\beta
p_{_P}^\theta+64\varepsilon_{J/\psi}^\mu\varepsilon_{_V}^{\ast\alpha}p_{J/\psi}^\nu
p_{_P}^\beta p_{_V}^\theta)\nonumber
\\
&&+E^\nu_1(m_q,u,v)\varepsilon_{\mu\nu\alpha\beta}(-64\varepsilon_{J/\psi}^\mu\varepsilon_{_V}^{\ast\alpha}
p_{_V}^\beta p_{J/\psi}\cdot
p_{_P}+64\varepsilon_{J/\psi}^\mu\varepsilon_{_V}^{\ast\alpha}
p_{_P}^\beta p_{J/\psi}\cdot
p_{_V}-64\varepsilon_{J/\psi}^\mu\varepsilon_{_V}^{\ast\beta}
p_{J/\psi}^\alpha p_{_V}\cdot p_{_P})]\}\nonumber \\
\end{eqnarray}
where $p_1'=p_1, p_2'=p_1-p_4-p_5, p_4'=p_6$.

For amplitude $\mathcal{A}^{2c2a}$, we have
\begin{eqnarray}
\mathcal{A}^{2c2a}&=&C^{2c2a}\widetilde{H}^{2c2a}(m_q,u,v)\Phi_{J/\psi}\Phi_{V}\Phi_{P}
\end{eqnarray}
with
\begin{eqnarray}
&&C^{2c2a}=\text{Tr}(T^aT^bT^c)\text{Tr}(T^bT^aT^c)\nonumber \\
&&\widetilde{H}^{2c2a}(m_q,u,v)=-{i\pi^2\over
(2\pi)^4}g_s^6({1\over 4N_C})^3{1\over (p_4+p_5)^2}\nonumber
\\
&&m_Qm_q\{E_0(m_q,u,v)\varepsilon_{\mu\nu\alpha\beta}[32\varepsilon_{J/\psi}^\mu
\varepsilon_{_V}^{\ast\nu}p_{_P}^\alpha p_{_V}^\beta p_2'\cdot
p_{J/\psi}+32\varepsilon_{J/\psi}^\mu
\varepsilon_{_V}^{\ast\alpha}p_{J/\psi}^\nu p_{_V}^\beta p_2'\cdot
p_{_P}-32\varepsilon_{J/\psi}^\mu
\varepsilon_{_V}^{\ast\alpha}p_{J/\psi}^\nu p_{_P}^\beta p_2'\cdot
p_{_V}\nonumber \\
&&+32\varepsilon_{J/\psi}^\mu
\varepsilon_{_V}^{\ast\nu}p_{_P}^\alpha p_{_V}^\beta p_1'\cdot
p_{J/\psi}+32\varepsilon_{J/\psi}^\mu
\varepsilon_{_V}^{\ast\alpha}p_{J/\psi}^\nu p_{_V}^\beta p_1'\cdot
p_{_P}-32\varepsilon_{J/\psi}^\mu
\varepsilon_{_V}^{\ast\alpha}p_{J/\psi}^\nu p_{_P}^\beta p_1'\cdot
p_{_V}\nonumber \\
&&-32\varepsilon_{J/\psi}^\nu
\varepsilon_{_V}^{\ast\alpha}p_2'^\mu p_{_V}^\beta p_{J/\psi}\cdot
p_{_P}-32\varepsilon_{J/\psi}^\nu
\varepsilon_{_V}^{\ast\alpha}p_1'^\mu p_{_V}^\beta p_{J/\psi}\cdot
p_{_P}-32\varepsilon_{J/\psi}^\nu
\varepsilon_{_V}^{\ast\alpha}p_2'^\mu p_{_P}^\beta p_{J/\psi}\cdot
p_{_V}\nonumber \\
&&-32\varepsilon_{J/\psi}^\nu
\varepsilon_{_V}^{\ast\alpha}p_1'^\mu p_{_P}^\beta p_{J/\psi}\cdot
p_{_V}-32\varepsilon_{J/\psi}^\nu
\varepsilon_{_V}^{\ast\beta}p_2'^\mu p_{J/\psi}^\alpha p_{_P}\cdot
p_{_V}-32\varepsilon_{J/\psi}^\nu
\varepsilon_{_V}^{\ast\beta}p_1'^\mu p_{J/\psi}^\alpha p_{_P}\cdot
p_{_V}]\nonumber \\
&&+[E_{1\theta}(m_q,u,v)\varepsilon_{\mu\nu\alpha\beta}
(64\varepsilon_{J/\psi}^\mu\varepsilon_{_V}^{\ast\nu}p_{_P}^\alpha
p_{_V}^\beta
p_{J/\psi}^\theta+64\varepsilon_{J/\psi}^\mu\varepsilon_{_V}^{\ast\alpha}p_{J/\psi}^\nu
p_{_V}^\beta
p_{_P}^\theta-64\varepsilon_{J/\psi}^\mu\varepsilon_{_V}^{\ast\alpha}p_{J/\psi}^\nu
p_{_P}^\beta p_{_V}^\theta)\nonumber
\\
&&+E^\nu_1(m_q,u,v)\varepsilon_{\mu\nu\alpha\beta}(64\varepsilon_{J/\psi}^\mu\varepsilon_{_V}^{\ast\alpha}
p_{_V}^\beta p_{J/\psi}\cdot
p_{_P}+64\varepsilon_{J/\psi}^\mu\varepsilon_{_V}^{\ast\alpha}
p_{_P}^\beta p_{J/\psi}\cdot
p_{_V}+64\varepsilon_{J/\psi}^\mu\varepsilon_{_V}^{\ast\beta}
p_{J/\psi}^\alpha p_{_V}\cdot p_{_P})]\}\nonumber \\
\end{eqnarray}
where $p_1'=p_1, p_2'=p_1-p_4-p_5, p_4'=p_6$.

For amplitude $\mathcal{A}^{2c2b}$, we have
\begin{eqnarray}
\mathcal{A}^{2c2b}&=&C^{2c2b}\widetilde{H}^{2c2b}(m_q,u,v)\Phi_{J/\psi}\Phi_{V}\Phi_{P}
\end{eqnarray}
with
\begin{eqnarray}
&&C^{2c2b}=\text{Tr}(T^aT^bT^c)\text{Tr}(T^bT^aT^c)\nonumber \\
&&\widetilde{H}^{2c2b}(m_q,u,v)=-{i\pi^2\over
(2\pi)^4}g_s^6({1\over 4N_C})^3{1\over (p_4+p_5)^2}\nonumber
\\
&&m_Qm_q\{E_0(m_q,u,v)\varepsilon_{\mu\nu\alpha\beta}[32\varepsilon_{J/\psi}^\mu
\varepsilon_{_V}^{\ast\nu}p_{_P}^\alpha p_{_V}^\beta p_2'\cdot
p_{J/\psi}+32\varepsilon_{J/\psi}^\mu
\varepsilon_{_V}^{\ast\alpha}p_{J/\psi}^\nu p_{_V}^\beta p_2'\cdot
p_{_P}-32\varepsilon_{J/\psi}^\mu
\varepsilon_{_V}^{\ast\alpha}p_{J/\psi}^\nu p_{_P}^\beta p_2'\cdot
p_{_V}\nonumber \\
&&+32\varepsilon_{J/\psi}^\mu
\varepsilon_{_V}^{\ast\nu}p_{_P}^\alpha p_{_V}^\beta p_1'\cdot
p_{J/\psi}+32\varepsilon_{J/\psi}^\mu
\varepsilon_{_V}^{\ast\alpha}p_{J/\psi}^\nu p_{_V}^\beta p_1'\cdot
p_{_P}-32\varepsilon_{J/\psi}^\mu
\varepsilon_{_V}^{\ast\alpha}p_{J/\psi}^\nu p_{_P}^\beta p_1'\cdot
p_{_V}\nonumber \\
&&+32\varepsilon_{J/\psi}^\nu
\varepsilon_{_V}^{\ast\alpha}p_2'^\mu p_{_V}^\beta p_{J/\psi}\cdot
p_{_P}+32\varepsilon_{J/\psi}^\nu
\varepsilon_{_V}^{\ast\alpha}p_1'^\mu p_{_V}^\beta p_{J/\psi}\cdot
p_{_P}+32\varepsilon_{J/\psi}^\nu
\varepsilon_{_V}^{\ast\alpha}p_2'^\mu p_{_P}^\beta p_{J/\psi}\cdot
p_{_V}\nonumber \\
&&+32\varepsilon_{J/\psi}^\nu
\varepsilon_{_V}^{\ast\alpha}p_1'^\mu p_{_P}^\beta p_{J/\psi}\cdot
p_{_V}+32\varepsilon_{J/\psi}^\nu
\varepsilon_{_V}^{\ast\beta}p_2'^\mu p_{J/\psi}^\alpha p_{_P}\cdot
p_{_V}+32\varepsilon_{J/\psi}^\nu
\varepsilon_{_V}^{\ast\beta}p_1'^\mu p_{J/\psi}^\alpha p_{_P}\cdot
p_{_V}]\nonumber \\
&&+[E_{1\theta}(m_q,u,v)\varepsilon_{\mu\nu\alpha\beta}
(64\varepsilon_{J/\psi}^\mu\varepsilon_{_V}^{\ast\nu}p_{_P}^\alpha
p_{_V}^\beta
p_{J/\psi}^\theta+64\varepsilon_{J/\psi}^\mu\varepsilon_{_V}^{\ast\alpha}p_{J/\psi}^\nu
p_{_V}^\beta
p_{_P}^\theta-64\varepsilon_{J/\psi}^\mu\varepsilon_{_V}^{\ast\alpha}p_{J/\psi}^\nu
p_{_P}^\beta p_{_V}^\theta)\nonumber
\\
&&+E^\nu_1(m_q,u,v)\varepsilon_{\mu\nu\alpha\beta}(-64\varepsilon_{J/\psi}^\mu\varepsilon_{_V}^{\ast\alpha}
p_{_V}^\beta p_{J/\psi}\cdot
p_{_P}-64\varepsilon_{J/\psi}^\mu\varepsilon_{_V}^{\ast\alpha}
p_{_P}^\beta p_{J/\psi}\cdot
p_{_V}-64\varepsilon_{J/\psi}^\mu\varepsilon_{_V}^{\ast\beta}
p_{J/\psi}^\alpha p_{_V}\cdot p_{_P})]\}\nonumber \\
\end{eqnarray}
where $p_1'=p_1, p_2'=p_1-p_4-p_5, p_4'=p_6$.

\end{document}